\documentclass{aa} 

\usepackage{graphicx}
\usepackage{tabularx}
\usepackage{placeins}
\usepackage[english]{babel}
\usepackage{array}
\usepackage{capt-of}
\usepackage[normalem]{ulem}

\newcolumntype{L}[1]{>{\raggedright\let\newline\\\arraybackslash\hspace{0pt}}m{#1}}
\newcolumntype{C}[1]{>{\centering\let\newline\\\arraybackslash\hspace{0pt}}m{#1}}
\newcolumntype{R}[1]{>{\raggedleft\let\newline\\\arraybackslash\hspace{0pt}}m{#1}}

\usepackage{txfonts}
\usepackage[colorlinks=true, allcolors=blue]{hyperref}
\def\kms{km~s$^{-1}$}
\def\nH{$n_\mathrm{H}$}
\def\dz{$\zeta_{\mathrm{H}_2}$}
\def\G0{$G_0$}

\def\Vs{$V_\mathrm{s}$}
\graphicspath{{folder/}{Figures/}}

\begin{document}

   \title{Modeling accretion shocks at the disk-envelope interface}

   \subtitle{Sulfur chemistry}

   \author{
          M. L. van Gelder\inst{1}
          \and
          B. Tabone\inst{1}
          \and
          E. F. van Dishoeck\inst{1,2}
          \and
          B. Godard\inst{3,4}
          }
\institute{
         Leiden Observatory, Leiden University, PO Box 9513, 2300RA Leiden, The Netherlands \\
         \email{vgelder@strw.leidenuniv.nl}
         \and
         Max Planck Institut f\"ur Extraterrestrische Physik (MPE), Giessenbachstrasse 1, 85748 Garching, Germany
         \and
         Observatoire de Paris, Universit\'e PSL, Sorbonne Universit\'e, LERMA, 75014 Paris, France
         \and
         Laboratoire de Physique de l'\'Ecole normale sup\'erieure, ENS, Universit\'e PSL, CNRS, Sorbonne Universit\'e, Universit\'e de Paris, 75005 Paris, France
             }

   \date{Received XXX; accepted XXX}

% \abstract{}{}{}{}{} 
% 5 {} token are mandatory
 
  \abstract
  % context heading (optional), leave it empty if necessary 
   {As material from an infalling protostellar envelope hits the forming disk, an accretion shock may develop which could (partially) alter the envelope material entering the disk. Observations with the Atacama Large Millimeter/submillimeter Array (ALMA) indicate that emission originating from warm SO and SO$_2$ might be good tracers of such accretion shocks.}
  % aims heading
   {The goal of this work is to test under what shock conditions the abundances of gas-phase SO and SO$_2$ increase in an accretion shock at the disk-envelope interface.}
  % methods heading
   {Detailed shock models including gas dynamics are computed using the Paris-Durham shock code for non-magnetized $J$-type accretion shocks in typical inner envelope conditions. The effect of pre-shock density, shock velocity, and strength of the ultraviolet (UV) radiation field on the abundance of warm SO and SO$_2$ is explored. Compared with outflows, these shocks involve higher densities ($\sim10^{7}$~cm$^{-3}$), lower shock velocities ($\sim$ few~\kms), and large dust grains ($\sim0.2$~$\mu$m) and thus probe a different parameter space.}
  % results heading
   {Warm gas-phase chemistry is efficient in forming SO under most $J$-type shock conditions considered. 
	In lower-velocity ($\sim3$~\kms) shocks, the abundance of SO is increased through subsequent reactions starting from thermally desorbed CH$_4$ toward H$_2$CO and finally SO. 
    In higher velocity ($\gtrsim4$~\kms) shocks, both SO and SO$_2$ are formed through reactions of OH and atomic S. 
    The strength of the UV radiation field is crucial for SO and in particular SO$_2$ formation through the photodissociation of H$_2$O.
   Thermal desorption of SO and SO$_2$ ice is only relevant in high-velocity ($\gtrsim5$~\kms) shocks at high densities ($\gtrsim10^7$~cm$^{-3}$).
   Both the composition in the gas phase, in particular the abundances of atomic S and O, and in ices such as H$_2$S, CH$_4$, SO, and SO$_2$ play a key role in the abundances of SO and SO$_2$ that are reached in the shock.
%   In strongly magnetized environments, the ion-neutral decoupling is weak resulting in non-physical long $C$-type shocks.
   % $CJ$-type shocks under intermediate strong magnetic fields are physically possible, but computing these is difficult due to numerical method becoming unstable under these conditions.
   }
   {Warm emission from SO and SO$_2$ is a possible tracer of accretion shocks at the disk-envelope interface as long as a local UV field is present. 
   Observations with ALMA at high-angular resolution could provide further constraints given that other key species for the gas-phase formation of SO and SO$_2$, H$_2$S and H$_2$CO, are also covered. 
   Moreover, the {\it James Webb} Space Telescope will give access to other possible slow, dense shock tracers such as H$_2$, H$_2$O, and [S\,{\sc i}]~$25~\mu$m.}


   \maketitle
%
%________________________________________________________________

\section{Introduction}
\label{sec:introduction}
It is currently still unknown how much reprocessing of material occurs during the collapse from a prestellar core to a protostar and disk. Two scenarios can be considered: the inheritance scenario where the chemical composition is conserved from cloud to disk, or the (partial) reset scenario where the envelope material is modified during its trajectory from envelope to disk. Some inheritance is suggested by the rough similarity between cometary and interstellar ices \citep{Mumma2011,Drozdovskaya2018,Drozdovskaya2019}. On the other hand, some modification of envelope material is expected due to the increase of temperature and stellar ultraviolet (UV) radiation during infall \citep[e.g.,][]{Aikawa1999,Visser2009}. Moreover, as the infalling envelope hits the disk, an accretion shock may develop which will alter the composition of the material flowing into the disk, and therefore (partially) reset the chemistry. In the most extreme case, reset implies complete vaporization of all molecules back to atoms with subsequent reformation. Milder versions of reset include sputtering of ices, and gas and ice chemistry modifying abundances.

%In this paper, accretion shocks are modeled under various physical conditions to attain the effect on the composition and, in particular, of suggested tracers such as SO and SO$_2$.
%In this paper, low-velocity ($<10$~\kms) accretion shocks in dense environments ($10^{5-8}$~cm$^{-3}$) are modeled, focusing on SO and SO$_2$ as possible tracers of such shock.

Already in the earliest Class~0 phase of the low-mass star-formation process, an accretion disk around the young protostar is formed \citep[e.g.,][]{Tobin2012,Murillo2013}. The impact of infalling envelope material onto the disk can cause a shock, raising temperatures of gas and dust at the disk-envelope interface to values much higher than from heating by stellar photons alone \citep{Draine1983,Neufeld1994}. These low-velocity ($\sim$ few \kms) accretion shocks at high densities ($\sim10^7$~cm$^{-3}$) are widely found in models and simulations \citep[e.g.,][]{Cassen1981,Li2013,Miura2017}, and often invoked in Solar System formation to explain phenomena like noble gas trapping \citep{Owen1992}. They are most powerful in the earliest stages when the disk is still small ($\lesssim 10$~AU). However, that infalling material may eventually end up in the protostar and not in the disk. As the embedded disk reaches a size of (several) tens of AU, the envelope material entering the disk will remain there and eventually form the building blocks of planets \citep[e.g.,][]{Harsono2018,Manara2018,Tychoniec2020}. 

These accretion shocks have not yet been undisputedly detected. Typical optical shock tracers such as [O\,{\sc i}]~$6300~\AA$ and [S\,{\sc ii}]~$6371~\AA$ are more sensitive to high-velocity shocks \citep[\mbox{$\gtrsim20$~\kms}, e.g.,][]{Podio2011,Banzatti2019} and difficult to detect in protostellar environments due to the high extinction by the surrounding envelope. Observations of mid-infrared (MIR) shock tracers such as H$_2$O, high-J CO, [S\,{\sc i}]~$25~\mu$m, and [O\,{\sc i}]~$63~\mu$m suffer from outflow contamination in low-spatial resolution space-based observations \citep[e.g.,][]{Kristensen2012,Nisini2015,Riviere-Marichalar2016} and from the Earth's atmosphere for ground-based observations; here the launch of the {\it James Webb} Space Telescope (JWST) will provide a solution. At sub-mm wavelengths, classical diagnostics of shocks include sputtering or grain-destruction products such as CH$_3$OH and SiO \citep[e.g.,][]{Caselli1997,Schilke1997,Gusdorf2008_1,Gusdorf2008_2,Guillet2009,Suutarinen2014}. However, usually these species are only observed in relatively high-velocity outflows and jets through broad emission lines \citep[e.g.,][]{Tychoniec2019,Taquet2020,Codella2020}. 

Sulfur bearing species such as SO and SO$_2$ have been suggested as possible accretion shock tracers. An enhancement of warm SO emission at the centrifugal barrier, the interface between envelope and disk, has been observed in the L1527 Class~0/I system with the Atacama Large Millimeter/submillimeter Array (ALMA) at $\sim100$~AU scales \citep{Sakai2014,Sakai2017}. A similar rotating structure in SO has been seen toward the Class~I system Elias~29 \citep{Oya2019}. Equivalently, warm SO$_2$ emission possibly related to an accretion shock was observed toward the B335 Class~0 protostar \citep{Bjerkeli2019} and a few Class~I sources \citep{Oya2019,Elizabeth2019}. However, both SO and SO$_2$ are also seen related to outflow activity, both in large-scale outflows \citep[e.g.,][]{Codella2014,Taquet2020} as in disk winds on smaller scales \citep[e.g.,][]{Tabone2017,Lee2018}. 
%Warm SO$_2$ related to thermal ice sublimation was also observed at IR wavelengths \citep[e.g.,][]{Dungee2018}. 
Moreover, high-angular resolution observations ($\sim30$~AU) of the Class~I source TMC1A show narrow SO emission lines coming from a ring-shaped morphology which may be linked to the warm inner envelope \citep{Harsono2021}. Emission from warm SO and SO$_2$ is thus not an unambiguous tracer of accretion shocks; comparison to shock models is necessary to make robust conclusions.

In shocks, the gas temperature readily rises to $>100$~K \citep{Draine1983,Flower2003,Godard2019}, enough to ignite gas-phase formation of SO and SO$_2$ \citep[e.g.,][]{Prasad1980,Hartquist1980}. A key species for this is the OH radical, which is efficiently produced in shocks through the endothermic reaction between H$_2$ and atomic oxygen, but also through photodissociation of H$_2$O. The OH radical reacts with atomic sulfur to form SO and subsequently to form SO$_2$. %Hence, at temperatures above $\gtrsim100$~K and in the presence of atomic sulfur and oxygen in the gas phase, SO and SO$_2$ may be formed in the gas phase in accretion shocks. 
Moreover, SO can also be formed from the SH radical reaction with atomic oxygen \citep{Hartquist1980}. Since the gas-phase chemistry of SO and SO$_2$ in shocks is dependent on radicals such as OH and SH, the abundance of these species is not solely determined through the temperature in the shock but also through the strength of the local UV field.

Alternatively, thermal desorption of SO and SO$_2$ ice in shocks can enhance the gas-phase abundances of SO and SO$_2$. 
In interstellar ices, OCS is the only securely identified sulfur-bearing species \citep{Geballe1985,Palumbo1995}, with a tentative detection for SO$_2$ \citep{Boogert1997,Zasowski2009}. H$_2$S is suggested as the main sulfur carrier in the ices \citep[e.g.,][]{Vidal2017}, but so far only upper limits could be derived \citep{Jimenez-Escobar2011}. To date, SO ice has not yet been detected. All aforementioned species are observed in cometary ices such as 67P/Churyumov–Gerasimenko \citep{Calmonte2016,Rubin2019,Altwegg2019}.
%However, only SO$_2$ ice has been tentatively detected in interstellar ices \citep{Boogert1997}.
\citet{Miura2017} modeled low-velocity (2~\kms) accretion shocks taking into account the dust dynamics but without any gas-phase chemistry included and could only explain the observations of \citet{Sakai2014} if SO was thermally desorbed while adopting dust grains that were smaller than 0.1~$\mu$m. Sputtering seems to be irrelevant as this becomes only efficient at larger shock velocities \citep[$>10$~\kms; e.g.,][]{Aota2015}. 
However, to thermally desorb sulfur-bearing ices such as SO and SO$_2$ from the grains, the dust temperature has to reach values on the order of 40--70~K.

%To explain the observations of warm SO, \citet{Miura2017} modeled low velocity (2~\kms) accretion shocks, taking into account the dust dynamics. They could only explain the observations of \citet{Sakai2014} if SO was thermally desorbed and if the dust grains were smaller than typical ISM sizes of 0.1~$\mu$m. Sputtering seems to be irrelevant as this becomes only efficient at larger shock velocities \citep[$>10$~\kms; e.g.,][]{Aota2015}. However, to thermally desorb SO and SO$_2$ from the grains the dust temperature has to reach value on the order of 40--60~K at typical inner envelope densities ($n\sim10^{6-8}$~cm$^{-3}$). 
%
%In weaker shocks, the dust temperature might not increase significantly enough to thermally desorb SO and SO$_2$. Nevertheless, the gas temperature can still rise to $>100$~K, enough to ignite gas-phase formation of these species \citep{Prasad1980}. A key species for this is the OH radical, which is efficiently produced in shocks through the endothermic reaction between H$_2$ and atomic oxygen \citep{Cohen1983}. The OH radical reacts with atomic sulfur to form SO and subsequently to form SO$_2$. Hence, at temperatures above $\gtrsim100$~K and in the presence of atomic sulfur and oxygen in the gas phase, SO and SO$_2$ may be formed in the gas phase in accretion shocks.

In this paper, we present a grid of irradiated low-velocity ($<10$~\kms) $J$-type accretion shock models at typical inner envelope densities ($10^{5}-10^{8}$~cm$^{-3}$). This is a different parameter space than usually explored in, for example, higher-velocity ($>10$~\kms) outflows at lower densities ($\sim10^4$~cm$^{-3}$). 
%, computed using the Paris-Durham shock code \citep{Flower2003,Lesaffre2013,Godard2019}. 
The goal is to test under what shock conditions the abundance of warm SO and SO$_2$ is significantly increased. 
%A new part of parameter space, in particular low-velocity ($<10$~\kms) shocks in high-density environments ($n \gtrsim 10^{5}$~cm$^{-3}$), is explored. 
In Sect.~\ref{sec:model}, the model and input parameters are introduced. We present the results of our analysis in Sect.~\ref{sec:results}, after which we discuss these results in Sect.~\ref{sec:discussion}. Our conclusions are summarized in Sect.~\ref{sec:summary}.

\section{Accretion shock model}
\label{sec:model}

%\mvg{Restructure: 2.3 $->$ 2.1, merge and shorten a lot 2.1 with 2.2 in 2.1. Refer to Fig 1 in Sect.in new Section 2.1 and Fig.2.2.}
\subsection{Shock model}
\label{subsec:model}
The accretion shock at the disk-envelope interface is modeled using a slightly modified version of the Paris-Durham shock code \citep{Flower2003,Lesaffre2013,Godard2019}. This publicly available numerical tool\footnote{\url{https://ism.obspm.fr}} computes the dynamical, thermal, and chemical structure of stationary plane parallel shock waves. The models therefore represent a continuous inflow of envelope material onto the disk and thus a continuous sequence of shocks. The location of the accretion shock is expected to move outward as the disk grows and evolves \citep{Visser2010}. 

In this paper, only stationary non-magnetized $J$-type accretion shocks are presented. The presence of a magnetic field could result in $C$ or $CJ$-type shocks \citep[e.g.,][]{Draine1980,Flower2003}, see Appendix~\ref{app:mag_shock} for details. However, the length of $C$-type accretion shocks is strongly dependent on the initial conditions and the size of dust grains and may reach envelope scales of $\sim1000$~AU, which is not consistent with the emission of SO and SO$_2$ seen on $<100$~AU scales \citep[e.g.,][]{Sakai2014,Sakai2017,Elizabeth2019,Oya2019}. $CJ$-type shocks could be relevant in intermediate magnetized inner envelopes \citep[$\sim0.1$~mG;][Appendix~\ref{app:mag_shock}]{Hull2017}, but to compute $CJ$-type shocks across our specific parameter space (low velocities, high densities), the Paris-Durham code needs to be tuned, which is beyond the scope of this work. Nevertheless, the results of $CJ$ and $C$ type shocks are not expected to be different from $J$-type shocks in terms of chemistry.

In the following sections, the main physical and chemical quantities and properties that are included in the model are introduced. In particular, the most relevant physical and chemical processes and updates on the version of the code presented by \citet{Godard2019} are highlighted. 

\subsubsection{Geometry}
\label{subsubsec:geometry}
\begin{figure}
\includegraphics[width=1.0\linewidth]{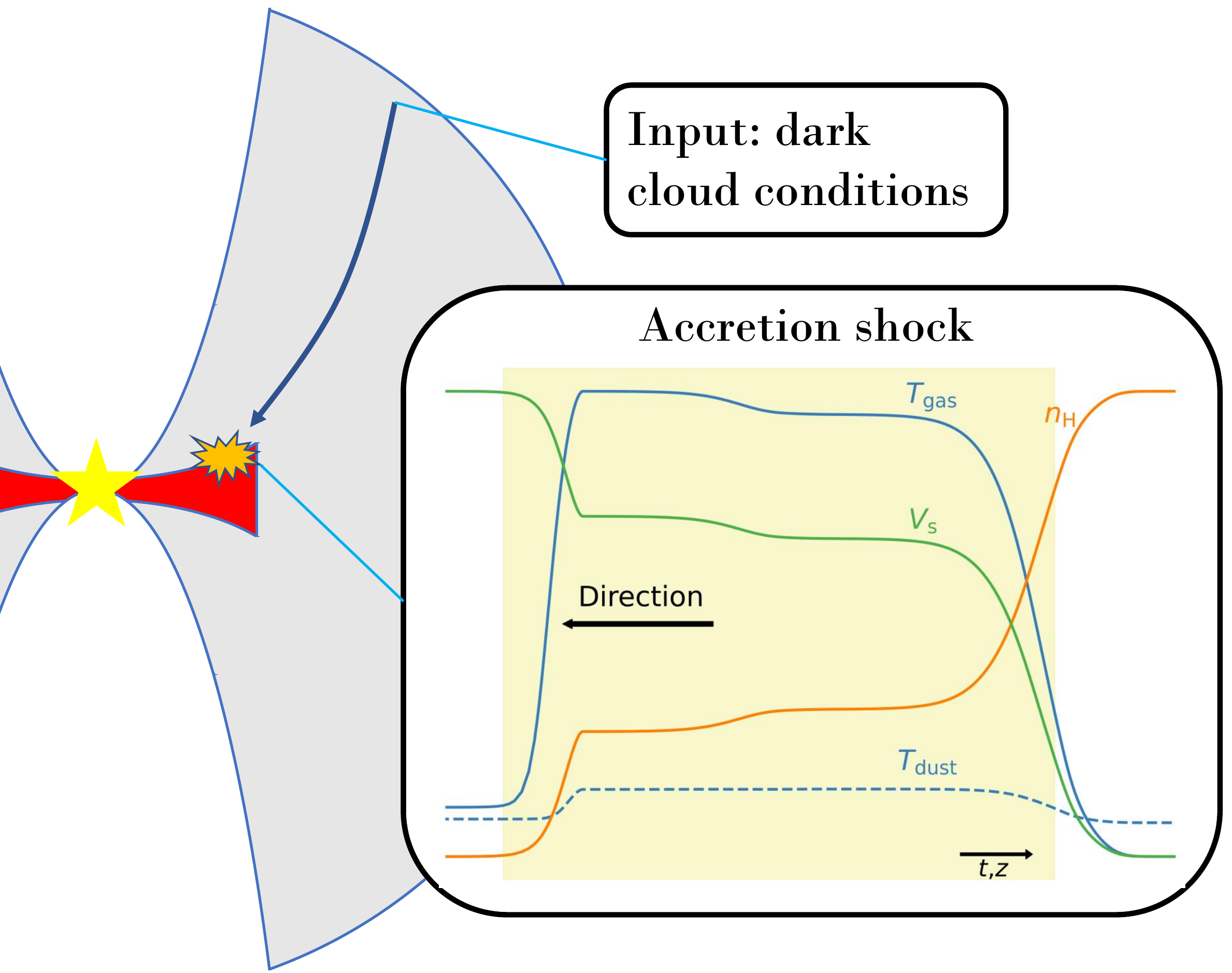}
\caption{Physical structure of a protostellar envelope. 
Starting with dark cloud conditions, 
%the pre-shock material is irradiated for $100$~years with an incident radiation field consisting of both the protostellar UV radiation field (strength parametrized by \G0) and the infrared radiation originating from the disk (constant for all models). Subsequently, 
an accretion shock model is calculated. The shock profile in the bottom panel is shown on a logarithmic scale, with the yellow region indicating the shocked region.}
\label{fig:physical_structure}
\end{figure}

The shock is calculated following a stationary plane parallel structure, see Fig.~\ref{fig:physical_structure}. A shock front propagates with a velocity \Vs\ in the direction of negative distance $z$.
% under the influence of a transverse magnetic field with strength $B$. 
The entire structure is irradiated with an isotropic radiation field equal to the isotropic standard interstellar radiation field (ISRF) scaled with a factor \G0\ \citep{Mathis1983}, as was implemented by \citet{Godard2019}. 
In protostellar systems, the UV radiation can originate both from the accretion onto the protostar itself or from shocks in the bipolar jets \citep[e.g.,][]{vanKempen2009,Yildiz2012,Benz2016}. Moreover, the extinction $A_\mathrm{V}$ between the source of UV irradiation and the shock at the disk-envelope interface can be as high as $\sim10$~mag \citep[e.g.,][]{Drozdovskaya2015} and is assumed to be taken into account with the choice of \G0. 
Therefore, the exact source of UV radiation is irrelevant here as long as a reasonable range in $G_0$ is covered and since the radiation field is assumed to be isotropic.
A diluted black body of 100~K representing far-infrared (FIR) dust emission from the accretion disk is added to the radiation field. Assuming an emitting region of $4.5$~AU at a distance of $100~$AU, the black body is diluted with a factor $W=5\times10^{-4}$ \citep[for details, see Appendix~A of][]{Tabone2020}, and is constant for all models. Additionally, secondary UV photons created through the collisional excitation of H$_2$ by electrons produced by cosmic ray ionization are included. 

\subsubsection{Radiative transfer and photon processes}
%The attenuation of the radiation field on the length scales of the shock is assumed to be negligible. However, 
Photodissociation and photoionization processes of both the gas and dust are taken into account. The cross sections are taken from the Leiden database \citep{Heays2017}. To save computation time, the photoreaction cross sections were only used for the photodissociation of CH, CH$_3$, CH$_4$, NH, CN, O$_2$, OH, H$_2$O, H$_2$CO, SH, H$_2$S, CS, OCS, SO, and SO$_2$, and the photoionization of C, S, CH, CH$_3$, CH$_4$, O$_2$, OH, H$_2$O, H$_2$CO, SH, H$_2$S, and SO. Photodissociation and photoionization of all other species are also included in the chemistry, but using the analytical expression of \citet{Heays2017}. Due to the self-shielding properties of H$_2$, CO, and N$_2$, photodissociation of these species is assumed to be negligible. The destruction of atoms and molecules by far-ultraviolet photons between $911~\AA$ and $2400~\AA$ are calculated including both continuum and discrete processes.

%The attenuation of the radiation field in the plane-parallel shock wave is solved assuming only absorption processes and neglecting emission and diffusion processes \citep{Godard2019}. Both the gas and dust contribute to the absorption coefficient. The photodissociation and photoionization cross sections of gas-phase species are taken from the Leiden database \citep{Heays2017}. However, to save computation time, the cross section of the gas was calculated taking only into account the photodissociation of CH, CH$_3$, CH$_4$, NH, CN, O$_2$, OH, H$_2$O, H$_2$CO, SH, H$_2$S, CS, OCS, SO, and SO$_2$, and the photoionization of C, S, CH, CH$_3$, CH$_4$, O$_2$, OH, H$_2$O, H$_2$CO, SH, H$_2$S, and SO. Photodissociation and photoionization of all other species are not included in the radiative transfer, but are included in the chemistry. Due to the self-shielding properties of H$_2$, CO, and N$_2$, photodissociation of these species is assumed to be negligible. The destruction of atoms and molecules by FUV photons between $911~\AA$ and $2400~\AA$ are calculated including both continuum and discrete processes. 

\subsubsection{Grains and PAHs}
The grains are assumed to have grown in protostellar envelopes to larger sizes than the typical interstellar medium (ISM) distribution \citep[e.g.,][]{Guillet2007,Miotello2014,Harsono2018,Galametz2019}. Here, we assume that all grains have an initial core size of $0.2~\mu$m \citep{Guillet2020}, and are either neutral, or single negatively or positively charged. Moreover, the dust grains are assumed to be dynamically coupled to the gas.
%The charge of grains can be altered via various mechanisms such as photoelectric ejection, the attachment of an electron, or a charge exchange between a dust grain and ions. 
Grain-grain interactions are not included since vaporization and shattering of dust grains are only relevant for $J$-type shocks moving at high velocities \citep[$> 20$~\kms;][]{Guillet2009}. Dust coagulation could potentially be relevant in changing the grain size distribution \citep{Guillet2011,Guillet2020}, but is not included here.

Other large species are polycyclic aromatic hydrocarbons (PAHs), which are modeled here as single particles with a size of $6.4~\AA$ \citep{Flower2003}. In most studies of interstellar shocks, the abundance of PAHs is set to the typical value of the ISM \citep[$\sim10^{-6}$ with respect to the proton density $n_\mathrm{H} = 2n(\mathrm{H}_2) + n(\mathrm{H}) + n(\mathrm{H}^+)$;][]{Draine2007}. However, given that PAHs freeze out in dense clouds, the abundance of PAHs in the gas is set to $\sim10^{-8}$ with respect to \nH\ \citep{Geers2009}. 
%This means that the charge carrying capacity of PAHs also drops by two orders of magnitude.

\subsubsection{Chemical network}
The chemical network adopted in this work consists of 143 species which can take place in about a thousand reactions, both in the gas phase and in the solid state. The network is largely similar to that used by \citet{Flower2015} and \citet{Godard2019}, with the addition of important sulfur bearing ices such as SO and SO$_2$ and their desorption and adsorption reactions. Moreover, the binding energies of all ice species were updated \citep[e.g., 1800~K, 3010~K, and 2290~K for SO, SO$_2$, and H$_2$S, respectively;][]{Penteado2017}. For a density of $10^8$~cm$^{-3}$, this results in a sublimation temperature of 37~K, 62~K, and 47~K for SO, SO$_2$, and H$_2$S, respectively. For simplicity, grain surface reactions have been disabled since the effect of these reactions on timescales of the shock is negligible. 
%The default input abundances are presented in Appendix~\ref{app:input_abundances}, where the total elemental abundances are set equal to those of \citet{Flower2003}. 
Sputtering of ices is not relevant here since in $J$-type shocks all material is coupled in one fluid.
%Dust processes are not included in the chemical network.

\subsubsection{Heating and cooling}
The thermal balance is calculated consistently throughout the shock. The main sources of gas heating in non-magnetized $J$-type shocks are through viscous stresses and compression of the medium at the shock front where $T_\mathrm{gas}$ increases adiabatically over a few mean free paths. A second form of heating occurs through irradiation coming from the incident local UV radiation field and through gas-dust thermal coupling. 

The gas is cooled through atomic lines of C, N, O, S, and Si, ionic lines of C$^+$, N$^+$, O$^+$, S$^+$, and Si$^+$, and rotational and rovibrational lines of H$_2$, OH, H$_2$O, NH$_3$, $^{12}$CO, and $^{13}$CO \citep{Flower2003,Lesaffre2013,Godard2019}. All cooling by atomic and ionic lines, as well as rovibrational lines of H$_2$ are calculated in the optically thin limit. The cooling of OH is calculated using an analytical cooling function \citep{Lebourlot2002}, which holds up to densities of $10^{10}$~cm$^{-3}$. However, for NH$_3$, local thermodynamic equilibrium (LTE) effects become relevant already for densities $\gtrsim10^{8}$~cm$^{-3}$. The cooling through pure rotational lines of NH$_3$ has therefore been recalculated numerically using the \textsc{RADEX} software \citep{vanderTak2007} in the optically thin limit, see Appendix~\ref{app:NH3_cool}. Cooling by rovibrational lines of H$_2$O, $^{12}$CO, and $^{13}$CO is computed using the tabulated values of \citet{Neufeld1993} which take into account opacity effects. Additionally, the gas can be cooled though collisions with dust grains which subsequently radiate away the heat.

The temperature of the dust is calculated from the equilibrium between heating through absorption of photons, thermal collisions with the gas which can both heat and cool the dust, and cooling through infrared emission \citep{Godard2019}. The FIR radiation field of the 100~K black body is of particular importance for the heating for dust grains in the pre and post-shock regimes as it sets a minimum dust temperature of $\sim22$~K. A dust temperature of $<20$~K would imply that even the most volatile species like CO and N$_2$ are frozen out onto dust grains. However, observations of embedded protostellar systems suggest that CO and N$_2$ are not frozen out in the inner regions of such systems \citep[e.g.,][]{vantHoff2018}. A higher dust temperature would result in more ice species being desorbed prior to the shock, but the effect on the shock structure and chemistry is negligible.

\begin{figure*}
\includegraphics[width=0.49\linewidth]{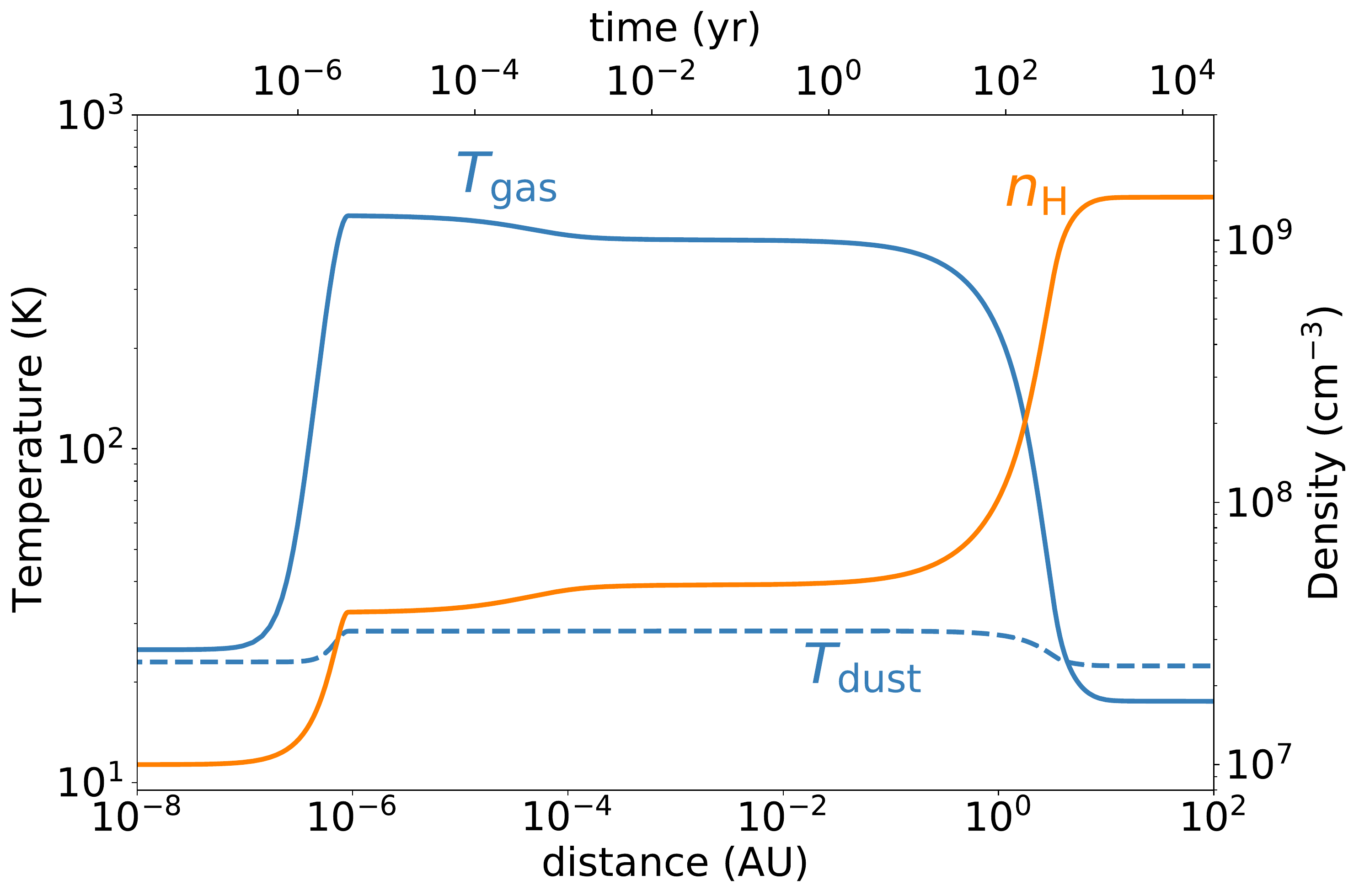}
\includegraphics[width=0.49\linewidth]{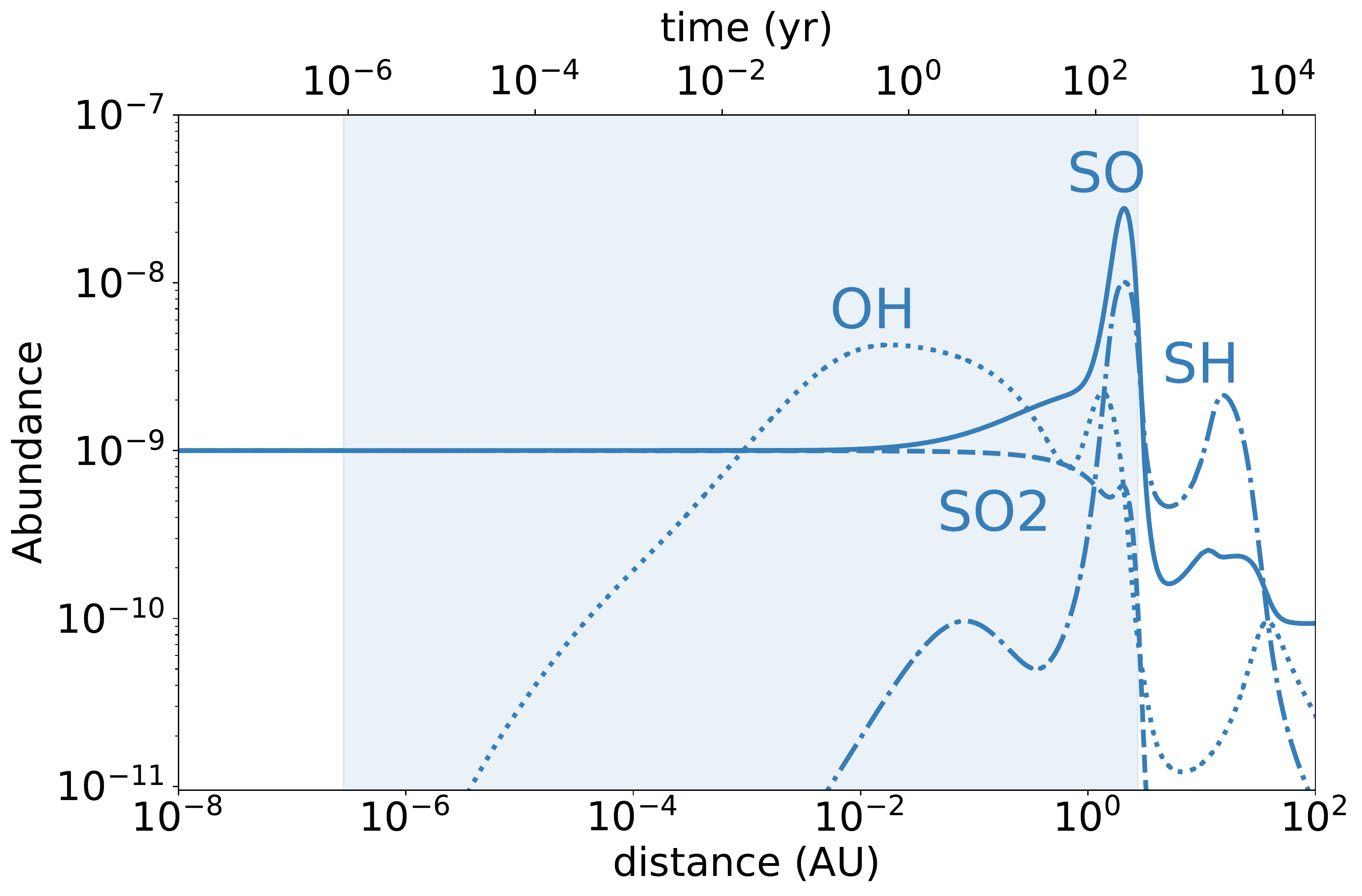}
\caption{
{\it Left:} Temperature and density structure in the fiducial $J$-type shock. The shock starts at 0~AU and ends at about 5~AU. The dust temperature does not increase enough for thermal sublimation of SO and SO$_2$. 
%{\it Right:} Total cooling rate (solid) as well as the main molecular coolants, H$_2$ (dashed), CO (dotted), and NH$_3$ (dot-dashed) in the shock. Initially, the cooling dominated by H$_2$. As the temperature drops, cooling by H$_2$ drops and CO becomes the dominant coolant. Toward the end of the shock, NH$_3$ which has thermally desobred from the ice takes over the cooling of the gas.  The region where $T_\mathrm{gas} > 50$~K is indicated with the shaded blue region. 
{\it Right:} Abundance profiles of SO (solid) and SO$_2$ (dashed) and the key species for gas-phase formation of these species: SH (dot-dashed) and OH (dotted). The region where $T_\mathrm{gas} > 50$~K is indicated with the shaded blue region.
}
\label{fig:fidu_shock_T_SO_SO2}
\end{figure*}

\begin{table}
\centering
\caption{Fiducial model parameters and explored parameter range}
\label{tab:phys_params}
\begin{tabular}{llll}
\hline \hline
 & Units & Fiducial & Range \\ %& Remarks \\ 
\hline
\nH\ & cm$^{-3}$ & $10^{7}$ & $10^{5} - 10^{8}$ \\ %& Initial hydrogen density \\
\dz\ & s$^{-1}$ & $10^{-17}$ & -- \\ %& Cosmic ray ionization rate \\
\G0\ & Mathis & 1 & $10^{-3} - 10^{2}$ \\ %& Radiation field strength \\
%B & $\mu$G & $10^{2}$ & $10^{-1} - 10^{3}$ \\ %& Initial magnetic field strength \\
%\bB & -- & 0 & -- \\ %& Initial magnetic field strength \\
\Vs\ & \kms & 3 & $1 - 10$ \\ %& Shock velocity \\
\hline
\end{tabular}
%\tablefoot{The strength of the magnetic field is determined as $B = b \sqrt{n_\mathrm{H}}$~$\mu$G.}
\end{table}

\subsection{Input parameters}
\label{subsuec:inputparam}
The range of physical parameters explored in this paper is presented in Table~\ref{tab:phys_params}. Important physical parameters are the initial proton density \nH, the cosmic ray ionization rate of H$_2$ \dz,  the strength of the local incident radiation field, and the shock velocity \Vs. 
Depending on the mass of the source and size of the disk, the density of infalling envelope material impacting the disk can be $\sim10^{5}-10^{8}$~cm$^{-3}$ \citep[e.g.,][]{Harsono2015}. Therefore, the investigated range of \nH\ covers $\sim10^{5}-10^{8}$~cm$^{-3}$, with the fiducial value at $10^7$~cm$^{-3}$.
The radiation field is parameterized as a combination of the ISRF multiplied with a factor \G0\ \citep{Mathis1983} and infrared emission from the disk (which is constant for all models). 
The strength of the local UV radiation field at the disk-envelope interface is highly uncertain and (among other things) dependent on the distance to the source of UV radiation (see Sect~\ref{subsubsec:geometry}). Without extinction, the local \G0\ value could be as high as $10^4$ at $\sim100$~AU \citep{Visser2012}. However, taking into account the extinction the local strength will likely not be much higher than a \G0\ of $\sim100$. Therefore, a large range in \G0\ is considered: $10^{-3}-10^{2}$, with a fiducial value of $G_0 = 1$.
%The range of \G0\ considered is $10^{-3}-10^{2}$, with the fiducial value at $G_0 = 1$. 
The fiducial \Vs\ is set to $3$~\kms, which is roughly equal to the velocity of the infalling envelope at $100$~AU for a $0.5$~M$_\odot$ star. Shock velocities of $1-10$~\kms\ are considered, which correspond to an infalling envelope at $10-1000$~AU. The initial gas temperature is set to 25~K.
%No transverse magnetic field is imposed on the shock.
%Grain growth is included implicitly by assuming the grains have already grown to an average size of 0.2~$\mu$m in the outer parts of the envelope \citep[e.g.,][]{Miotello2014,Harsono2018,Galametz2019}. Moreover, freezeout of PAHs is taken into account by lowering the abundance of PAHs to $10^{-8}$ with respect to \nH\ \citep[e.g., two orders of magnitude lower than the ISM value of $10^{-6}$;][]{Draine2007,Geers2009}. 

Input abundances of smaller molecules and atoms are set to match typical low-mass dark cloud conditions \citep{vanderTak2003,Boogert2015,Navarro2020,Tafalla2021,Goicoechea2021}. The initial gas-phase abundance of atomic S and O is set to $10^{-6}$ with respect to \nH, with most of the oxygen and carbon reservoirs locked up in refractory grain cores or ices such as H$_2$O and CH$_4$ \citep[$10^{-4}$ and $10^{-6}$, respectively;][]{Boogert2015}. H$_2$S ice is assumed to be the dominant sulfur carrier with an abundance of $\sim2\times10^{-5}$ \citep[e.g.,][]{Vidal2017, Navarro2020}. The abundance of both SO and SO$_2$ ice was set to $10^{-7}$ \citep[e.g.,][]{Boogert2015,Rubin2019,Altwegg2019}, with the initial gas-phase abundance at $10^{-9}$ \citep{vanderTak2003}. The full list of input abundances is presented in Appendix~\ref{app:input_abundances}.

\begin{figure}
\includegraphics[width=1\linewidth]{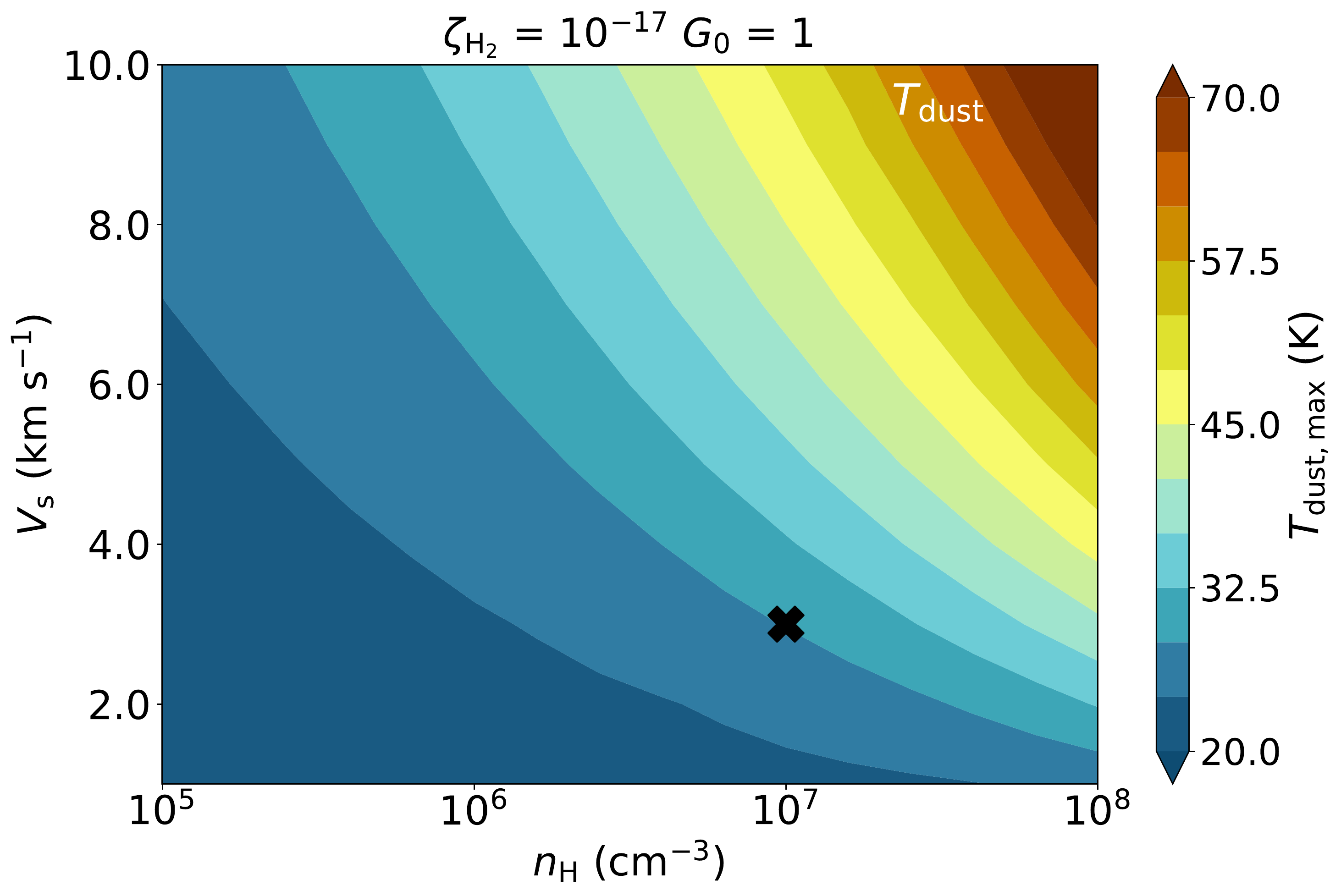}
\caption{Maximum dust temperature reached (in color) in shock models as function of initial \nH\ and \Vs. All other physical parameters are kept constant to the fiducial values and listed on top of the figure. The black cross indicates the position of the fiducial model. 
%The region where the total length of the shock is $>100$~AU is highlighted with hashed region.
}
\label{fig:Vs_nH_Tdust}
\end{figure}

\begin{figure*}
\includegraphics[width=0.49\linewidth]{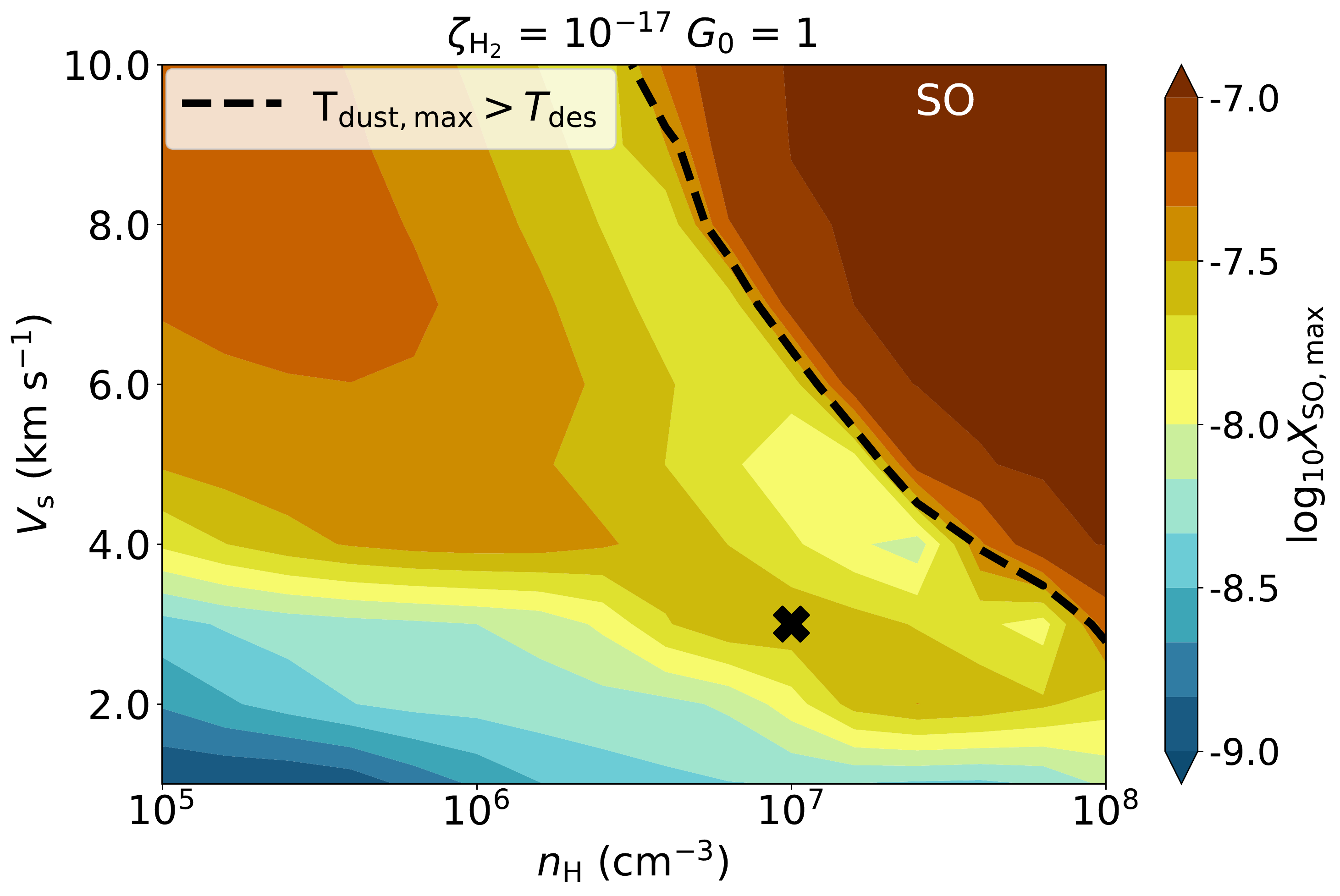}
\includegraphics[width=0.49\linewidth]{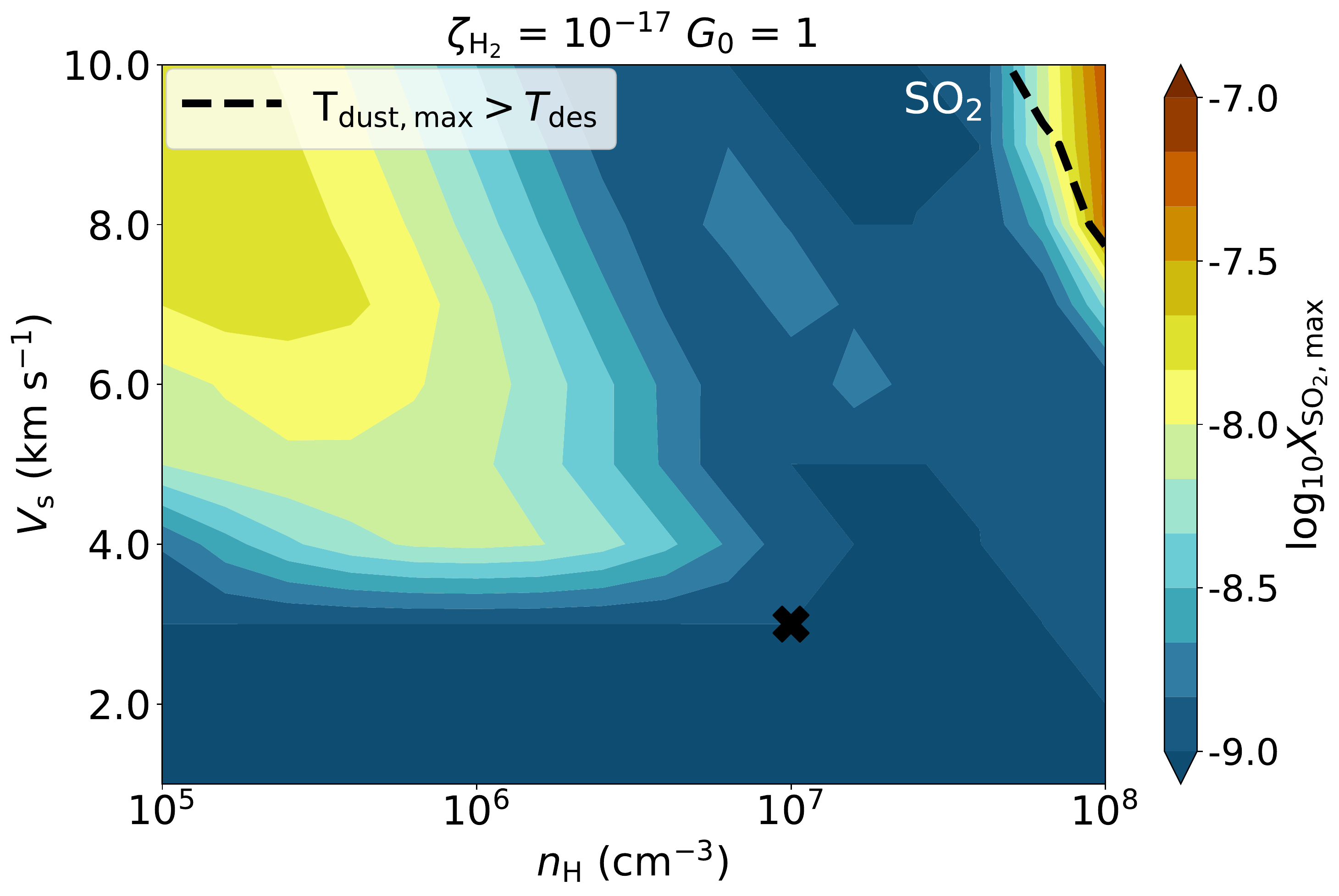}
\caption{Maximum abundance reached (in color) of SO (left) and SO$_2$ (right) in shock models as function of initial \nH\ and \Vs. All other physical parameters are kept constant to the fiducial values and listed on top of the figure. The black cross indicates the position of the fiducial model. The dashed black line shows the ice line, i.e., where 50\% of the ice is thermally desorbed into the gas in the shock. 
%The region where the total length of the shock is $>100$~AU is highlighted with hashed region.
}
\label{fig:Vs_nH_SO_SO2}
\end{figure*} 

\section{Results}
\label{sec:results}
\subsection{Temperature and density}
The gas and dust temperature, $T_\mathrm{gas}$ and $T_\mathrm{dust}$, respectively, and density structure of the fiducial shock model are shown in the left of Fig.~\ref{fig:fidu_shock_T_SO_SO2}. 
At the start of the shock, $T_\mathrm{gas}$ increases to $\sim500$~K, which is in agreement with the analytic expression for a weakly magnetized and fully molecular shock derived by \citet{Lesaffre2013} from the Rankine-Hugoniot relations,
\begin{align}
T_\mathrm{gas,max} \approx 53 V_\mathrm{s}^2,
\label{eq:gas_temp}
\end{align}
where \Vs\ is the initial shock velocity in units of \kms. 
%The gas and dust get compressed at the shock front, initially increaseing the density \nH\ by about a factor $\sim2-3$. 
Following the jump in temperature, the gas is cooled down through radiation of rotational and rovibrational molecular lines and atomic lines. The most dominant coolant right after the shock front is H$_2$, while cooling by (optically thick) pure rotational and rovibrational lines of CO and H$_2$O take over as the temperature drops below $\lesssim400$~K. Cooling of the gas through gas-dust thermal coupling is negligible. Cooling through NH$_3$ in the optically thin limit is significant for those shocks with $V_\mathrm{s} \gtrsim 6$~\kms\ since in those shocks the abundance of NH$_3$ is significantly increased through high-temperature gas-phase chemistry. The end of the shock is defined as the distance where $T_\mathrm{gas}$ drops below 50~K, which corresponds to $\sim3$~AU for the fiducial model. In the post-shock, the thermal balance between gas and dust is determined by heating of the dust through the FIR radiation field, gas-grain collisions that transfer the heat from dust to gas, and cooling of the gas through predominantly CO rotational transitions.

The dust temperature is very relevant for these shocks since it increases to values where thermal desorption of ices occurs. In the shock, the material compresses, increasing the density by up to two orders of magnitude (i.e., to about 10$^{9}$~cm$^{-3}$ in Fig.~\ref{fig:fidu_shock_T_SO_SO2}). Due to the thermal coupling between the gas and dust, $T_\mathrm{dust}$ increases to $\sim30$~K for the fiducial model, enough for thermal sublimation of volatile species such as CH$_4$, but not for more strongly bound ices such as H$_2$S, H$_2$O, and CH$_3$OH. However, as shown in Fig.~\ref{fig:Vs_nH_Tdust}, $T_\mathrm{dust}$ can become as large as $>70$~K in the highest-velocity ($\sim10$~\kms) shocks in the densest ($10^{8}$~cm$^{-3}$ media, resulting in thermal sublimation of several ices including H$_2$S, SO, and SO$_2$. 
%For $G_\mathrm{0} = 1$, the maximum dust temperature reached in the shock as function of initial \Vs\ and \nH\ is presented in Fig.~\ref{fig:Vs_nH_Tdust}.
%For constant \Vs, $T_\mathrm{dust}$ naturally increases with increasing initial \nH. Additionally, larger \Vs\ results in a stronger compression in the shock and thus a stronger increase in \nH. As a results, the maximum dust temperature reached in the shock increases from the bottom left of Fig.~\ref{fig:Vs_nH_Tdust} toward the top right. 
Dust heating through UV radiation is only significant for strongest irradiated environments with $G_\mathrm{0} \gtrsim 50$ and only up to $T_\mathrm{dust}\lesssim25$~K. Since photodesorption of ices is negligible on timescales of the shock (see Eq.~\eqref{eq:tau_photodes}), thermal desorption is the dominant mechanism in releasing ices to the gas phase.

\subsection{Chemistry of SO and SO$_2$}
The abundance profiles of SO and SO$_2$ in the fiducial model are presented in the right part of Fig.~\ref{fig:fidu_shock_T_SO_SO2}. Toward the end of the shock, as the gas and dust are cooling down, a significant increase in the abundances of SO is visible up to the $\sim5\times10^{-8}$ level. The abundance of SO$_2$ does not increase above the dark cloud abundance of $10^{-9}$ \citep{vanderTak2003}. The maximum abundance reached in the shock for SO and SO$_2$ as function of different initial \nH\ and \Vs\ is presented in Fig.~\ref{fig:Vs_nH_SO_SO2}. In the following subsections different parts of the parameter space are highlighted using the relevant chemical formation pathways in Fig.~\ref{fig:S_chem_shock}.

\subsubsection{Low-velocity shocks: $\sim3$~\kms}
\begin{figure*}
\includegraphics[width=1\linewidth]{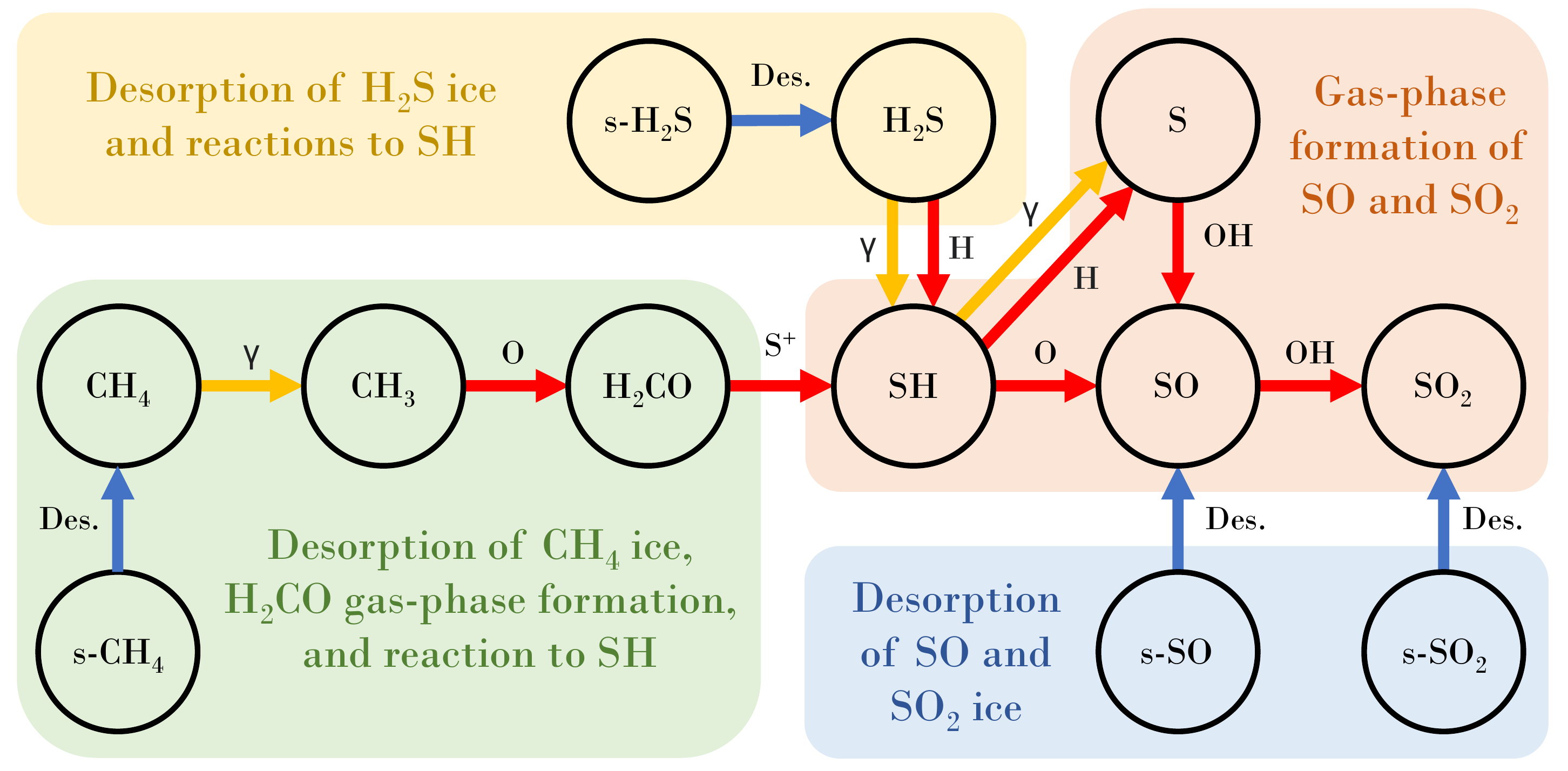}
\caption{Chemical reactions leading to the formation of gas-phase SO and SO$_2$. Species denoted as s-X are located in the ice mantles of dust grains. Reverse reactions are not shown for clarity. Red arrows denote chemical reactions with the annotated species, yellow arrows photodissociation reactions, and blue arrows ice desorption via either thermal or non-thermal mechanisms. Relevant chemical pathways are highlighted with a colored background. }
\label{fig:S_chem_shock}
\end{figure*}

In low-velocity ($\sim3$~\kms) shocks, thermal desorption of SO and SO$_2$ ice does not occur since $T_\mathrm{dust}$ only rises to $\lesssim35$~K. Any increase in abundance thus originates from gas-phase chemistry. Here, the relevant chemical reaction leading to the formation of SO is,
\begin{align}
\rm SH + O & \rightarrow \rm SO + H. \label{reac:SH_O}
\end{align}
The main formation pathway here of SH is via the green route of Fig.~\ref{fig:S_chem_shock}, that is, through the reaction of H$_2$CO with S$^+$. Since H$_2$CO only desorbs directly from the ice when $T_\mathrm{dust} \gtrsim 65$~K, H$_2$CO is formed through the gas-phase reaction of the CH$_3$ radical with atomic O. To get a significant amount of CH$_3$ in the gas, CH$_4$ ice needs to desorb \citep[at $T_\mathrm{dust} \gtrsim 25$~K;][]{Penteado2017} and subsequently photodissociate into CH$_3$ and atomic H. The S$^+$ ion originates from the photo-ionization of atomic S. The strongest increase in abundance through this route at intermediate densities ($n_\mathrm{H} \sim 10^7$~cm$^{-3}$). It is important to note that H$_2$CO is a representative here of any hydrocarbon reacting with S$^+$ to form SH. 

Some SO$_2$ is also formed in low-velocity shocks through a reaction of SO with OH (see Reaction~\eqref{reac:SO_OH}). However, the abundance of SO$_2$ does not increase above the initial dark cloud abundance of $10^{-9}$ because no significant amount of OH is available at these shock velocities.

\subsubsection{High-velocity shocks: $>4$~\kms}
For shocks propagating at higher velocities ($\gtrsim4$~\kms), both SO and SO$_2$ are efficiently formed through gas-phase chemistry, see Fig.~\ref{fig:Vs_nH_SO_SO2}. After the start of the shock, H$_2$O is readily formed through reactions of OH with H$_2$ \citep[][see also Fig.~\ref{fig:Vs_nH_H2S_H2O_SiO_H2CO}]{Flower2010}. As the shock cools down to below $\lesssim300$~K, the production of H$_2$O stops, but OH is still being formed through photodissociation of H$_2$O. This increases the OH abundance in the gas which stimulates the formation of SO and SO$_2$ through the red route of Fig.~\ref{fig:S_chem_shock},
\begin{align}
\rm S + OH & \rightarrow \rm SO + H,  \label{reac:S_OH} \\
\rm SO + OH & \rightarrow \rm SO_2 + H. \label{reac:SO_OH}
\end{align}
The highest abundances are achieved in less dense environments ($\lesssim 10^6$~cm$^{-3}$), with the maximum abundance of both SO and SO$_2$ dropping when moving toward intermediate densities ($\sim 10^{6}-10^{7}$~cm$^{-3}$). In the latter conditions, NH$_3$ becomes a dominant coolant in the tail of the shock. Therefore, the shock cools down more quickly and hence the region of favorable conditions for SO and SO$_2$ formation \mbox{($100 \lesssim T_\mathrm{gas} \lesssim 300$~K)} is reduced and their maximum abundances are lower.
%In less dense media ($\lesssim 10^7$~cm$^{-3}$), an increase of abundance of both SO and SO$_2$ is seen in Fig.~\ref{fig:Vs_nH_SO_SO2}, which gets stronger toward higher velocity shocks (\gtrsim at lower densities. For this part of the parameter space, the conditions in the tail of the shock, as the gas cools down below 300~K, are favorable for reactions~\eqref{reac:S_OH} and \eqref{reac:SO_OH}. 

%The maximum abundance of both SO and SO$_2$ drops in media with intermediate densities ($n_\mathrm{H} \sim 10^{6-7}$~cm$^{-3}$). In these conditions, NH$_3$ is efficiently formed though gas-phase chemistry and a dominant coolant in the tail of the shock. Therefore, the region of favorable conditions for SO and SO$_2$ formation is decreased and hence the maximum abundance attained is lower. 

At the highest densities ($n_\mathrm{H} > 10^{7}$~cm$^{-3}$), thermal desorption of sulfur-bearing ices becomes relevant for the chemistry (i.e., the blue route in Fig.~\ref{fig:S_chem_shock}). SO ice thermally sublimates as the dust temperature reaches $\sim37$~K, leading to a strong increase in its gas-phase abundance (see Fig.~\ref{fig:Vs_nH_SO_SO2}). Thermal desorption of SO$_2$ only occurs for the highest velocity shocks in the densest media. However, desorption of H$_2$S ice (at $T_\mathrm{dust}\sim47$~K) is also relevant for the gas-phase chemistry as subsequent photodissociation and reactions with atomic H lead to an increased abundance of SH and atomic S (i.e., the yellow route of Fig.~\ref{fig:S_chem_shock}).

\subsubsection{Dependence on UV radiation field}
As shown in Reactions~\eqref{reac:SH_O}-\eqref{reac:SO_OH} and Fig.~\ref{fig:S_chem_shock}, gas-phase formation of SO and SO$_2$ is dependent on radicals such as OH and SH. These radicals can be created in various ways, including high-temperature chemical reactions \citep{Prasad1980,Hartquist1980}. However, a more dominant physical process for enhancing the abundance of OH and SH is the photodissociation of H$_2$O, H$_2$S, and interestingly also CH$_4$. This process is dependent on the strength of the local UV radiation field, which is parametrized here with \G0.

\begin{figure}
\includegraphics[width=1\linewidth]{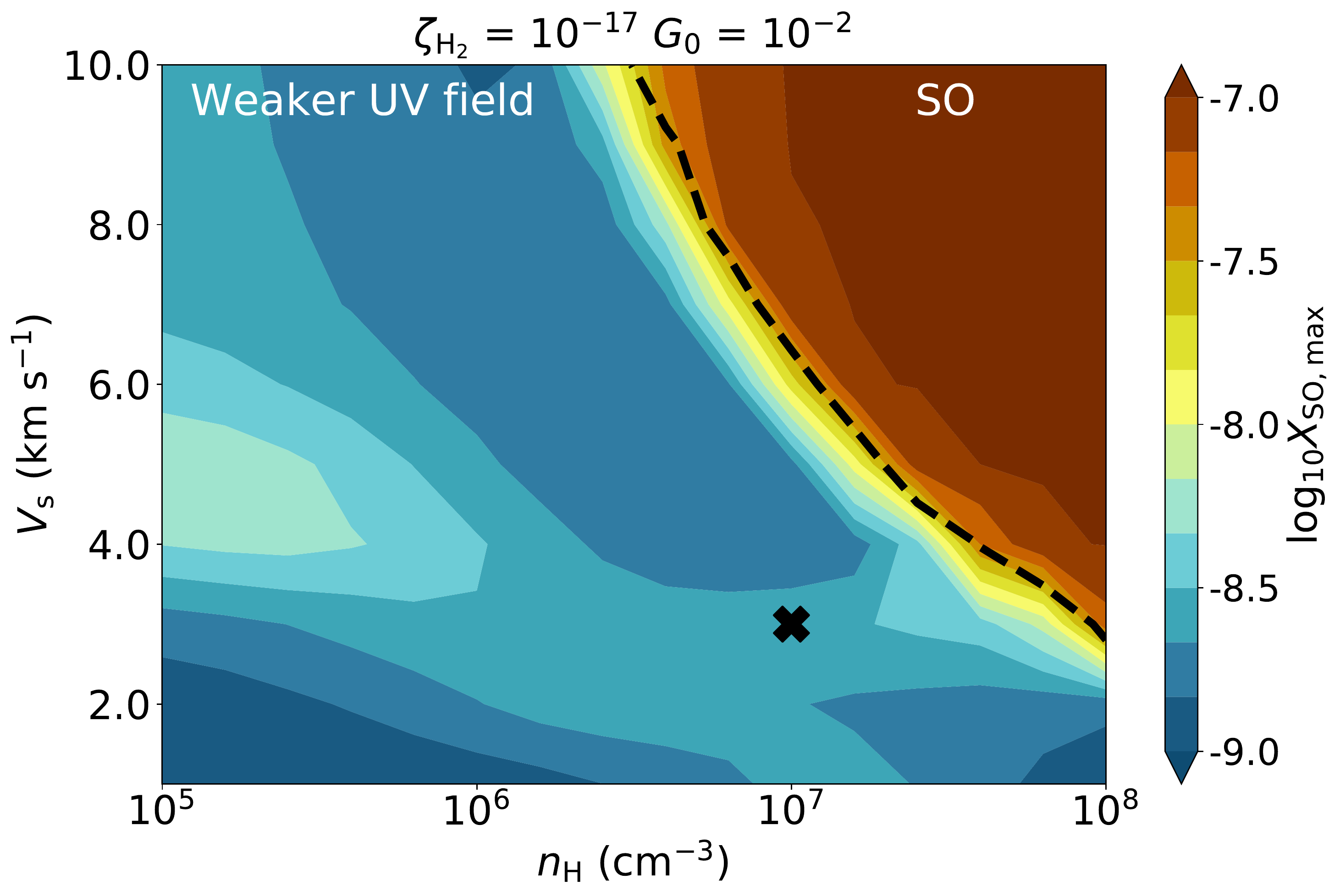}
\caption{Similar figure as Fig.~\ref{fig:Vs_nH_SO_SO2}, but now for SO in shocks with a weaker UV field ($G_0 = 10^{-2}$).  
%The region where the total length of the shock is $>100$~AU is highlighted with hashed region.
}
\label{fig:Vs_nH_SO_G01e-2}
\end{figure}

In Fig.~\ref{fig:Vs_nH_SO_G01e-2}, the maximum abundance of SO as function of \Vs\ and \nH\ is presented for $G_0 = 10^{-2}$. Similar figures of both SO and SO$_2$ for various \G0\ are presented in Appendix~\ref{app:SO_SO2_G0}.
Decreasing the strength of the UV field decreases the amount of SO and SO$_2$ produced in high-velocity ($\gtrsim4$~\kms) shocks due to less OH being produced through H$_2$O photodissociation. On the other end, increasing $G_0$ shifts the spot for efficient SO and SO$_2$ formation in the shock to higher densities to create the favorable conditions for their chemistry. Moreover, for $G_0 \gtrsim 10$ photodissociation of thermally desorbed H$_2$S (i.e., when $n_\mathrm{H}\gtrsim10^7$~cm$^{-3}$ and $V_\mathrm{s}\gtrsim4$~\kms) significantly increases the abundance of atomic S (i.e., yellow route of Fig.~\ref{fig:S_chem_shock}) which leads to an increase in abundances of SO and in particular SO$_2$ (see Fig.~\ref{fig:Vs_nH_SO2_diffG0}).

\begin{table}
\centering
\caption{Input abundances of key Sulfur-bearing species}
\label{tab:S_input_abun}
\begin{tabular}{lcccccc}
\hline \hline
Case & S & SO & SO$_2$ & s-SO & s-SO$_2$ & s-H$_2$S  \\
\hline
Fiducial & 1(-6) & 1(-9) & 1(-9) & 1(-7) & 1(-7) & 2(-5)\\
Low-S & 1(-8) & 1(-9) & 1(-9) & 1(-7) & 1(-7) & 2(-5)  \\ 
High-S & 2(-5) & 1(-9) & 1(-9) & 1(-7) & 1(-7) & 1(-7)  \\
\hline
\end{tabular}
\tablefoot{a(b) represents $\mathrm{a\times10^b}$. Species denoted as s-X are located in the ice mantles of dust grains. All other abundances are the same as those presented in Appendix~\ref{app:input_abundances}.} 
\end{table}

\begin{figure*}
\includegraphics[width=0.49\linewidth]{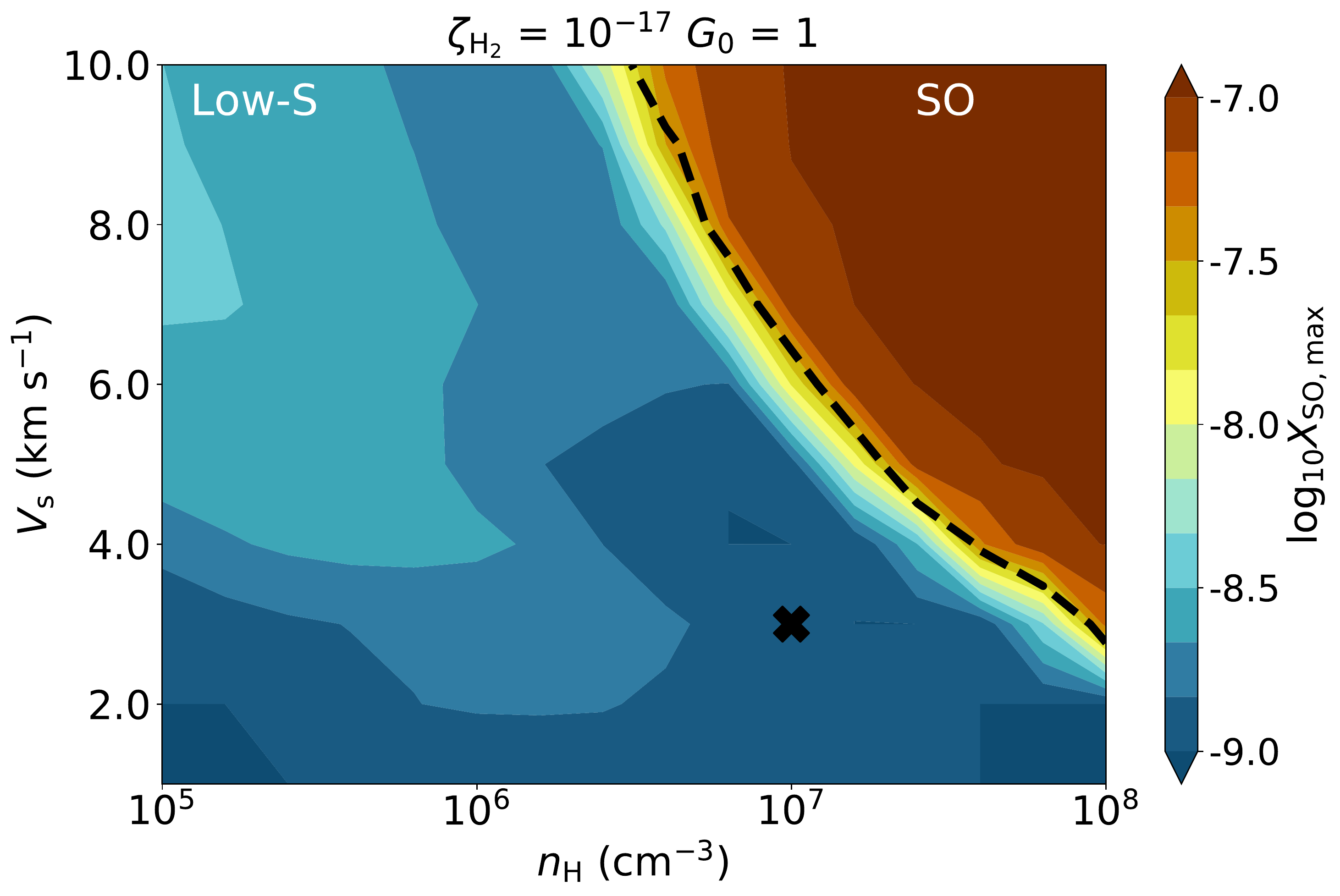}
\includegraphics[width=0.49\linewidth]{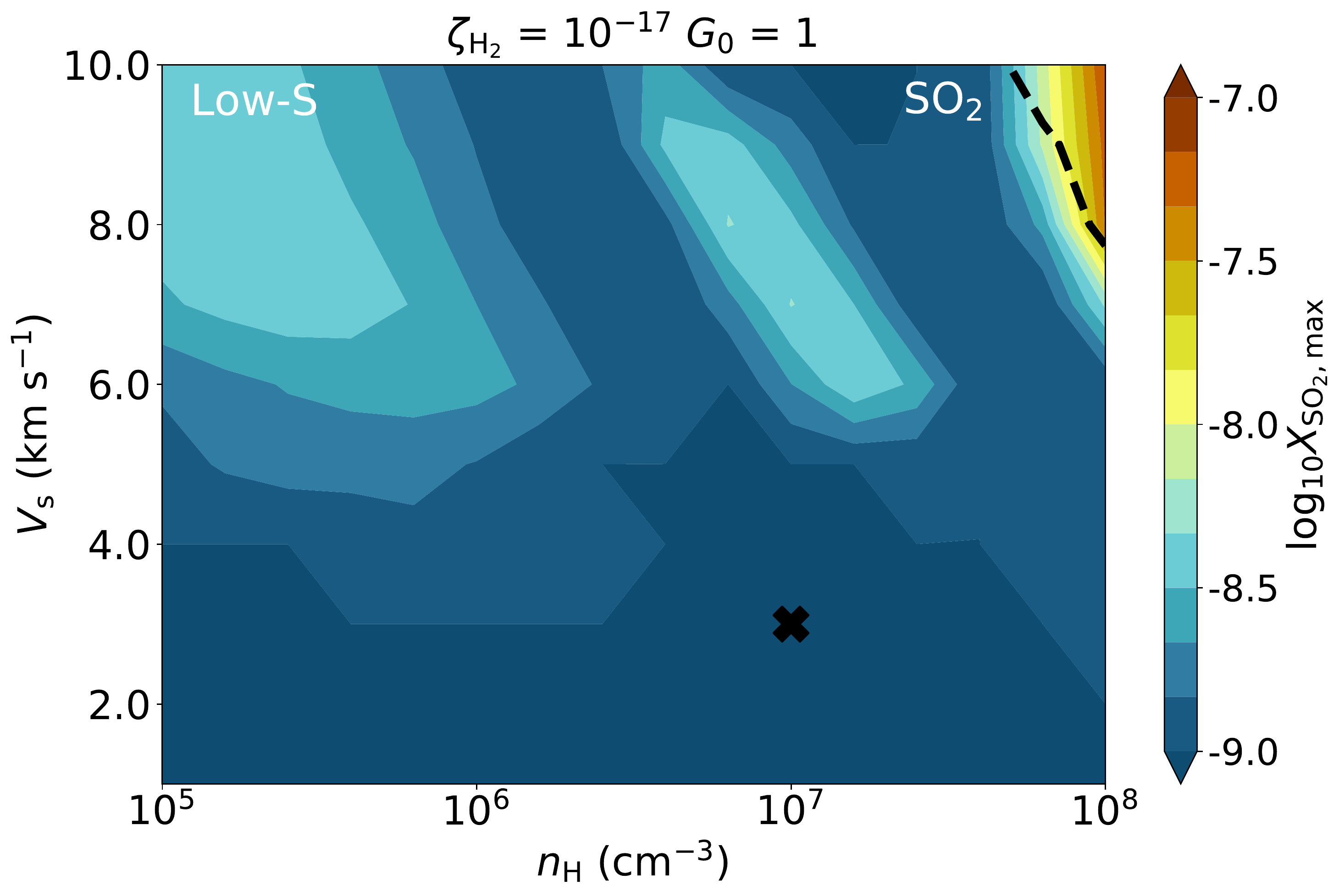}
\includegraphics[width=0.49\linewidth]{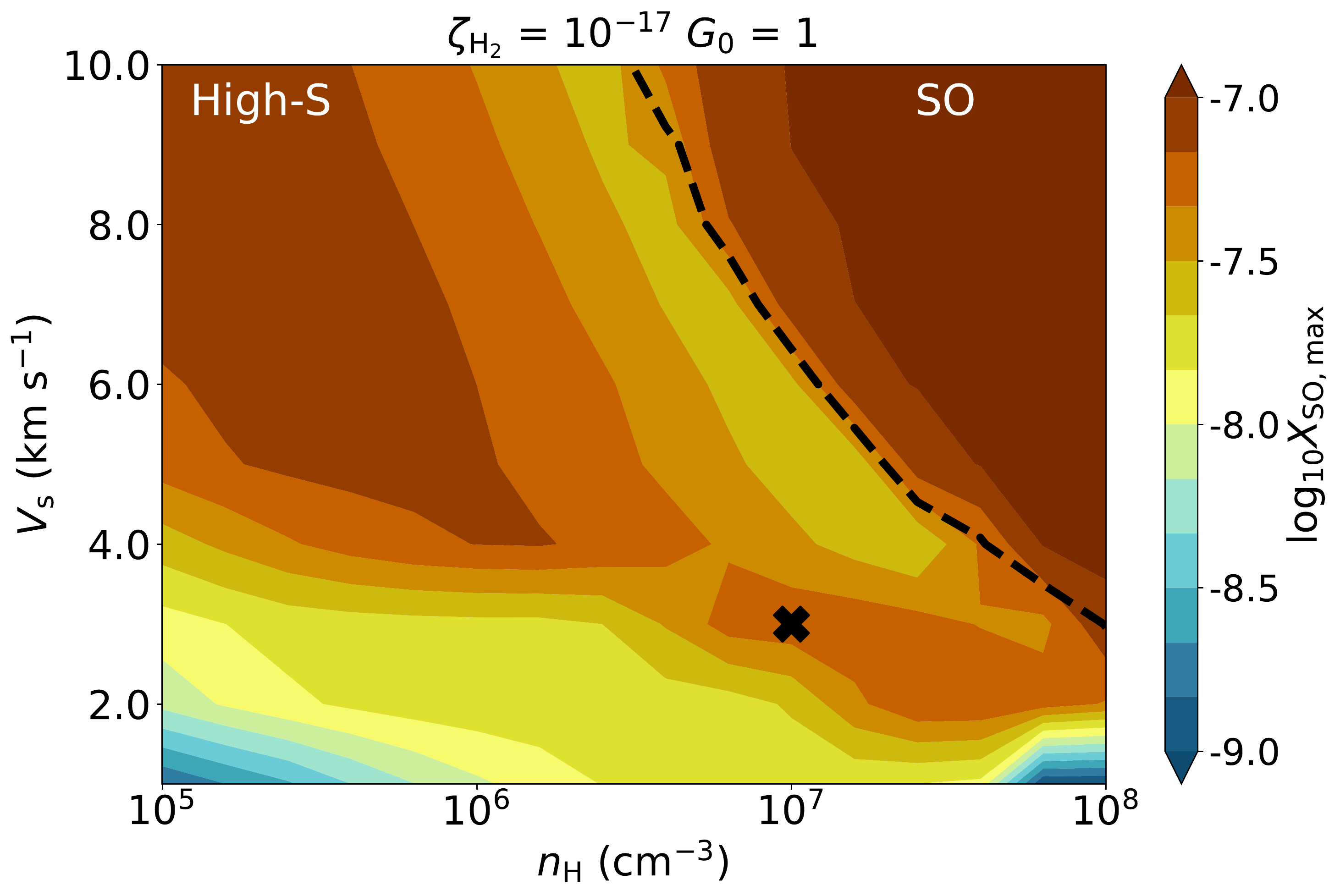}
\includegraphics[width=0.49\linewidth]{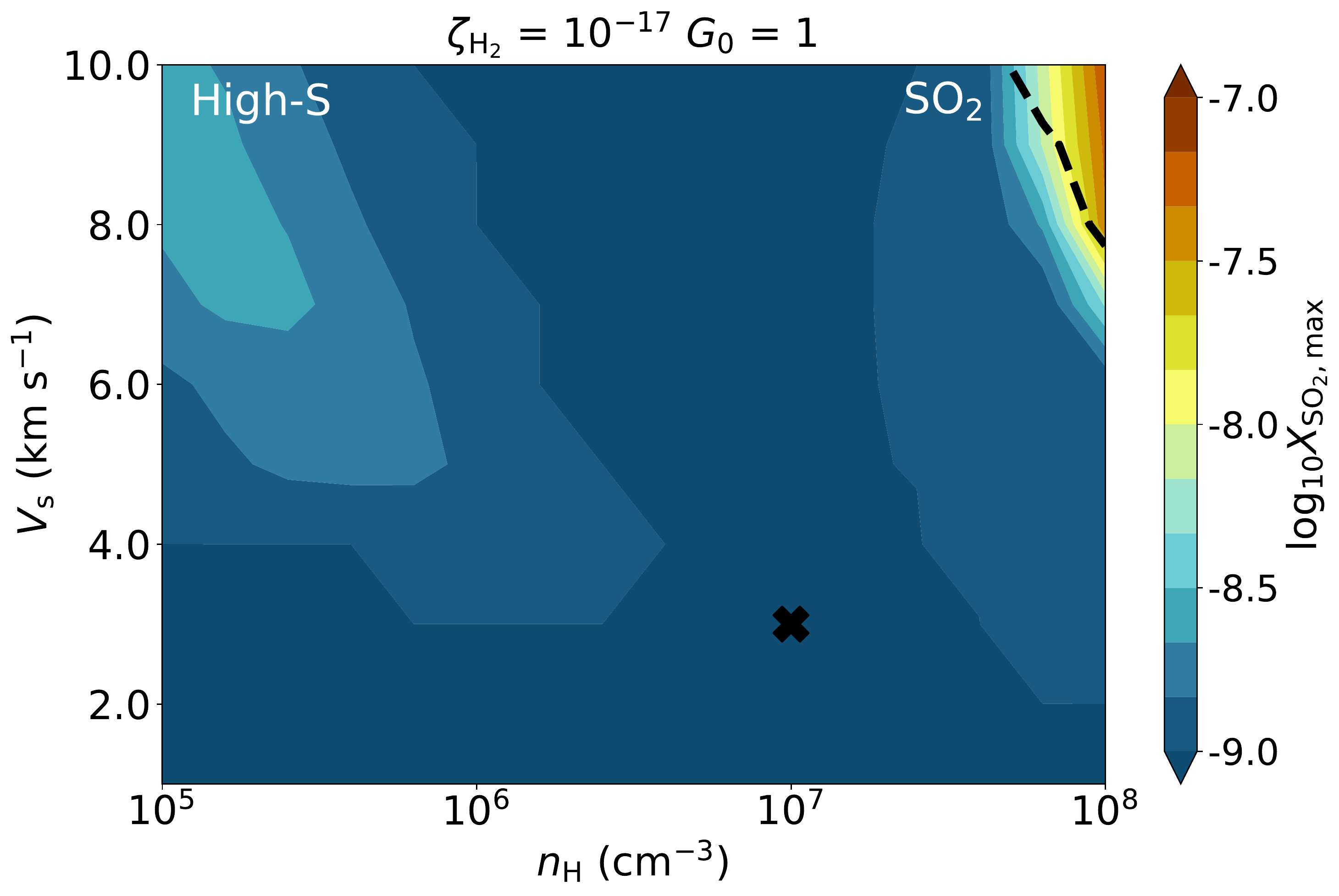}
\caption{Maximum abundance reached (in color) of SO (left) and SO$_2$ (right) in shock models as function of initial \nH\ and \Vs\ for an initial gas-phase atomic S abundance of $10^{-8}$ (top row) and $10^{-5}$ (bottom row). All other physical parameters are kept constant to the fiducial values and listed on top of the figure. The black cross indicates the position of the fiducial model. The dashed black line shows the ice line, i.e., where 50\% of the ice is thermally desorbed into the gas in the shock. 
%The region where the total length of the shock is $>100$~AU is highlighted with hashed region.
}
\label{fig:Vs_nH_SO_SO2_diffSabun}
\end{figure*}

At lower shock velocities ($V_\mathrm{s} \lesssim 4$~\kms), water is not formed in the shock and hence photodissociation of H$_2$O does not increase the abundance of SO and SO$_2$. The increase of SO abundance is here due to the green route of Fig.~\ref{fig:S_chem_shock}, which is initialized by thermal desorption and subsequent photodissociation of CH$_4$. 
This formation route is most relevant for $G_0 = 1$; for lower \G0, the UV radiation field is not strong enough to photodissociate significant amounts of CH$_4$. 
For higher \G0, higher densities are required for the efficient production of SO through the green route.
%For higher \G0, CH$_4$ has thermally desorbed and converted to SH in the pre-shock environment since for these strong UV fields the grain is additionally heated. Hence, the green route of Fig.~\ref{fig:S_chem_shock} occurs already efficiently in the pre-shock medium and no increase in SO and SO$_2$ abundance is seen in the shock.
%. In the shock itself, this SH reacts in the shock to form either H$_2$S or atomic S and is therefore less available to form SO as the shock cools down.

\subsubsection{Dependence on initial conditions}
\label{subsubsec:init_conditions}
The abundance of all species, including SO and SO$_2$, is dependent on the initial abundances. The interstellar ice abundances of SO and SO$_2$ are not well known, with only a tentative detection for SO$_2$ \citep{Boogert1997,Zasowski2009}. Here,the ice abundances of SO and SO$_2$ are estimated at the $10^{-7}$ level based on what is found in cometary ices \citep{Calmonte2016,Rubin2019,Altwegg2019}. However, they are only relevant in the shocks where thermal desorption of these ices occurs (i.e., the blue route of Fig.~\ref{fig:S_chem_shock} and to the right of the dashed line in Fig.~\ref{fig:Vs_nH_SO_SO2_diffSabun}). For those shocks, increasing or decreasing the ice abundances with an order of magnitude will also result in an increase or decrease of the gas-phase abundances with an order of magnitude, given that the ice abundance is dominant over what is achieved through gas-phase chemistry alone. 

For shocks where the abundance of SO and SO$_2$ is mainly increased through gas-phase chemistry (i.e., to the left of the dashed line in Fig.~\ref{fig:Vs_nH_SO_SO2_diffSabun}), the maximum abundance reached is mostly directly dependent on the initial abundance of atomic sulfur. To test the effect of initial atomic S abundance on the maximum abundances of SO and SO$_2$, shock models are calculated for a lower gas-phase atomic sulfur abundance (Low-S: $X_\mathrm{S} = 10^{-8}$) and assuming that almost all sulfur is atomic and in the gas phase (High-S: $X_\mathrm{S} = 2\times 10^{-5}$). The initial abundances for these cases are presented in Table~\ref{tab:S_input_abun}. The resulting maximum SO and SO$_2$ abundances for these models are shown in Fig.~\ref{fig:Vs_nH_SO_SO2_diffSabun}. 

%\begin{table*}
%\centering
%\caption{Input abundances of dominant Sulfur-bearing species}
%\label{tab:S_input_abun}
%\begin{tabular}{lllllllllll}
%\hline \hline
% & S & SO & SO$_2$ & H$_2$S & OCS & CS & SO* & SO$_2$* & H$_2$S* & OCS*  \\
%\hline
%Fiducial & 1(-6) & 1(-9) & 1(-9) & 1(-9) & 2(-9) & 3(-9) & 1(-7) & 1(-7) & 2(-5) & 2(-7) \\
%Low-S & 1(-8) & 1(-9) & 1(-9) & 1(-9) & 2(-9) & 3(-9) & 1(-7) & 1(-7) & 2(-5) & 2(-7) \\ 
%High-S & 2(-5) & 1(-9) & 1(-9) & 1(-9) & 2(-9) & 3(-9) & 1(-7) & 1(-7) & 1(-7) & 2(-7) \\
%\hline
%\end{tabular}
%\tablefoot{a(b) stands for $\mathrm{a\times10^b}$. Species in ice mantles on dust grains are indicated with a *.} 
%\end{table*}

In the Low-S case, the maximum abundance of SO drops for both low and high velocity shocks (excluding higher densities where the abundance is increased through thermal ice sublimation). This is straightforward to interpret since both S and S$^{+}$ are main formation precursors of SO in both the red and green formation routes of Fig.~\ref{fig:S_chem_shock}, respectively. At lower densities ($\lesssim10^{6}$~cm$^{-3}$), the maximum SO$_2$ abundance drops in a similar way since less SO is formed. However, at intermediate densities ($\sim10^{7}$~cm$^{-3}$), slightly more SO$_2$ is formed in high velocity shocks compared to Fig.~\ref{fig:Vs_nH_SO_SO2}. This is because atomic S and SO are `competing' to react with OH to form respectively SO and SO$_2$. Dropping the initial atomic S abundance results in more SO$_2$ being formed, also because the majority of the gas-phase SO here originates from thermal ice sublimation. 

In the High-S case, not only the chemistry but also the thermal structure of the shock is slightly altered since cooling through atomic S becomes significant. Furthermore, more SO is formed through both the red and green routes of Fig.~\ref{fig:S_chem_shock}. Interestingly, the maximum abundance of SO$_2$ attained decreases compared to Fig.~\ref{fig:Vs_nH_SO_SO2} since with more atomic S in the gas phase to react with OH, less OH is available to react with SO toward SO$_2$.

The initial abundances of other species are also important for the chemistry of SO and SO$_2$. Increasing or decreasing the amount of atomic oxygen results in respectively more or less OH and H$_2$O formed in the shock and thus directly affects the abundances of SO and SO$_2$. Moreover, reactions with atomic carbon are a main destruction pathway for SO \citep[forming CS;][]{Hartquist1980}, hence an increased initial atomic carbon abundance results in lower abundances of both SO and SO$_2$ in the shock.

In reality, the initial dark cloud abundances may also be altered during the infall from envelope to disk since both the temperature and the strength of the UV radiation field increase \citep[e.g.,][]{Aikawa1999,Visser2009,Drozdovskaya2015}. 
To take this into account, a pre-shock model should be calculated. In Appendix~\ref{app:pre-shock_mod}, the effect of calculating such a pre-shock model on the abundance of SO attained in the shock is presented. Any alterations on the pre-shock conditions are only important for the chemistry of SO and SO$_2$ if the timescale over which the pre-shock is calculated is longer than about 10\% of the photodesorption timescale. For typical infalling envelopes, this is not expected to be relevant.
%The effect of running a pre-shock model on the initial conditions of the shock is presented in Appendix~\ref{app:pre-shock_mod}.

\subsection{Effect of grain size and PAHs}
%\subsection{PAHs and grain size}
%The dust grains in protostellar envelopes coagulate, resulting a on average larger grains than the typical ISM distribution \citep[e.g.,][]{Miotello2014,Harsono2018,Galametz2019}. Moreover, PAHs are frozen out onto dust grains, reducing the gas-phase abundance by up to two orders of magnitude \citep{Geers2009}. 

\begin{figure}
\includegraphics[width=\linewidth]{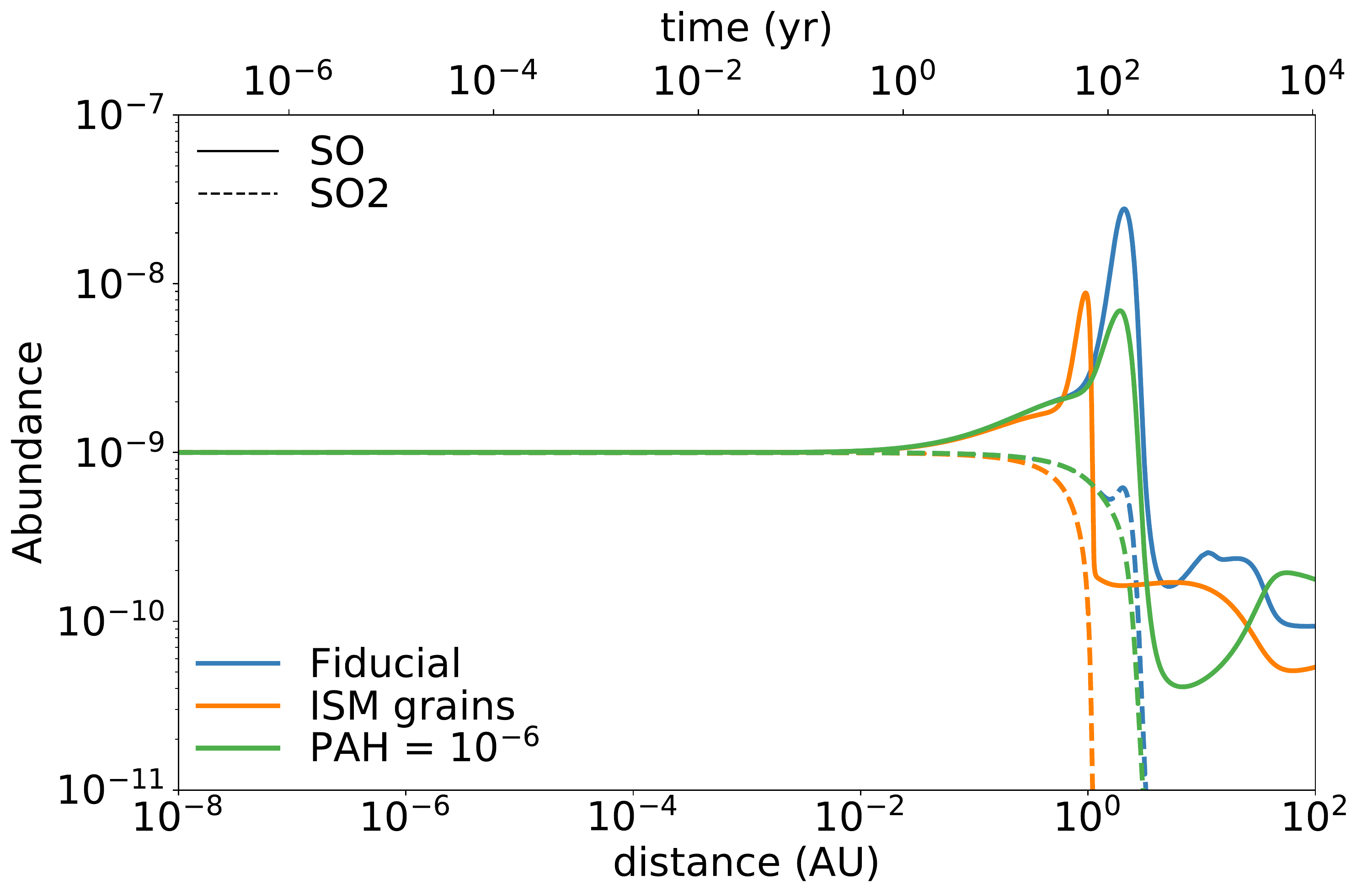}
\caption{
Abundance profile of SO (solid) and SO$_2$ (dashed) for the fiducial model (blue) with a PAH abundance of $10^{-8}$ using a grain size of $0.2$~$\mu$m.
%The figure is zoomed in on the tail of the shock where the gas and dust dust are cooling down.
Overplotted are the same model using a typical ISM grain size of 0.03~$\mu$m (orange) and a model with a PAH abundance of $10^{-6}$ (green). All other parameters are kept at their fiducial value (see Table~\ref{tab:phys_params}). 
%The vertical lines mark the distance where $T_\mathrm{gas}$ drops below 50~K.
}
\label{fig:PAH_grain_dependence}
\end{figure}

The size of the dust grains has an effect on both the physical structure of the shock and on the chemistry \citep[e.g.,][]{Guillet2007,Miura2017}. In Fig.~\ref{fig:PAH_grain_dependence}, the abundance profiles of SO and SO$_2$ are shown for the fiducial model and for the same model while adopting a typical ISM grain size distribution \citep[computed using a single size of $\sim0.03$~$\mu$m;][]{Mathis1977,Godard2019}. 
%Already in the pre-shock environment the effect of smaller dust grains is apparent; the abundances of SO and SO$_2$ are significantly higher than for the fiducial model. This originates from the photodesorption rate of ices being inversely proportional to the grain size. For smaller grains, more H$_2$O and H$_2$S are photodesorbed and subsequently photodissociated into OH and SH, respectively, in the pre-shock region and available for reactions with atomic S and O to form SO and SO$_2$. 
The decrease in grain size leads to stronger coupling between gas and dust and thus to a slightly higher dust temperature ($\sim35$~K against $\sim30$~K for $0.2$~$\mu$m grains). 
Moreover, the increased gas-grain coupling results in the dust being a dominant coolant in the tail of the shock reducing the length of the shock ($\sim1$~AU against $\sim3$~AU for the fiducial model).
%Moreover, the shock is more quickly cooled down due the increased thermal radiation of the dust grains. 
Nevertheless, the abundance of SO is increased in the shock, but the maximum abundance reached is about a factor of $2-3$ lower compared to the fiducial model. No significant increase is seen for SO$_2$. \citet{Miura2017} found that the dust temperature easily reaches $\sim50$~K in low-velocity shocks taking into account dust aerodynamics for ISM-size dust grains. The full treatment of dust physics, including aerodynamic heating and grain-grain interactions \citep{Guillet2007,Guillet2009,Guillet2011,Guillet2020,Miura2017}, will likely move the ice line of Fig.~\ref{fig:Vs_nH_SO_SO2} toward lower \nH\ and \Vs, but is beyond the scope of this work.

Contrary to dust grains, the effect of PAHs on the physical structure of $J$-type shocks is negligible since PAHs do not contribute to the cooling and only little to the heating if a strong UV field is present. However, PAHs do play a key role in the ionization balance and are therefore highly relevant for the chemistry \citep{Flower2003}. In Fig.~\ref{fig:PAH_grain_dependence}, the abundances of SO and SO$_2$ are also shown for the fiducial model with a PAH abundance of $10^{-6}$ \citep[equal to the value derived in the local diffuse ISM;][]{Draine2007}. The maximum abundance of SO is a factor $3-4$ lower when the abundance of PAHs is at the ISM level because the PAHs are dominating the ionization balance. Consequently, the abundance of other ions, include S$^+$, drops by more than an order of magnitude. Given that the green route of Fig.~\ref{fig:S_chem_shock} is the dominant SO formation route here and dependent on S$^+$, less SO and subsequently SO$_2$ are formed. For high-velocity shocks, this effect is not apparent since Reactions~\eqref{reac:S_OH} and \eqref{reac:SO_OH} dominate the formation of SO and SO$_2$ and are independent of ions.
%In fact, the reaction of atomic sulfur with OH is the dominant formation route here for SO since OH is formed from photodissociation of H$_2$O and therefore in dependent of PAHs.

\subsection{Other molecular shock tracers: SiO, H$_2$O, H$_2$S, CH$_3$OH, and H$_2$}
Other classical molecular shock tracers include SiO, H$_2$O, H$_2$S, CH$_3$OH, and H$_2$. Abundance maps similar to Fig.~\ref{fig:Vs_nH_SO_SO2} of SiO, H$_2$O, H$_2$S, and CH$_3$OH are presented in Appendix~\ref{app:H2S_H2O_SiO_H2CO}. SiO is a tracer of shocks with significant grain destruction since then the silicon which is normally locked in dust grains can react with the OH radical to form SiO \citep[e.g.,][]{Caselli1997,Schilke1997,Gusdorf2008_1,Gusdorf2008_2,Guillet2009}. It is often observed in the high velocity bullets of jets originating from young Class~0 protostars \citep[e.g.,][]{Guilloteau1992,Tychoniec2019,Taquet2020}. However, the abundance of SiO is not significantly increased in our shock models because most of the Si remains locked up in silicates in the dust grains. 

Water is the most abundant ice in protostellar envelopes \citep[][]{Boogert2015,vanDishoeck2021}, but also frequently observed in outflow shocks \citep[e.g.,][]{Flower2010,Herczeg2012,Kristensen2012,Nisini2013,Karska2018}. Here, H$_2$O is efficiently produced in shocks with $V_\mathrm{s} \gtrsim 4$~\kms\ (see Fig.~\ref{fig:Vs_nH_H2S_H2O_SiO_H2CO}) and very important for the gas-phase chemistry of SO and SO$_2$. Thermal desorption of H$_2$O is negligible.

H$_2$S is suggested as a main carrier of sulfur in ices \citep[e.g.,][]{Vidal2017}, despite being undetected thus far \citep{Jimenez-Escobar2011}. In the gas phase it is generally observed close to Class~0 protostars  where the emission may originate from thermal desorption \citep{Tychoniec2021}, or in outflow shocks \citep[e.g.,][]{Holdship2016}. In our shock models, gas-phase formation of H$_2$S becomes efficient at slightly higher velocities than for water ($V_\mathrm{s} \gtrsim 5$~\kms, see Fig.~\ref{fig:Vs_nH_H2S_H2O_SiO_H2CO}). Moreover, for high-velocity shocks in dense media, H$_2$S ice is thermally sublimated.

Methanol is a tracer of lower-velocity ($<10$~\kms) outflow shocks where ices are sputtered off the dust grains \citep[e.g.,][]{Suutarinen2014}. However, since sputtering is not relevant in our single-fluid $J$-type accretion shock models, no significant increase of CH$_3$OH abundance is evident except at the highest densities ($\gtrsim10^8$~cm$^{-3}$) and velocities ($\gtrsim10$~\kms) where CH$_3$OH ice thermally sublimates.
 
\begin{figure}
\includegraphics[width=1\linewidth]{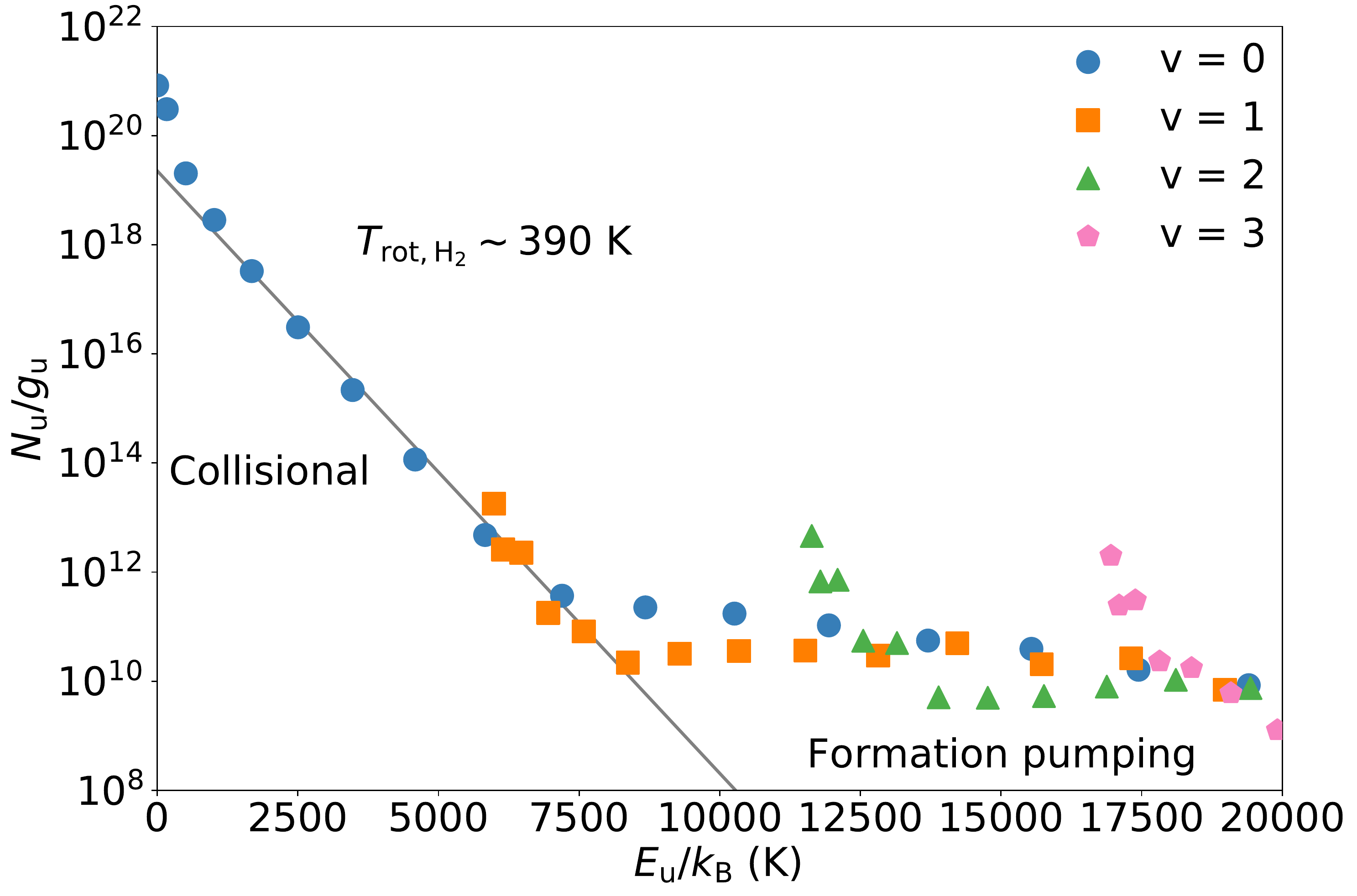}
\caption{Rotational diagram of H$_2$ created from the fiducial shock model. The levels of different vibrational states are indicated with different symbols and colors. Note that only the pure rotational levels with $E_\mathrm{u}/k_\mathrm{B} \lesssim 6000$~K are collisionally populated. The fit to the pure rotational levels is shown in gray line with the derived rotational temperature annotated.  
}
\label{fig:H2_rot}
\end{figure}

Emission from warm H$_2$ is generally observed toward shocks in protostellar jets \citep[e.g.,][]{McCaughrean1994}. Contrary to the molecules discussed above, the abundance of H$_2$ remains constant along the shock and only thermal excitation occurs. Hence, H$_2$ can be a powerfull diagnostic of accretion shocks in inferring the temperature from a rotational diagram. The population of the lowest 50 rotational and rovibrational levels of H$_2$ are computed during the shock \citep{Flower2003}.
%, taking into account radiative pumping by UV radiation \citep{Godard2019}
% and in very energetic shocks even dissociated \citep{Flower2010}. However, since the temperature of our accretion shocks does not reach the dissociation threshold of H$_2$ ($\sim10^4$~K), the abundance of H$_2$ remains constant along the shock. 
The rotational diagram of H$_2$ for the fiducial model is presented in Fig.~\ref{fig:H2_rot}. Since the temperature in this shock reaches about $\sim500$~K, levels up to $\sim6000$~K are collisionally populated. The population of all higher-$E_\mathrm{up}$ levels is set during the formation of H$_2$ on the dust grains following a Boltzmann distribution at 17248~K \citep[1/3 of the dissociation energy of H$_2$;][]{Black1987}. Using the rotational diagram of Fig.~\ref{fig:H2_rot}, a rotational temperature $\sim 390$~K can be derived for this shock.

\section{Discussion}
\label{sec:discussion}
\subsection{Comparison to SO and SO$_2$ with ALMA}
Emission of warm ($T_\mathrm{ex} > 50$~K) SO and SO$_2$ around the disk-envelope interface has been suggested as a possible tracer of accretion shocks \citep{Sakai2014,Sakai2017,Bjerkeli2019,Oya2019,Elizabeth2019}. Here, we have shown that the abundance of both SO and SO$_2$ can increase in shocks in typical inner envelope conditions. The abundance of SO is estimated to be $\sim10^{-8}$ in the suggested accretion shock around the L1527 Class 0/I protostar \citep{Sakai2014}, but this is uncertain since the density of H$_2$ was guessed. Furthermore, by itself this does not provide strong constraints on the physical conditions of the shock (see Fig.~\ref{fig:Vs_nH_SO_SO2}), other than that a weak UV field should be present if SO ice is not thermally sublimated ($G_0 > 10^{-2}$; Fig.~\ref{fig:Vs_nH_SO_diffG0}). However, it remains difficult to derive abundances with respect to \nH\ from observations. 

Recent ALMA observations show that the amount of warm SO$_2$ seems to increase with the bolometric luminosity \citep[i.e., stronger UV field;][]{Elizabeth2019}. This is in line with our results since the abundance of SO$_2$ only increases when a significant UV field is present (i.e., $G_0 \gtrsim 1$, see Fig.~\ref{fig:Vs_nH_SO2_diffG0}). Especially for higher velocity shocks ($>4$~\kms) at higher densities ($>10^7$~cm$^{-3}$) the maximum abundance of SO$_2$ increases with 2 orders of magnitude between $G_0 = 1$ and $G_0 = 100$. 
However, besides a significant UV field, a shock velocity of $V_\mathrm{s} \gtrsim 4$~\kms\ is necessary to increase the abundance of SO$_2$ and the maximum abundance reached in the shock is also dependent on the initial density \nH.

\begin{figure}
\includegraphics[width=\linewidth]{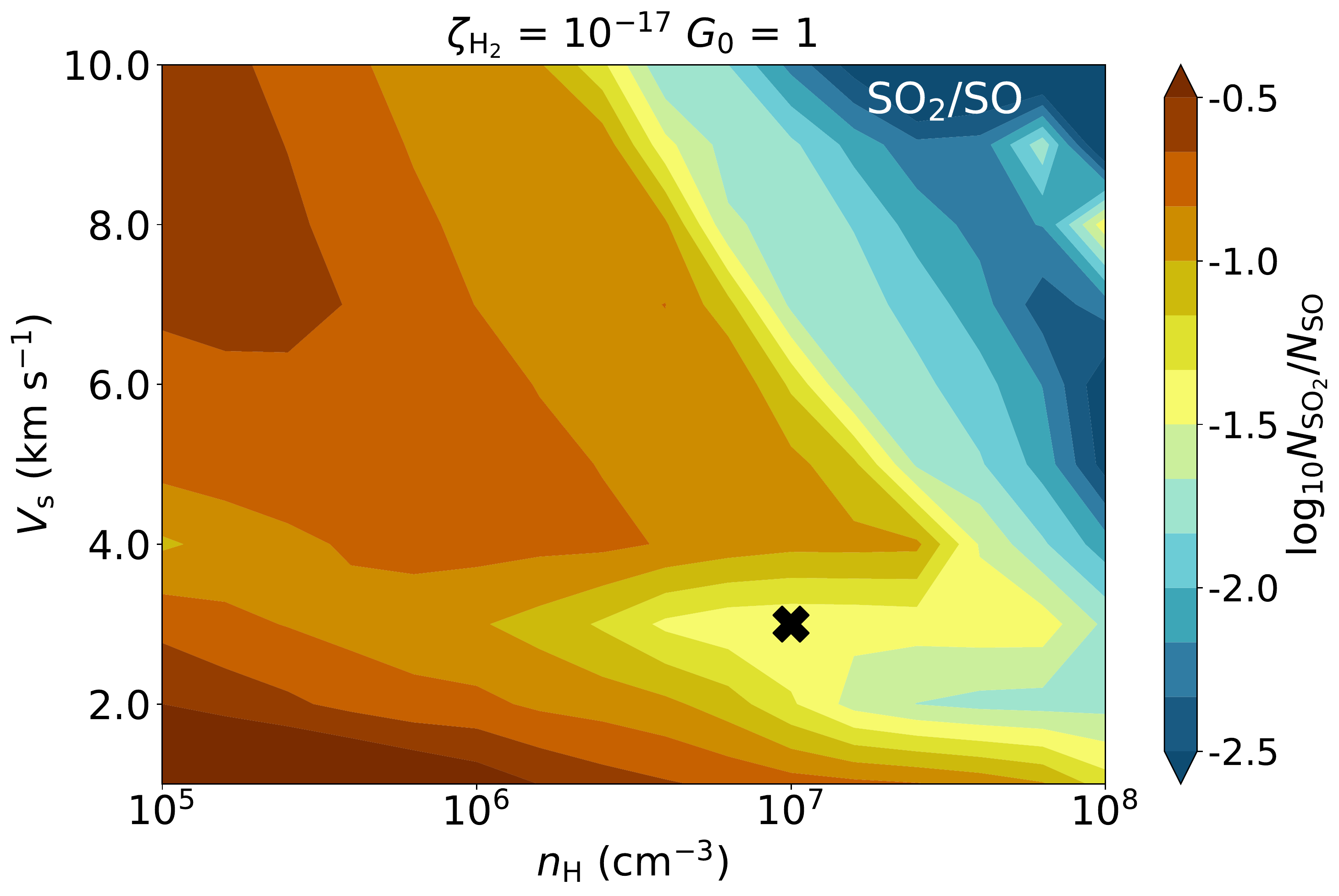}
\includegraphics[width=\linewidth]{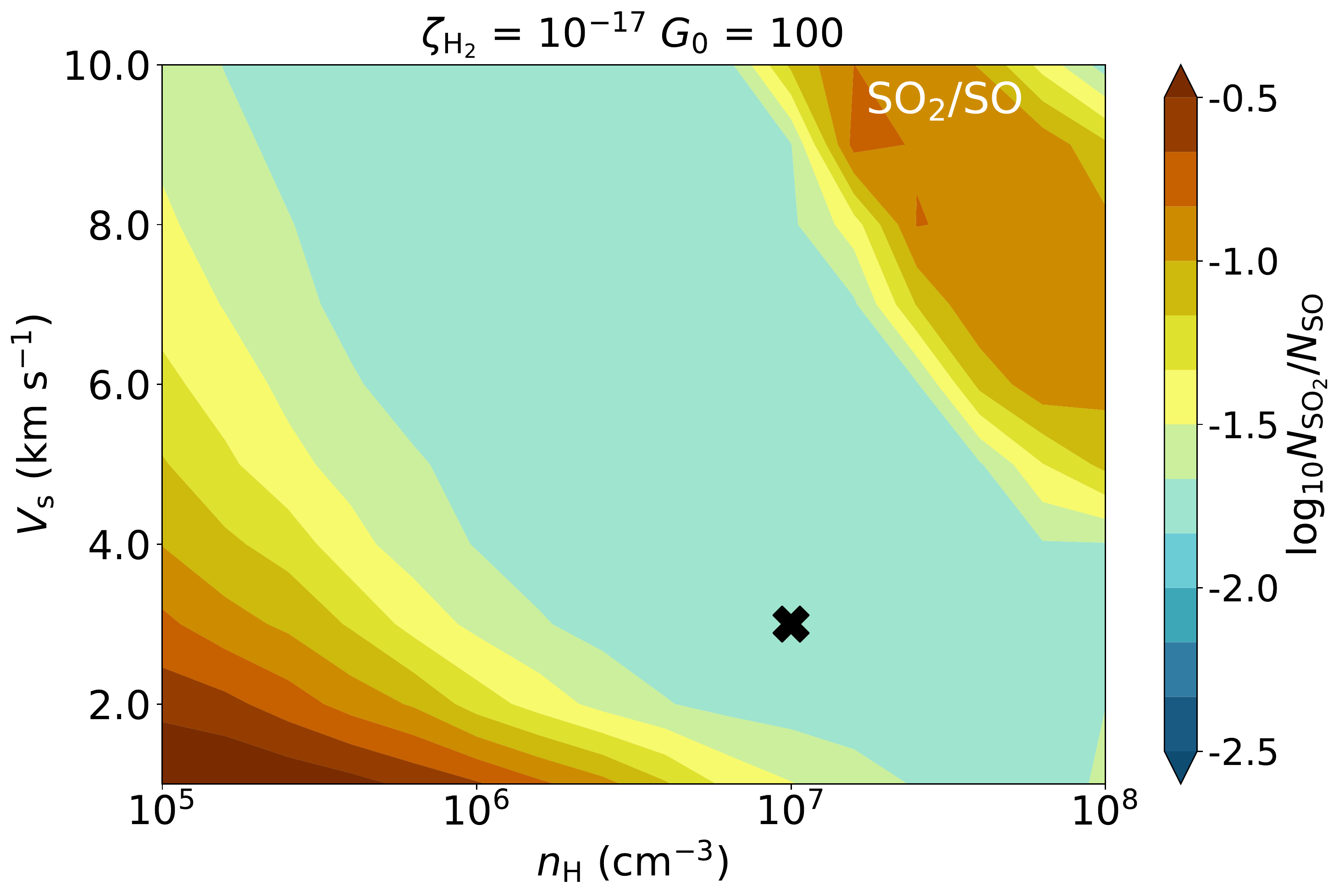}
\caption{
Ratio of column densities of SO$_2$ and SO as function of \nH\ and \Vs for $G_0 = 1$ (top) and $G_0 = 100$ (bottom). All other physical parameters are kept constant to the fiducial values and listed on top of the figure. The black cross indicates the position of the fiducial model.
}
\label{fig:Vs_nH_SO_SO2_column_ratio}
\end{figure}

Comparing the ratio of observed column densities to those derived from our shock models can more accurately constrain the physical conditions in a shock. 
In Fig.\ref{fig:Vs_nH_SO_SO2_column_ratio}, the column density ratio of SO$_2$/SO in the shock is presented as function of \Vs\ and \nH\ for $G_0 = 1$ and $G_0 = 100$. The only protostellar source where both SO and SO$_2$ have been detected related to a possible accretion shock is Elias~29 \citep{Oya2019}, where a column density ratio of $0.5^{+0.4}_{-0.2}$ is derived. 
When $G_0 = 1$, this would imply a low-density shock ($<10^6$~cm$^{-3}$) propagating at either low ($<2$~\kms) or high ($>7$~\kms) velocities, see upper panel in Fig.~\ref{fig:Vs_nH_SO_SO2_column_ratio}. For $G_0 = 100$, shocks propagating at higher velocities ($>4$~\kms) at high densities ($>10^7$~cm$^{-3}$) produce SO$_2$/SO column density ratios similar to that of Elias~29 \citep[][see lower panel in Fig.~\ref{fig:Vs_nH_SO_SO2_column_ratio}]{Oya2019}. 

Since both SO and SO$_2$ emission is also linked to outflow activity and passive heating in the inner envelope \citep{Codella2014,Tabone2017,Lee2018,Taquet2020,Harsono2021}, high spatial resolution observations with ALMA are necessary to both spatially and spectrally disentangle the different protostellar components and determine their emitting area. 
Moreover, to robustly test if the SO and SO$_2$ emission traces the accretion shock, key species for their gas-phase formation such as H$_2$S and H$_2$CO should be observed on similar scales. This also allows for more accurately constraining the physical conditions of a shock.
%Moreover, combining observations of SO and SO$_2$ with other tracers will better constrain the physical conditions. To robustly test SO and SO$_2$ as accretion shock tracers, key species for their formation such as H$_2$S and H$_2$CO should be observed on similar scales.

\subsection{Predicting H$_2$, H$_2$O, and [S\,{\sc i}] with JWST}
With the launch of JWST, near-infrared (NIR) and MIR shock tracers such as H$_2$, H$_2$O, and [S\,{\sc i}] can be used to search for accretion shocks. Although the abundance of H$_2$ does not increase in accretion shocks, it is still thermally excited and therefore a good shock tracer at NIR and MIR wavelengths \citep[e.g.,][]{McCaughrean1994}. 
%Using the rotational diagram of Fig.~\ref{fig:H2_rot}, a rotational temperature $\sim 390$~K can be derived for our fiducial model. 
Observing the rotational sequence of H$_2$ in the NIR and MIR could thus put constraints on the physical properties of the shock. For low-velocity shocks ($\lesssim5$~\kms), most of the strong H$_2$ transitions (i.e., 0-0~S(3) and 0-0~S(5)) are located in the JWST/MIRI range ($5-28$~$\mu$m).
%The sensitivity of JWST/MIRI is $\sim10^{-4}$~erg~cm$^{-2}$~s$^{-1}$~sr$^{-1}$ around the 0-0~S(3) and 0-0~S(5) lines of H$_2$ for an integration of $\sim1$~hour. For all shock models with $V\mathrm{s} > 2$~\kms, the estimated line intensity of the 0-0~S(3) and 0-0~S(5) lines of typically $\gtrsim5\times10^{-4}$~erg~cm$^{-2}$~s$^{-1}$~sr$^{-1}$ should result in a $>5\sigma$ detection. 
For higher velocity shocks ($\gtrsim5$~\kms), also vibrational levels of H$_2$ get populated and the strongest H$_2$ transitions shift toward the NIRSpec range of JWST ($1-5$~$\mu$m).

Warm water is also a well-known and important tracer of shocks \citep[e.g.,][]{Herczeg2012,Kristensen2012,Nisini2013,vanDishoeck2021}. Observing H$_2$O toward an accretion shock is a clear indication of a high-velocity shock ($\gtrsim4$~\kms, see Fig.~\ref{fig:Vs_nH_H2S_H2O_SiO_H2CO}). Warm water should be easily detectable with JWST as both the vibrational band around $6$~$\mu$m and several pure rotational transitions at $>10$~$\mu$m fall within the MIRI range ($5-28$~$\mu$m). Moreover, the detection of MIR lines of the OH radical can be an indication of H$_2$O photodissociation \citep{Tabone2021}, and therefore be spatially co-located with emission of warm SO and SO$_2$ in ALMA observations.

An alternative shock tracer in the JWST/MIRI band is the [S\,{\sc i}]~$25~\mu$m transition. It is often observed in protostellar outflows \citep[e.g.,][]{Neufeld2009,Lahuis2010,Goicoechea2012}, where the intensity of the line can range from a few \% of the H$_2$ 0-0~S(3) and 0-0~S(5) lines to a factor of a few higher. It is therefore uncertain whether [S\,{\sc i}]~$25~\mu$m will be detectable toward accretion shocks, also given that the sensitivity of MIRI drops at the highest wavelengths. Moreover, at $25$~$\mu$m, the angular resolution of JWST is $\sim1"$ (i.e., 300~AU at a distance of 300~pc) which complicates the spatial disentanglement of an accretion shock with outflow emission.

%\subsection{Thermal desorption vs gas-phase formation}
%\mvg{Short section on whether COMs are there (probably not...)}

%\subsection{Effect on inheritance vs. reset scenario}
%\mvg{What is the effect of the results in the paper on the inheritance vs. reset scenario?}

\section{Summary}
\label{sec:summary}
We have modeled low-velocity ($\leq 10$~\kms) non-magnetized \mbox{$J$-type} shocks at typical inner envelope densities ($10^{5}-10^{8}$~cm$^{-3}$) using the Paris-Durham shock code to test SO and SO$_2$ as possible molecular tracers of accretion shocks. The main conclusions of this work are as follows:
\begin{itemize}
\item In low-velocity ($\sim3$~\kms) shocks, SO can be efficiently formed at intermediate densities ($\sim10^7$~cm$^{-3}$) through the SH radical reacting with atomic O. Here, SH is formed through S$^+$ reacting primarily with H$_2$CO; though the latter species is a representative of any hydrocarbon. 

\item Both SO and SO$_2$ can be efficiently formed through reactions of atomic S with OH in high-velocity ($\gtrsim4$~\kms) shocks in low-intermediate dense environments ($\lesssim10^7$~cm$^{-3}$)
%, given that strength of the local UV radiation field is equal or larger than the ISRF
. The formation of SO and SO$_2$ occurs in the end of the shock, where the abundance of OH is enhanced through photodissociation of earlier formed H$_2$O.

\item Thermal desorption of SO ice can occur in high-velocity ($\gtrsim4$~\kms) shocks at high densities ($\gtrsim10^7$~cm$^{-3}$), while thermal desorption of SO$_2$ ice is only relevant at the highest shock velocities ($\gtrsim10$~\kms) and highest densities ($\gtrsim10^8$~cm$^{-3}$).

\item The chemistry of both SO and SO$_2$ through all formation routes is strongly linked to the strength of the local UV radiation field since the formation of both the OH and SH radicals is linked to photon processes such as photodissociation and photoionization. In particular, SO$_2$ is only formed in the shock through gas-phase chemistry if the strength of the local UV radiation field is greater or equal to the ISRF.

\item The composition of infalling material, in both gas-phase species like atomic O and S and in the ices such as H$_2$S, CH$_4$, SO, and SO$_2$, is highly relevant for the chemistry of SO and SO$_2$ in accretion shocks.

%\item The size of the dust grains has a strong impact on both the physical and chemical structure of an accretion shock. 
%For typical ISM size dust grains, photodesorption and subsequent photodissociation of key ices such as H$_2$O and H$_2$S already increases the abundance SO and SO$_2$ in the pre-shock medium, leaving little formation left in the shock itself. 
%Increasing the abundance of PAHs to the ISM level decreases the abundance of SO and SO$_2$ as the abundance of key ions such as S$^+$ is decreased.

%\item In strongly magnetized environments ($\sim1$~mG), the decoupling between ions and neutrals is too weak to create $C$-type shocks on disk scales ($<100$~AU). For intermediate magnetic field strengths ($\sim100$~$\mu$G), a shock on disk scales is possible and the abundances of both SO and SO$_2$ increase in the shock through similar chemistry as in $J$-type shocks.

\item Observations of warm SO and SO$_2$ should be complemented with key species for their formation such as H$_2$S and H$_2$CO. Moreover, the launch of JWST will add additional NIR and MIR accretion shock tracers such as H$_2$, H$_2$O, and [S\,{\sc i}]~$25~\mu$m.
\end{itemize} 

Our results highlight the key interplay between both physics and chemistry in accretion shocks. The abundance of gas-phase molecules such as SO and SO$_2$ is not solely determined through high-temperature gas-phase chemistry, but also thermal sublimation of ices and photodissociation through UV radiation play an important role. Additional high-angular resolution observations with ALMA are necessary to spatially disentangle the disk from the envelope in embedded systems in order to assess the physical properties of accretion shocks at the disk-envelope interface. Moreover, future facilities such as JWST will provide additional NIR and MIR shock tracers such as H$_2$, OH, and H$_2$O. It is crucial to attain the physical structure of accretion shocks since it determines the chemical composition of material that enters the disk and is eventually incorporated in planets.

\begin{acknowledgements}
The authors would like to thank the anonymous referee for their constructive comments on the manuscript. The authors would also like to thank G. Pineau des For{\^e}ts for valuable discussions about the project. Astrochemistry in Leiden is supported by the Netherlands Research School for Astronomy (NOVA). MvG and BT acknowledge support from the Dutch Research Council (NWO) with project number NWO TOP-1 614.001.751.
\end{acknowledgements}

%-------------------------------------------------------------------

\bibliographystyle{aa}
\bibliography{refs}

\clearpage

\begin{appendix}
\section{Updated cooling of NH$_3$}
\label{app:NH3_cool}
Typically, cooling by NH$_3$ in the optically thin limit is calculated using the empirical relation derived by \cite{Lebourlot2002}. However, this relation does not take into account that the populations get thermalized if the density gets above the critical density. Therefore, the cooling by NH$_3$ is recalculated using the RADEX software \citep{vanderTak2007}. Assuming all emission is optically thin (i.e., low column densities) and emitted through rotational lines, a grid of gas temperatures and H$_2$ densities is set. The full range explored is $5-5000$~K and $10^{2}-10^{12}$~cm$^{-3}$, respectively. The linewidth is set to $2$~\kms. The cooling (in erg~s$^{-1}$) for each grid point is calculated by summing over the emission of each spectral line and dividing that by the assumed column density.

Using the cooling grid computed with RADEX, a cooling function, $\Lambda$, with $T_\mathrm{gas}$ and $n_\mathrm{H_2}$ as free parameters can be derived. Minimum $\chi^2$ fitting of the grid computed with RADEX was used to calculate $\Lambda$, which resulted in,
\begin{align}
\Lambda_{\mathrm{NH}_3} = \Lambda_\infty \left(1 - \exp\left[\frac{-n_{\mathrm{H}_2}}{n_{\mathrm{0}}}\right]\right), \label{eq:NH3_cool}
\end{align}
%\begin{align}
%\Lambda_{\mathrm{NH}_3}~(\mathrm{erg~cm^{-3}~s^{-1}}) = \Lambda_\infty \left(1 - \exp\left[\frac{-n_{\mathrm{H}_2}}{n_{\mathrm{0}}}\right]\right) n_{\mathrm{NH}_3}, \label{eq:NH3_cool}
%\end{align}
with $\Lambda_\infty$ and $n_{\mathrm{0}}$ numerically derived as,
\begin{align}
\Lambda_\infty & = 2.85 \times 10^{-15} \left( 1 - \exp\left[\frac{-T_\mathrm{gas}}{81.7}\right] \right)^4, \\
n_{\mathrm{0}} & = 8.33 \times 10^{8} \left( 1 - \exp\left[ \frac{-T_\mathrm{gas}}{125} \right] \right).
\end{align}
$\Lambda_\infty$ is the cooling of NH$_3$ in the LTE limit, whereas $n_{\mathrm{0}}$ is a measure of a critical density above which LTE effects become relevant. Both parameters are dependent on $T_\mathrm{gas}$. 

\begin{figure}
\includegraphics[width=\linewidth]{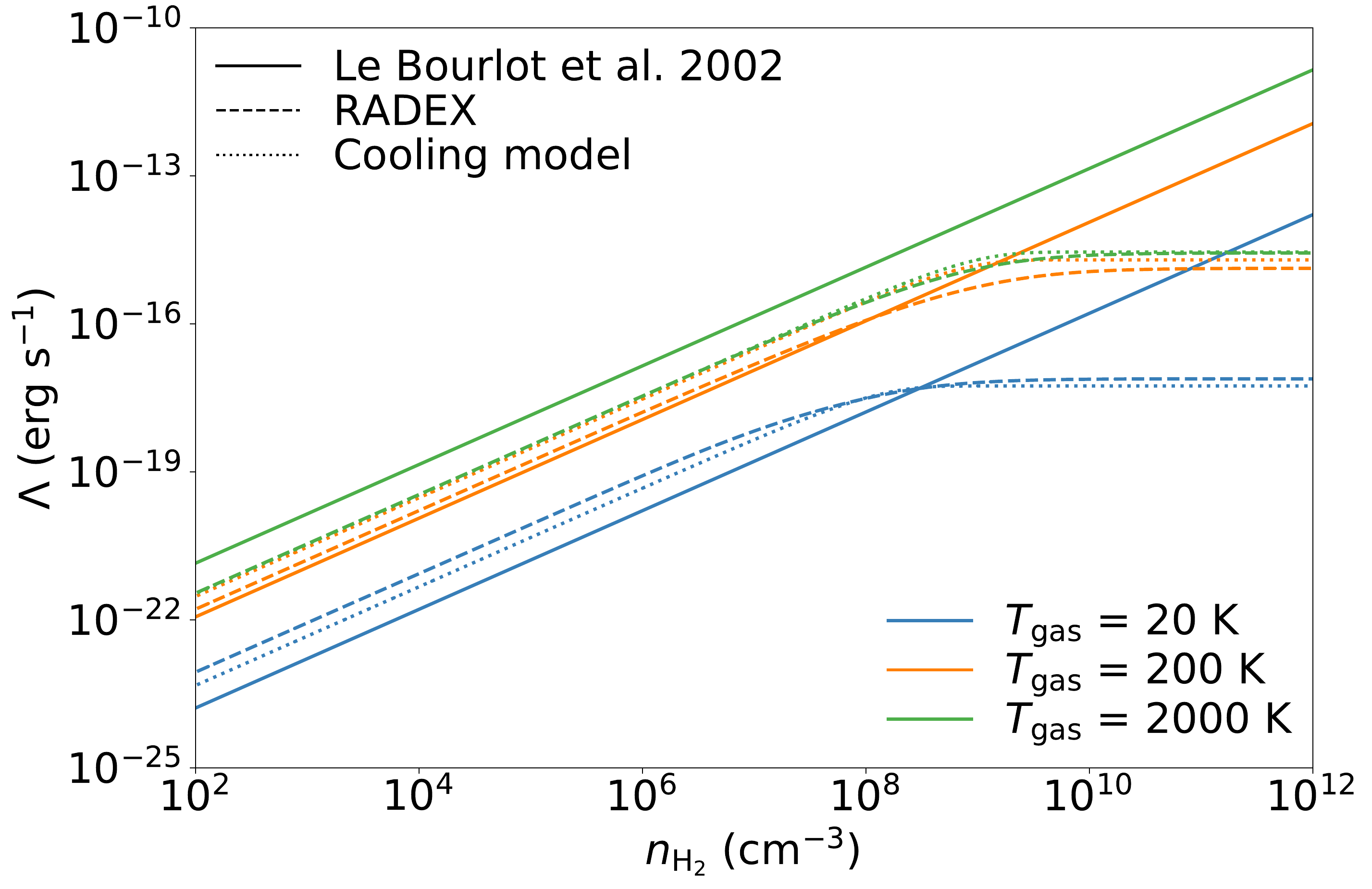}
\caption{
Cooling by NH$_3$ as function of $n_\mathrm{H_2}$ for various $T_\mathrm{gas}$ (colors). The empirical relation by \citet{Lebourlot2002} is shown as the solid lines, the RADEX computations in the dashed line, and the derived cooling function of Eq.~\eqref{app:NH3_cool} in the dotted line.
}
\label{fig:Cooling_NH3}
\end{figure}

The function in Eq.~\eqref{eq:NH3_cool} is presented in Fig.~\ref{fig:Cooling_NH3} for a few values of $T_\mathrm{gas}$. Overplotted are the RADEX calculations and the empirical relation of \citet{Lebourlot2002}. Indeed, at high densities ($\gtrsim 10^9$~cm$^{-3}$) the latter relation starts overproducing the cooling computed with RADEX by orders of magnitude. However, the updated cooling function represent the cooling by RADEX within a factor of 3 over the entire temperature and density range. Since densities of $\gtrsim10^9$~cm$^{-3}$ are frequently reached in our shock models, the updated cooling function was adopted in the updated version of the shock code.

In high-temperature shocks, significant amounts of NH$_3$ might be produced through gas-phase chemistry, enhancing the density and thus the optical depth of the spectral lines. 
Moreover, the density of NH$_3$ can be increased through ice sublimation in high-density shocks. If the lines of NH$_3$ become optically thick, Eq.~\ref{eq:NH3_cool} is no longer valid and cooling tables similar to those of \citet{Neufeld1993} should be calculated for NH$_3$ taking into account opacity effects. This is beyond the scope of this work.

%\mvg{figure comparing \citet{Lebourlot2002}, grid model, and derived equation to be added.}
%The full derivation of Eq~\eqref{eq:NH3_cool} is presented in Appendix~\ref{app:NH3_cool}.

\section{Input abundances}
%\subsection{Input abundances}
\label{app:input_abundances}
The input abundances used in the shock models are presented in Table~\ref{tab:input_abundances}. The total elemental abundances match those of  \citet{Flower2003}. Most of the oxygen, carbon, and sulfur budget is locked up in dust grains and in the ice mantles. The initial abundances of oxygen and nitrogen-bearing ices are equivalent to those presented by \citet{Boogert2015}, while the sulfur-bearing ice abundances are matched to those found in both protostellar envelopes and comets \citep{Calmonte2016,Rubin2019,Altwegg2019,Navarro2020}. The gas-phase abundances of the key sulfur-bearing molecules, SO, SO$_2$, CS, OCS, and H$_2$S, are taken from \citet{vanderTak2003}. All other molecular gas-phase abundances are set equal to those derived on cloud scales by \citet{Tafalla2021}. The gas-phase abundance of atomic S in protostellar envelopes is highly uncertain. It is set here at a $10^{-6}$ level, which is consistent with the models of \citet{Goicoechea2021}. The effect of initial atomic S abundance on the maximum abundances of SO and SO$_2$ attained in the shock is presented in Sect.~\ref{subsubsec:init_conditions}.

\begin{table}[h]
\centering
\caption{Dark cloud input abundances}
\label{tab:input_abundances}
\begin{tabular}{ll|ll}
\hline \hline
$\rm H$          & $5.0 \times 10^{-5}$   & $\rm Grains$     & $4.6 \times 10^{-11}$ \\
$\rm H_2$        & $5.0 \times 10^{-1}$   & s-$\rm H_2O$     & $1.0 \times 10^{-4}$  \\
$\rm He$         & $1.0 \times 10^{-1}$   & s-$\rm O_2$      & $1.0 \times 10^{-8}$  \\
$\rm C$          & $1.0 \times 10^{-8}$   & s-$\rm CO$       & $8.3 \times 10^{-6}$  \\
$\rm O$          & $1.0 \times 10^{-6}$   & s-$\rm CO_2$     & $1.3 \times 10^{-5}$  \\
$\rm CO$         & $8.5 \times 10^{-5}$   & s-$\rm CH_4$     & $1.5 \times 10^{-6}$  \\
$\rm C_2H$       & $2.0 \times 10^{-9}$   & s-$\rm N_2$      & $3.9 \times 10^{-6}$  \\
$\rm C_3H_2$     & $2.0 \times 10^{-10}$  & s-$\rm NH_3$     & $1.0 \times 10^{-6}$  \\
$\rm CH_3OH$     & $1.5 \times 10^{-9}$   & s-$\rm CH_3OH$   & $1.9 \times 10^{-5}$  \\
$\rm N$          & $1.0 \times 10^{-6}$   & s-$\rm H_2CO$    & $6.2 \times 10^{-6}$  \\
$\rm CN$         & $1.5 \times 10^{-9}$   & s-$\rm HCO_2H$   & $7.2 \times 10^{-6}$  \\
$\rm HCN$        & $3.0 \times 10^{-9}$   & s-$\rm OCS$      & $2.1 \times 10^{-7}$  \\
$\rm HNC$        & $1.0 \times 10^{-9}$   & s-$\rm H_2S$     & $1.8 \times 10^{-5}$  \\
$\rm N_2$        & $3.5 \times 10^{-5}$   & s-$\rm SO$       & $1.0 \times 10^{-7}$  \\
$\rm S$          & $1.0 \times 10^{-6}$   & s-$\rm SO_2$     & $1.0 \times 10^{-7}$  \\
$\rm H_2S$       & $1.0 \times 10^{-9}$   & c-$\rm O$        & $1.4 \times 10^{-4}$  \\
$\rm CS$         & $3.0 \times 10^{-9}$   & c-$\rm Si$       & $3.4 \times 10^{-5}$  \\
$\rm SO$         & $1.0 \times 10^{-9}$   & c-$\rm Mg$       & $3.7 \times 10^{-5}$  \\
$\rm SO_2$       & $1.0 \times 10^{-9}$   & c-$\rm Fe$       & $3.2 \times 10^{-5}$  \\
$\rm OCS$        & $2.0 \times 10^{-9}$   & c-$\rm C$        & $1.6 \times 10^{-4}$  \\
$\rm Si$         & $1.0 \times 10^{-10}$  & $\rm HCO^+$      & $1.5 \times 10^{-9}$  \\
$\rm Fe$         & $1.5 \times 10^{-8}$   & $\rm N_2H^+$     & $1.5 \times 10^{-10}$ \\
$\rm PAH$        & $1.0 \times 10^{-8}$  \\
%$\rm G$          & $4.6 \times 10^{-11}$ \\
\hline
\end{tabular}
\tablefoot{Abundances are with respect to \nH. Species in ice mantles on dust grains are indicated as s-X. Species in composing the cores of dust grains are denoted as c-X.}
\end{table}

%Atomic S at $10^{-6}$ level: \citet{Goicoechea2021}. Most S-bearing species in gas: \citet{vanderTak2003}. Most gas-phase species in clouds: \citet{Tafalla2021}. Ice abundances: \citet{Boogert2015,Navarro2020}.

%\subsection{Pre-shock calculations}
\section{Changing the pre-shock conditions}
\label{app:pre-shock_mod}
To test the effect of alteration of infalling dark cloud material in the inner envelope on an accretion shock, various pre-shock models are calculated. Since the increase of the temperature and protostellar radiation field are negligible in the outer parts of the envelope \citep[e.g.,][]{Drozdovskaya2015}, changes in temperature, composition, and ionization balance are most efficient in the inner $\sim1000$~AU of the envelope when the material is already approaching the disk and protostar. Given that the infall velocity goes as $v(r) = \sqrt{2GM/r}$, it takes about $\sim100$~years to move from $1000$~AU to disk scales ($\sim100$~AU) for a $0.5$~M$_\odot$ star. However, depending on the mass of the protostar and the size of the disk, this pre-shock timescale $t_\mathrm{pre-shock}$ might range from, for example, $\sim10-10^{4}$~years. Therefore, the effect of infall time (i.e., $t_\mathrm{pre-shock}$) on the accretion shock is explored here. The thermal balance is included in the pre-shock model, but for simplicity the density is assumed to remain constant.

%Therefore, the pre-shock model runs only for as long as the infall takes from $\sim1000$~AU to the size of the disk ($\sim100$~AU). Given that the infall velocity goes as $v(r) = \sqrt{2GM/r}$, this is about $\sim100$~years for a $0.5$~M$_\odot$ star.
%, assuming the infall velocity goes as $v(r) = \sqrt{2GM/r}$ for a $0.5$~M$_\odot$ star. 
%However, to test the effect of the duration of the pre-shock calculation on the shock itself, infall timescales of between 10 and 10$^{4}$ years are considered. %The thermal balance is included in the pre-shock model, and the density is assumed to remain constant in this pre-shock model.
%However, here we test infall timescales from 10 years to $10^{4}$ years.
%The density is assumed to remain constant in this pre-shock model.

%The duration of the pre-shock model is set to 100 years, equal to the time infalling material takes to get from $\sim1000$~AU to$\sim100$~AU, assuming the infall velocity goes as $v(r) = \sqrt{2GM/r}$ for a $0.5$~M$_\odot$ star. However, in this 100 years, no chemical equilibrium is reached, and thus the results may differ for different pre-shock durations. 

Given that the infalling material is ionized and photodissociated on a timescale of \citep{Heays2017},
\begin{align}
\tau_\mathrm{ionize/disso} \sim \frac{10-100}{G_0} \mathrm{~yrs},
\label{eq:tau_ionize}
\end{align}
the ionization balance is fully altered for all $t_\mathrm{pre-shock} \gtrsim 10$~years if the strength of the local UV field is $G_0 \gtrsim 1$. For models with weaker UV fields ($G_0 = 10^{-3}-10^{-1}$), the ionization balance is dependent on $t_\mathrm{pre-shock}$. 
Moreover, the ice composition may be altered on the photodesorption and adsorption timescales for 0.2~$\mu$m grains of \citep{Hollenbach2009,Godard2019},
\begin{align}
\tau_\mathrm{photodes} & \approx \frac{2\times10^5}{G_0} \mathrm{~yrs}, \label{eq:tau_photodes} \\ 
\tau_\mathrm{adsorb} & \approx \frac{10^{9}-10^{10}}{n_\mathrm{H}} \mathrm{~yrs}.
\end{align}
It is important to note here that if $\tau_\mathrm{ionize/disso} < \tau_\mathrm{adsorb}$, any photodesorbed species are dissociated before they can adsorb back on the grain. This increases the abundance of atomic and ionic species.

\begin{figure}
\includegraphics[width=\linewidth]{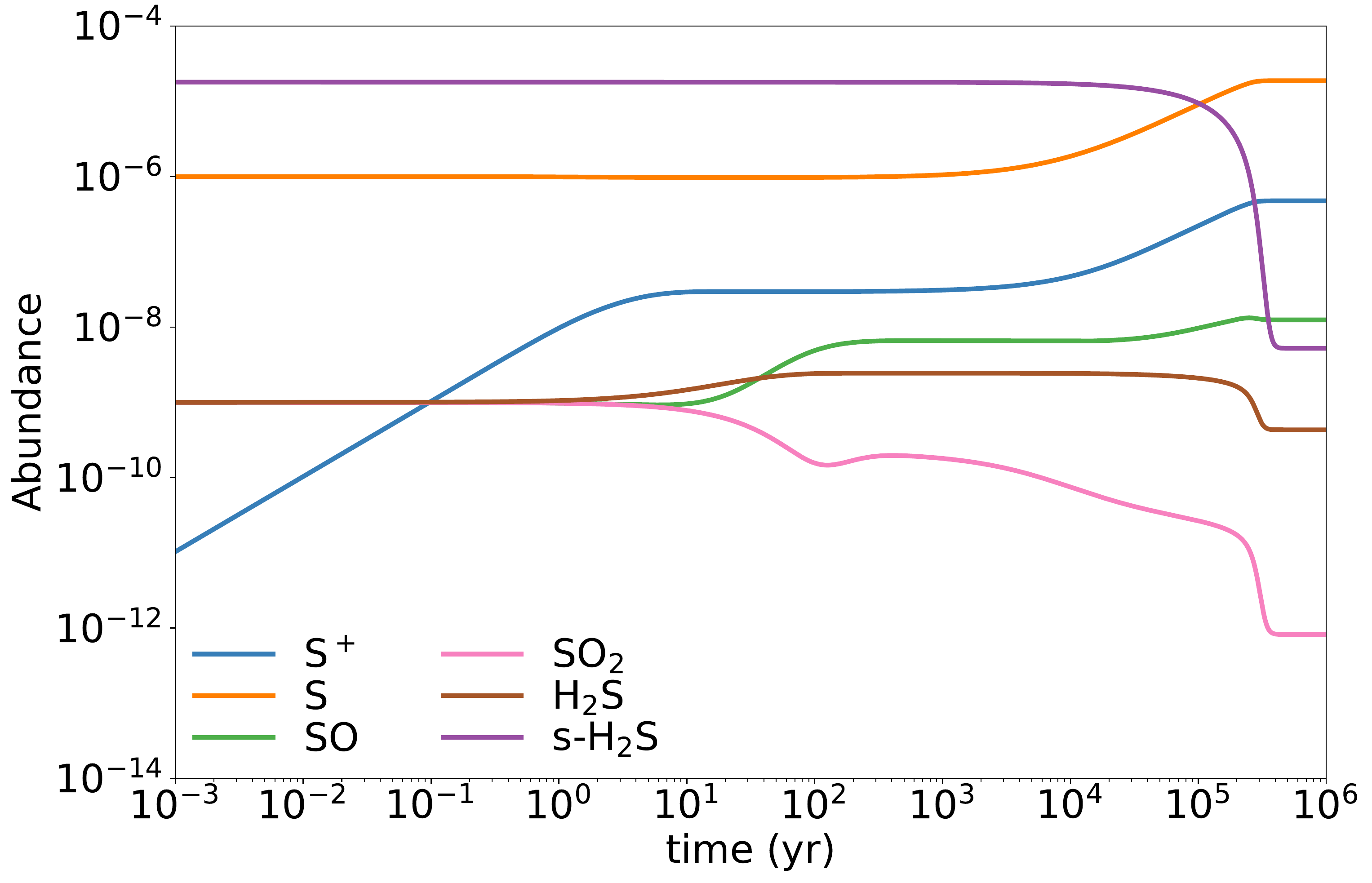}
\caption{Abundances of S$^+$, S, SO, SO$_2$, H$_2$S, and s-H$_2$S during a pre-shock calculation. All physical parameters are kept constant to the fiducial values.}
\label{fig:SS_fidu}
\end{figure}

The abundances of S$^+$, S, SO, SO$_2$, H$_2$S, and s-H$_2$S during a pre-shock calculation are presented in Fig.~\ref{fig:SS_fidu} for the fiducial model conditions (see Table~\ref{tab:phys_params}). Following Eq.~\eqref{eq:tau_ionize}, the ionization balance of S$^+$/S is set after $\sim10$~years. However, at longer timescales of $\sim10^{4}-10^{5}$~years, the abundances of both S$^+$ and S increase above the initial abundance of atomic S. On these timescales, H$_2$S ice is photodesorbed (see Eq.~\eqref{eq:tau_photodes}) and subsequently photodissociated, therefore enhancing the abundance of both atomic S and S$^+$. It is interesting to note that since the abundance of H$_2$S is much higher than that of atomic S in the gas phase, already on timescales of $0.1\tau_\mathrm{photodes}$ the abundance of atomic S is increased with a factor of a few. Meanwhile, the abundance of SO is increased through gas-phase chemistry via Reaction~\eqref{reac:S_OH}, and SO$_2$ is photodissociated into SO and atomic O.

\begin{figure*}
\includegraphics[width=0.49\linewidth]{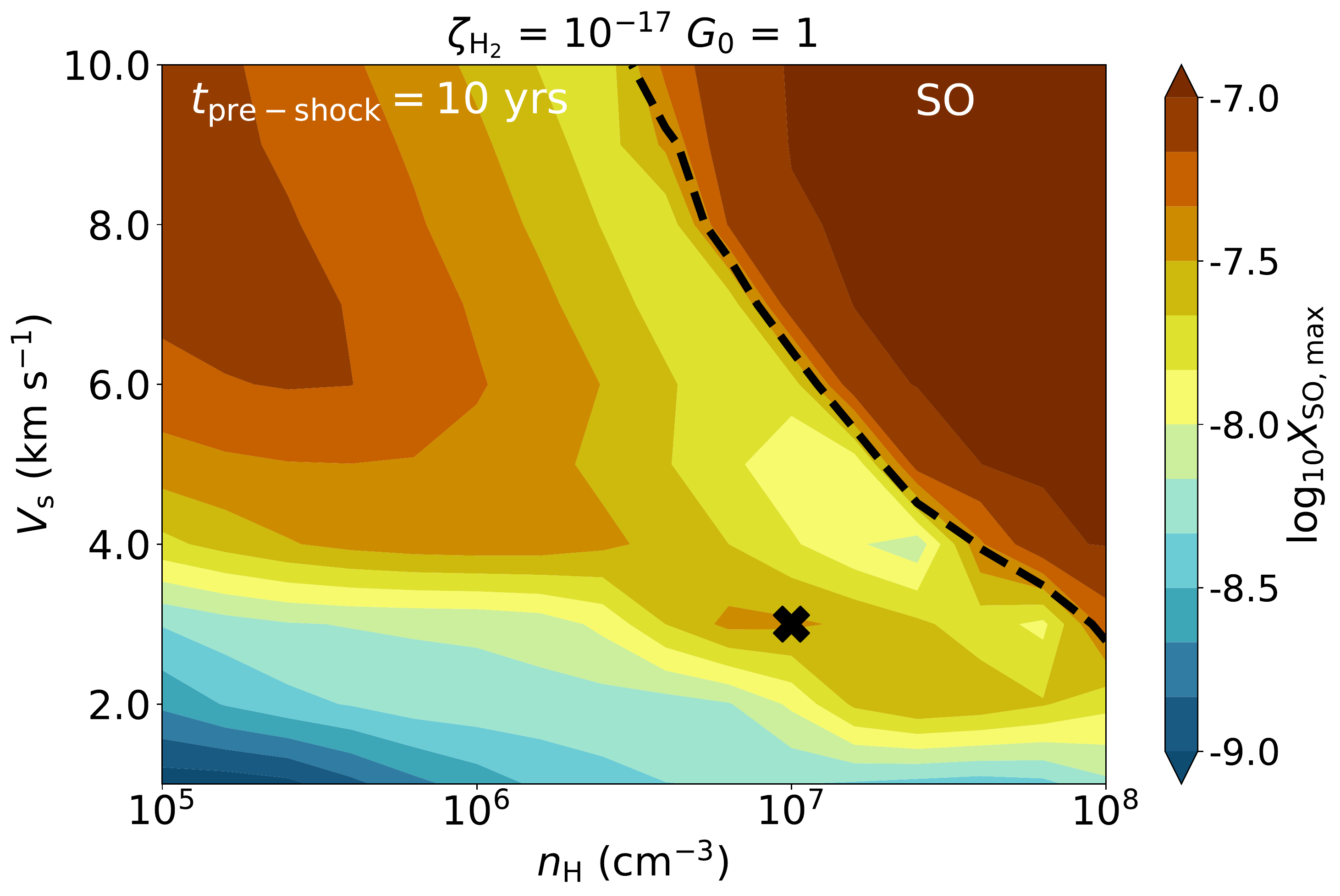}
\includegraphics[width=0.49\linewidth]{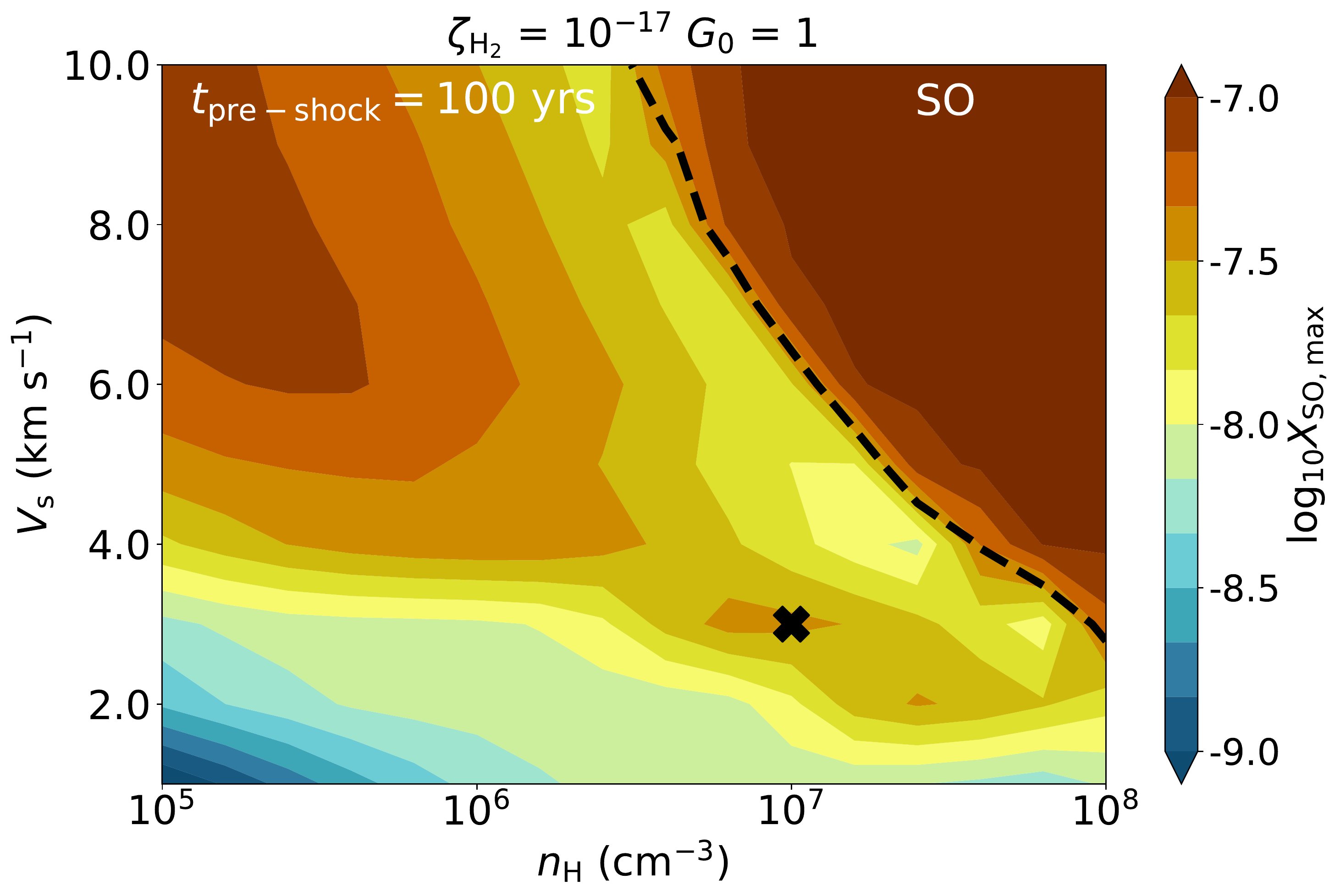}
\includegraphics[width=0.49\linewidth]{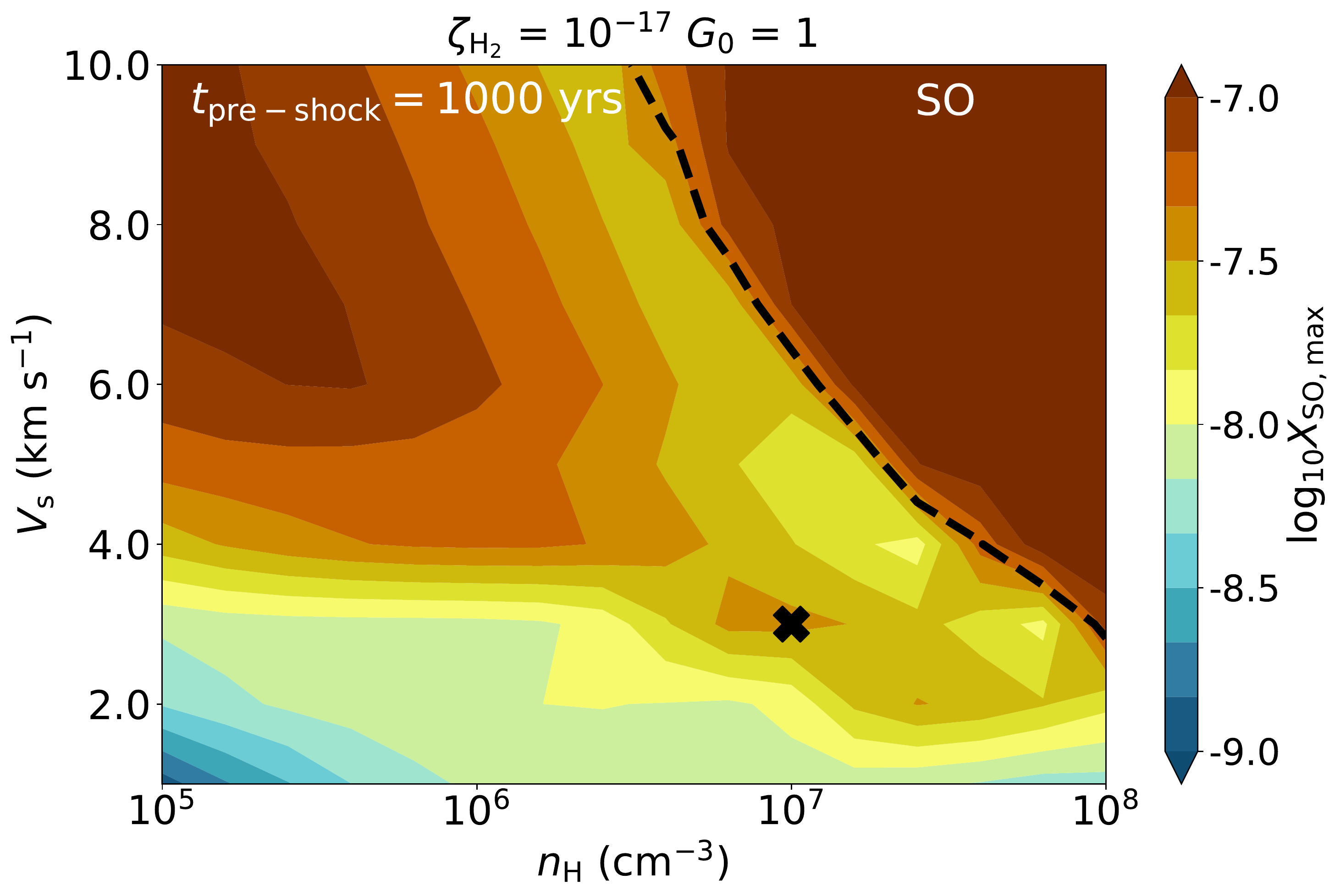}
\includegraphics[width=0.49\linewidth]{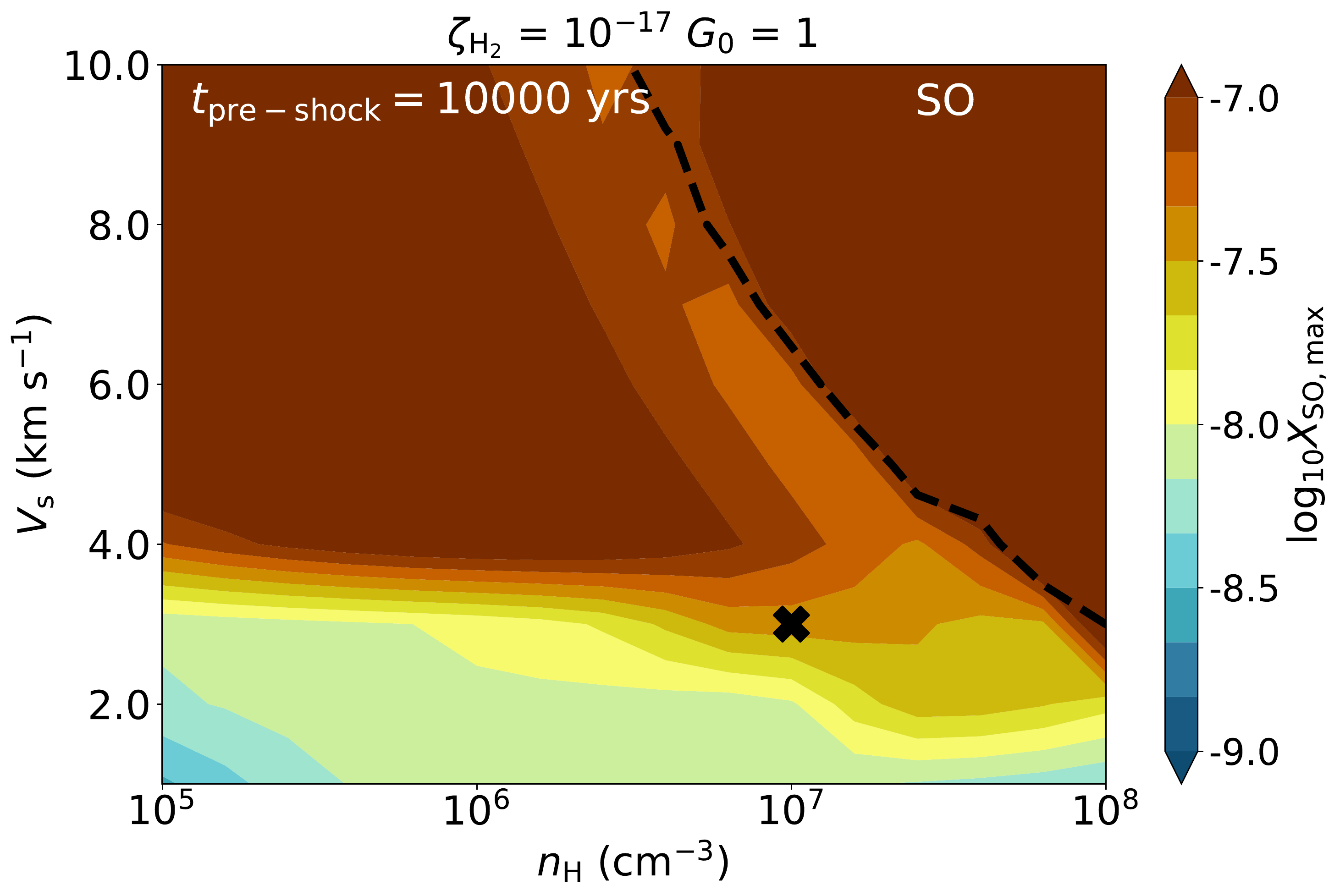}
\caption{
Maximum abundance reached (in color) of SO in shock models as function of initial \nH\ and \Vs after calculating a pre-shock model for 10 (top left), 100 (top right), 1000 (bottom left), and 10000 (bottom right) years. All other physical parameters are kept constant to the fiducial values and listed on top of the figure. The black cross indicates the position of the fiducial model. The dashed black line shows the ice line, i.e., where 50\% of the ice is thermally desorbed into the gas in the shock.
%Similar figure as Fig.~\ref{fig:Vs_nH_SO_SO2} but for SO after calculating a pre-shock model for 10 (top left), 100 (top right), 1000 (bottom left), and 10000 (bottom right) years. }
}
\label{fig:SS_SOgrid}
\end{figure*}

The effect of the length of the pre-shock model on the maximum abundances of SO in the subsequent accretion shock is presented in Fig.~\ref{fig:SS_SOgrid}. Up to $t_\mathrm{pre-shock}\lesssim 1000$~years, the maximum abundance reached in the shock is independent of the time over which the pre-shock model is calculated. In the $t_\mathrm{pre-shock} = 10000$~years case, photodesorption and subsequent photodissociation of H$_2$S increases the initial atomic S abundance and hence the maximum abundance of SO reached in the shock. For different strengths of the UV radiation field (i.e., different \G0), the timescale for photodesorption changes (i.e., Eq.~\eqref{eq:tau_photodes}). $t_\mathrm{pre-shock}$ does thus not affect the final abundances of SO (and SO$_2$) in the shock if less than roughly $\sim10\%$ of the photodesorption timescale $\tau_\mathrm{photodes}$, given that $\tau_\mathrm{adsorb} > \tau_\mathrm{ionize/disso}$. 

%\section{Timescales for chemical processes}
%All values for $k$ are given in $s^{-1}$ units
%\subsection{Desorption processes}
%Useful quantities: 
%\begin{align}
%f(H2) & = \frac{0.46 n(H) + n(H_2)}{n(H) + n(H_2)} \frac{n(H_2)}{n(H) + n(H_2)} \\
%\end{align}
%Rates: 
%\begin{align}
%k_{photodes} & = Y G_0 F_0 \pi <r_m^2> n_g f(X) n(X)^{-1} & \propto G_0 a_g^2 a_g^{-3} n(X) n(X)^{-1} & \propto G_0 a_g^{-1} \\
%k_{secphotodes} & = Y \frac{7.1\times10^{-11}}{n_g / n_H} \zeta (T_n/300.)^{\alpha} f(H2) Y n_g f(X) n(X)^{-1} & \propto \zeta a_g^{3} a_g^{-3} f(H_2) n(X) n(X)^{-1} & \propto \zeta \\
%k_{thrmdes} & = 1.6 \times 10^{11} \sqrt{\frac{E_b}{m(X)/m(H)}} e^{-E_b/T_g} & & \propto e^{-1/T_g} \\
%k_{crdes} & = Y \frac{\zeta}{10^{-17}} n_g \pi <r_m^2> e^{-(E_b-850)/70} f(X) n(X)^{-1} & \propto \zeta a_g^{-3} a_g^2 n(X) n(X)^{-1} & \propto \zeta a_g^{-1}
%\end{align}
%
%\subsection{Adsorption processes}
%\begin{align}
%k_{adsor} & = \frac{Y \pi n_g <r_m^2> \sqrt{\frac{8 k_B T_{g,eff}}{m(X)/\pi}}}{1.0 + 0.04(T_{g,eff} + T_g)^{\frac{1}{2}} + 2\times10^{-3} T_{g,eff} + 8\times10^{-6}T_{g,eff}^2)} & \propto a_g^2 a_g^{-3} & \propto a_g^{-1}
%\end{align}

%\section{Fiducial model shock profiles}
%\label{app:fidu_model}

\section{Abundance grids}
\label{app:abundance_grids}
\subsection{H$_2$S, H$_2$O, SiO, and CH$_3$OH for $G_0=1$}
\label{app:H2S_H2O_SiO_H2CO}
\begin{figure*}
\includegraphics[width=0.49\linewidth]{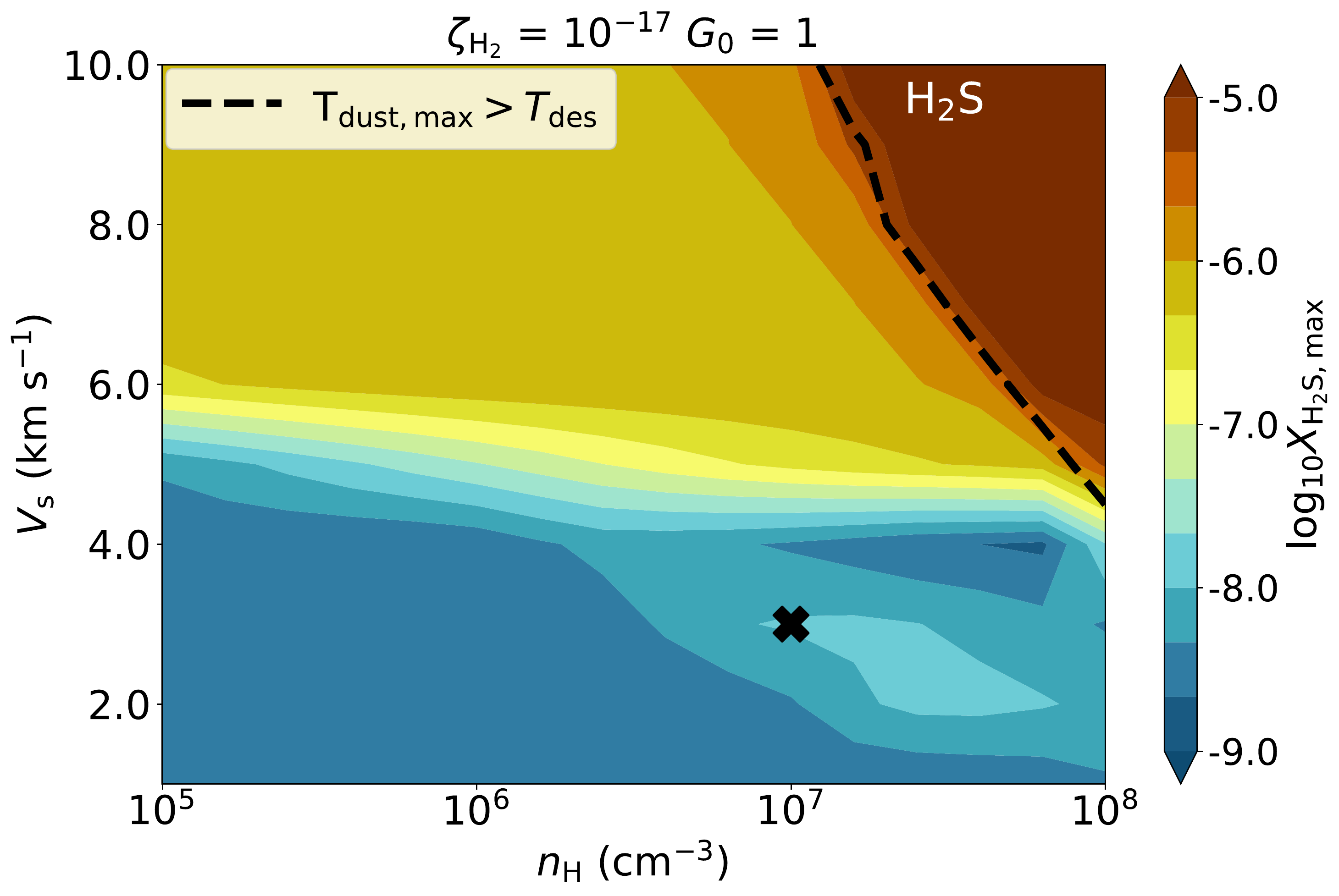}
\includegraphics[width=0.49\linewidth]{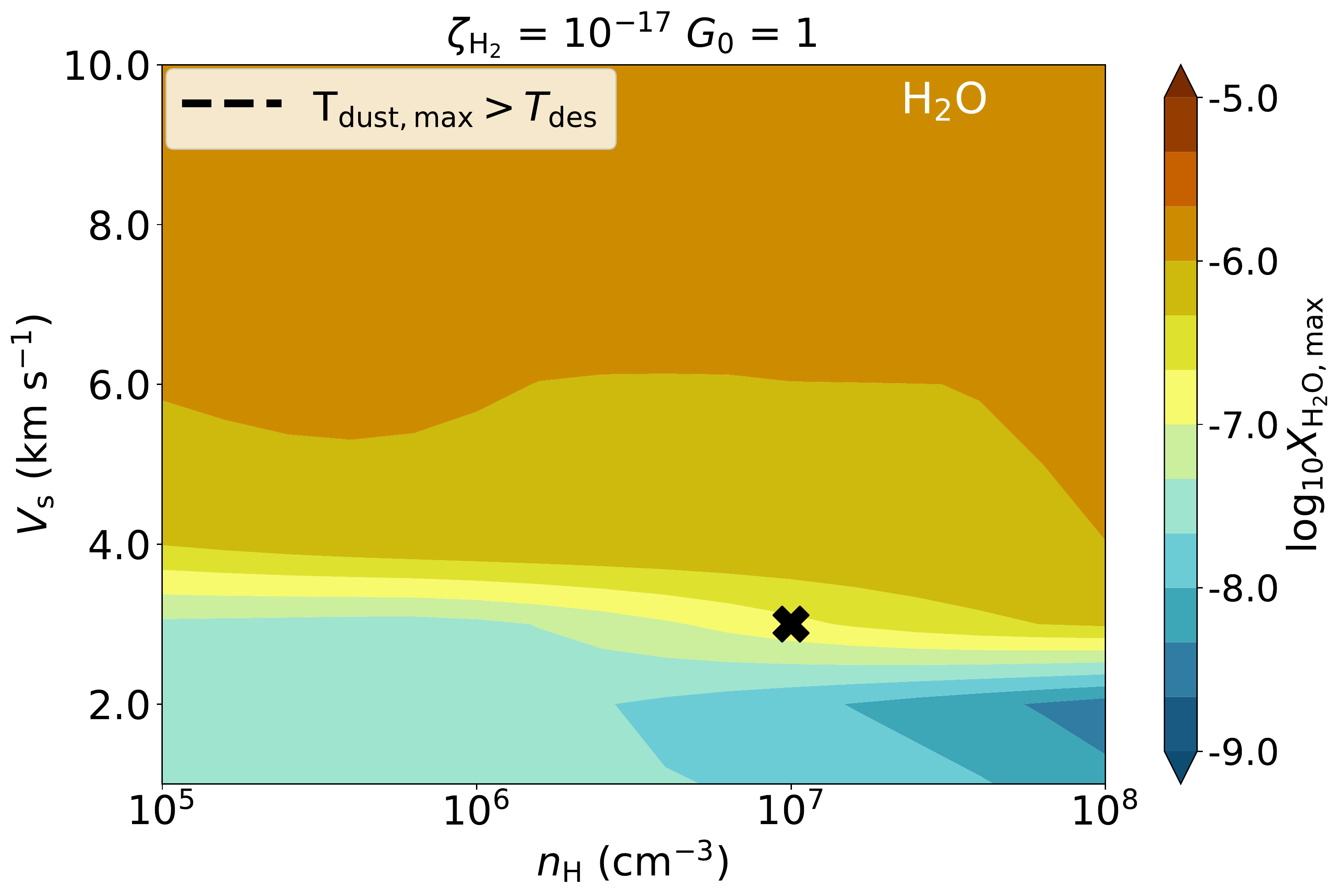}
\includegraphics[width=0.49\linewidth]{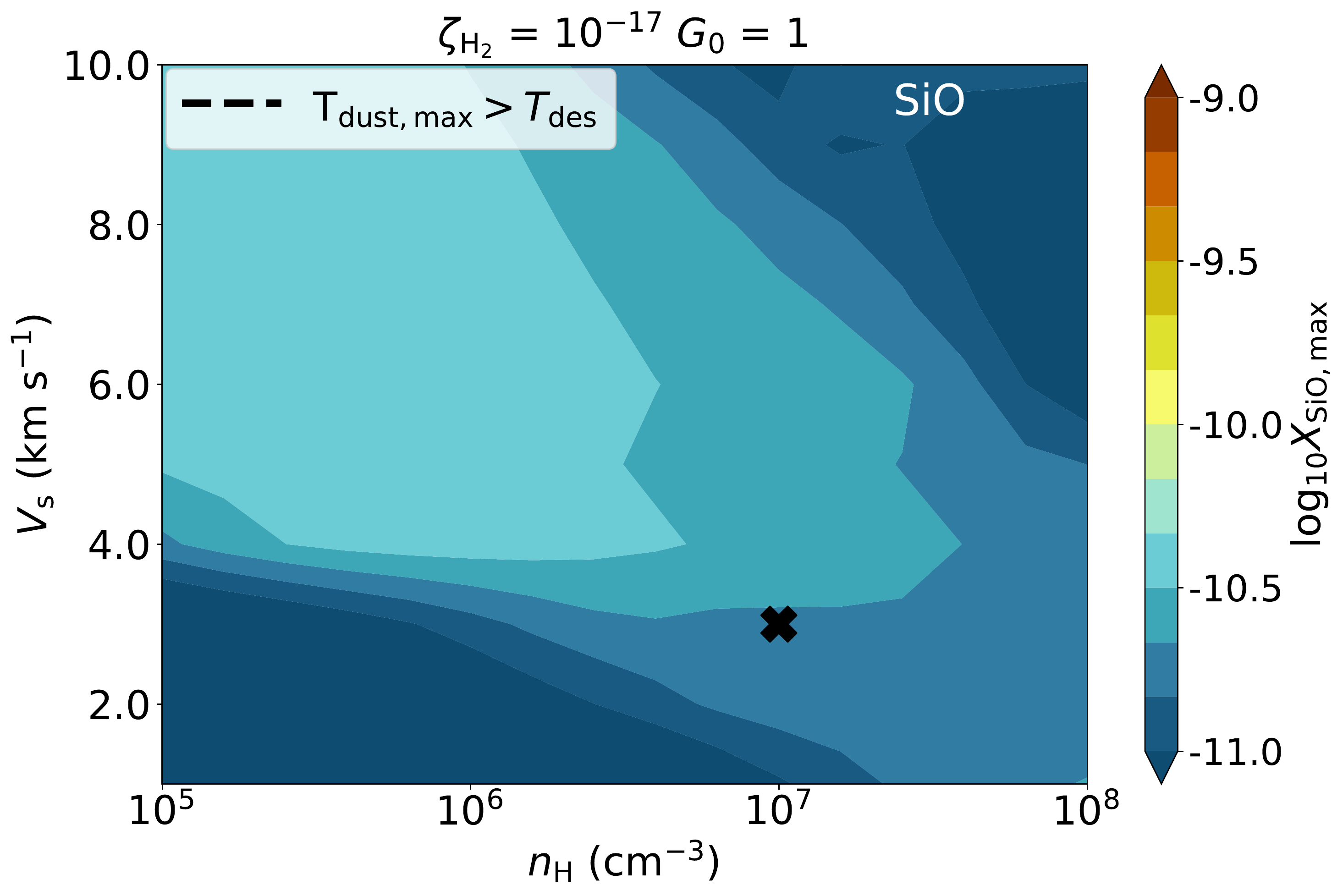}
\includegraphics[width=0.49\linewidth]{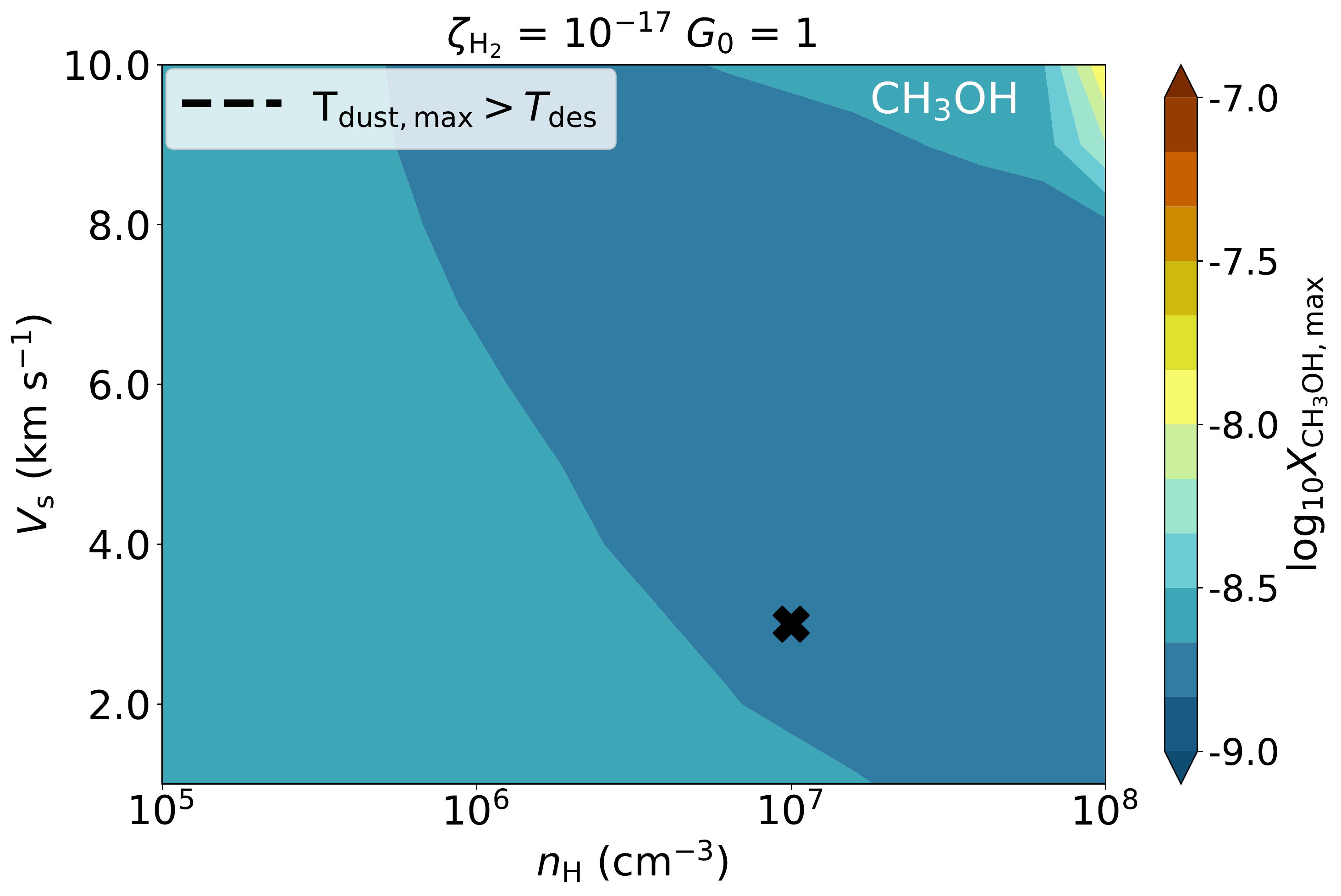}
\caption{Similar figure as Fig.~\ref{fig:Vs_nH_SO_SO2}, but now for H$_2$S (top left), H$_2$O (top right), SiO (bottom left), and CH$_3$OH (bottom right). All other physical parameters are kept constant to the fiducial values and listed on top of the figure. The black cross indicates the position of the fiducial model. The dashed black line shows the ice line, i.e., where 50\% of the ice is thermally desorbed into the gas in the shock. 
%The region where the total length of the shock is $>100$~AU is highlighted with hashed region.
}
\label{fig:Vs_nH_H2S_H2O_SiO_H2CO}
\end{figure*}

In Fig.~\ref{fig:Vs_nH_H2S_H2O_SiO_H2CO}, the maximum abundance for H$_2$S, H$_2$O, SiO, and CH$_3$OH are presented as function of initial \nH\ and \Vs. All other parameters are set to their fiducial value (e.g., $G_0 = 1$). 
%\begin{figure*}
%\includegraphics[width=0.49\linewidth]{{x(SiO)_x(H2O)_x(H2S)_nHVAR_dz1e-17_G01e-02_bB1e+00_Vs5.0}.pdf}
%\includegraphics[width=0.49\linewidth]{{x(SiO)_x(H2O)_x(H2S)_nH1e+07_dz1e-17_G0VAR_bB1e+00_Vs5.0}.pdf}
%\includegraphics[width=0.49\linewidth]{{x(SiO)_x(H2O)_x(H2S)_nH1e+07_dz1e-17_G01e-02_bBVAR_Vs5.0}.pdf}
%\includegraphics[width=0.49\linewidth]{{x(SiO)_x(H2O)_x(H2S)_nH1e+07_dz1e-17_G01e-02_bB1e+00_VsVAR}.pdf}
%\caption{Abundance profiles of gas-phase SiO, H$_2$O, and H$_2$S in the shock for different \nH\ (top left), \G0\ (top right), \bB\ (bottom left), and \Vs\ (bottom right). The fiducial is indicated in orange in all images. The $t=0$ point is set to the start of the shock. The color shaded areas indicate where $T_\mathrm{gas} > 100$~K. The vertical lines mark the end of the shock. \mvg{For now, jumps are C* and $CJ$-shocks not taken into account correctly. Have to check.}}
%\label{fig:SiO_H2O_H2Sprofiles}
%\end{figure*}

\subsection{SO and SO$_2$ for various $G_0$}
\label{app:SO_SO2_G0}
\begin{figure*}
\includegraphics[width=0.49\linewidth]{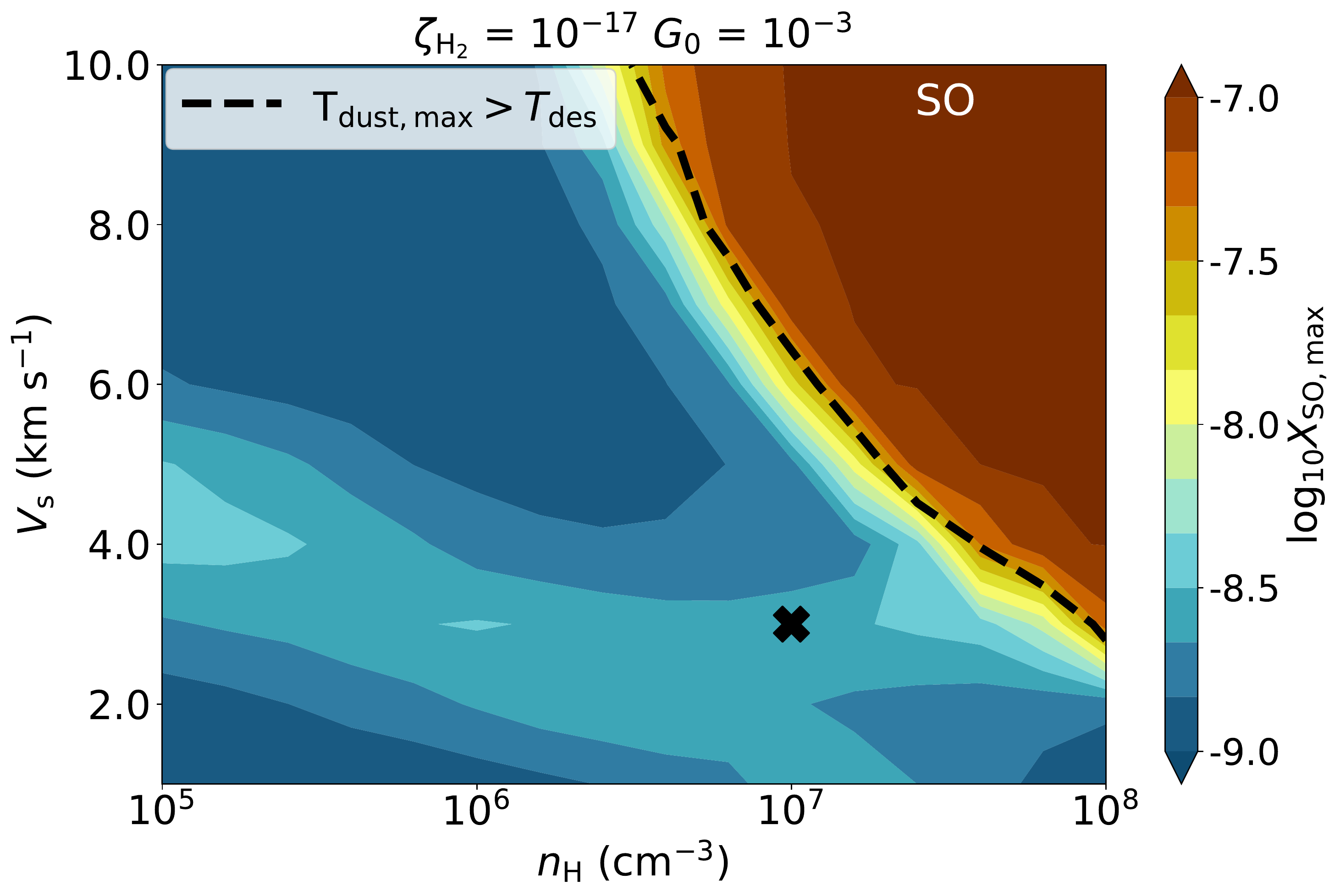}
\includegraphics[width=0.49\linewidth]{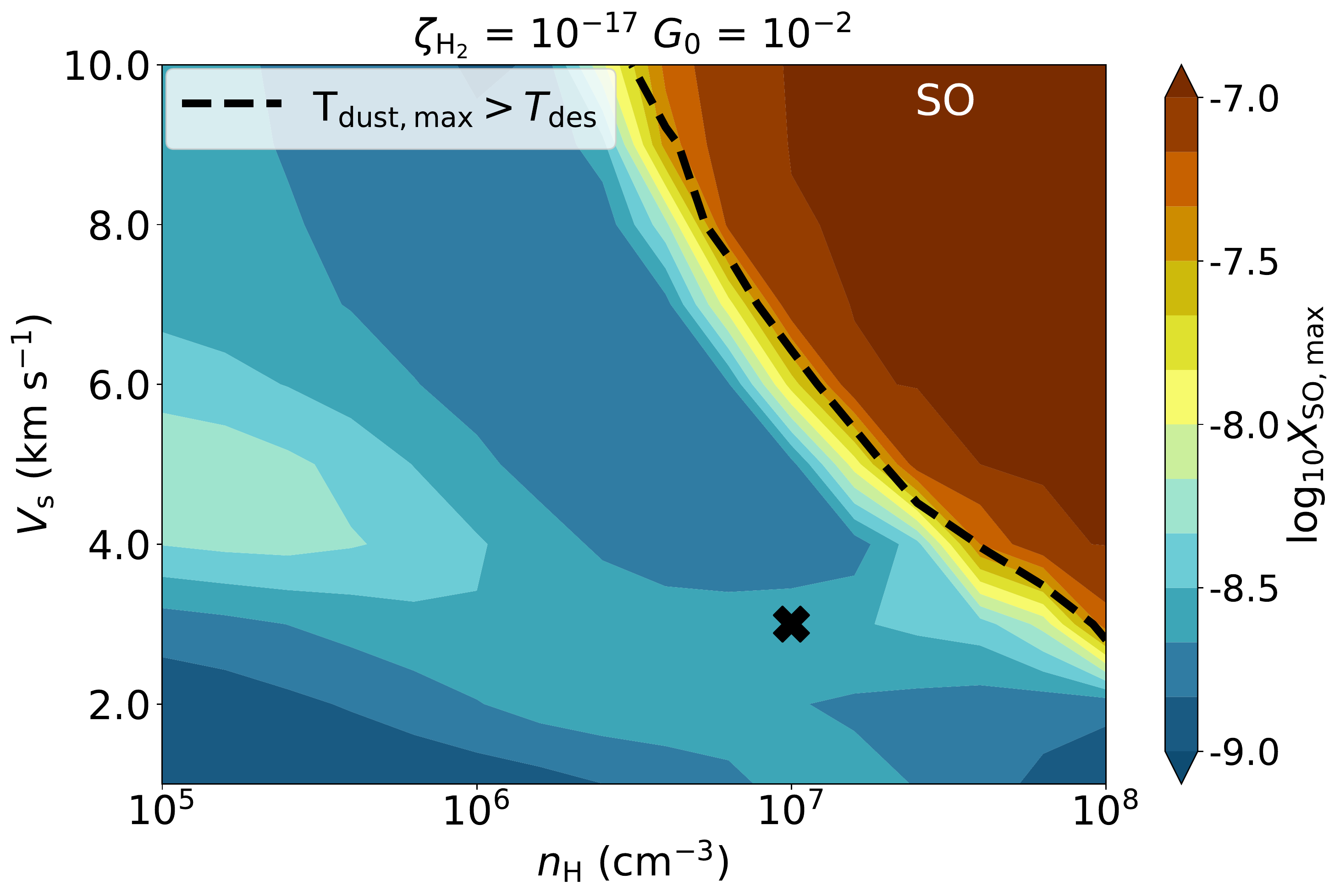}
\includegraphics[width=0.49\linewidth]{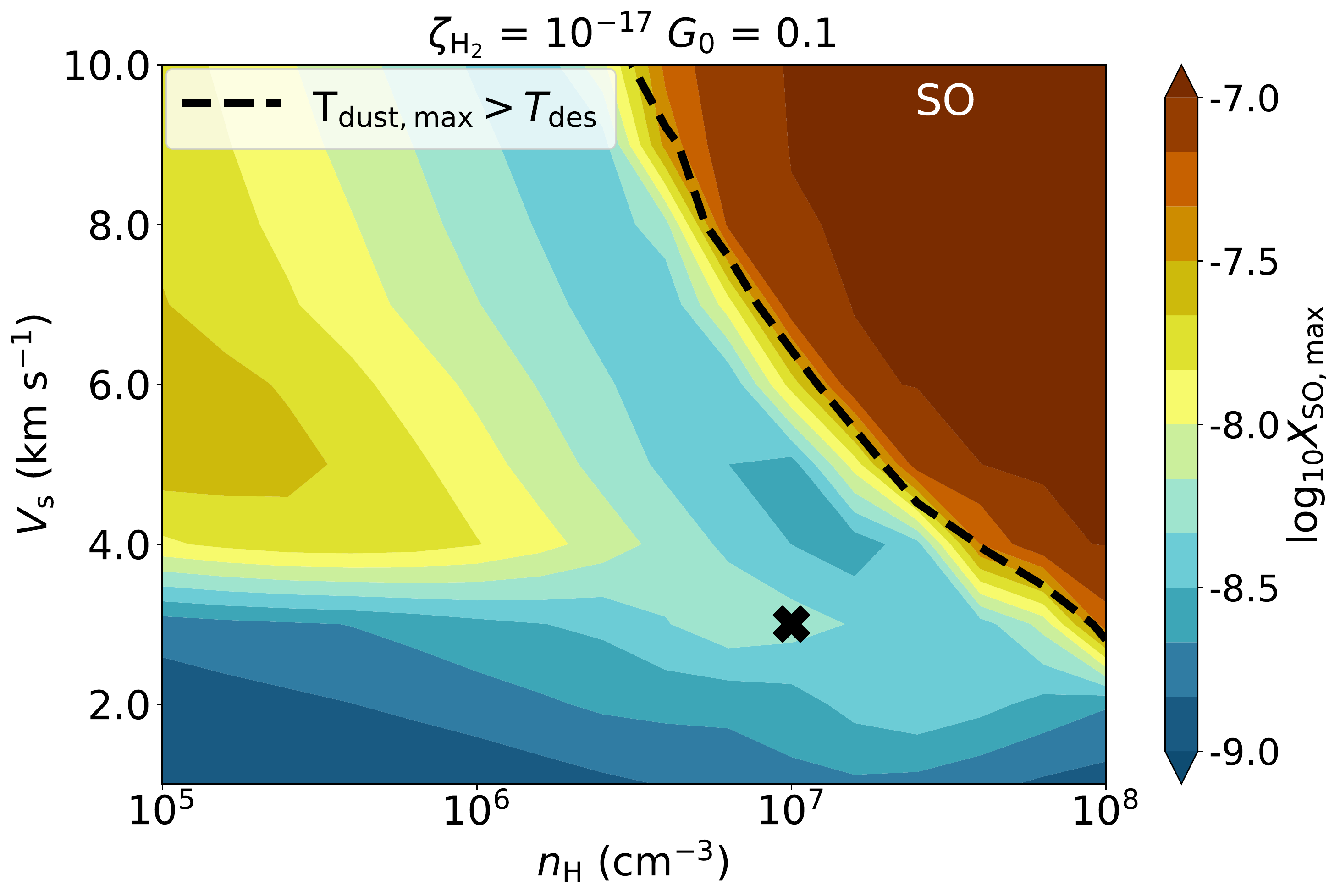}
\includegraphics[width=0.49\linewidth]{{SO_abun_species_max_Vs_vs_nH_dz1e-17_G01e+00_bB0e+00}.pdf}
\includegraphics[width=0.49\linewidth]{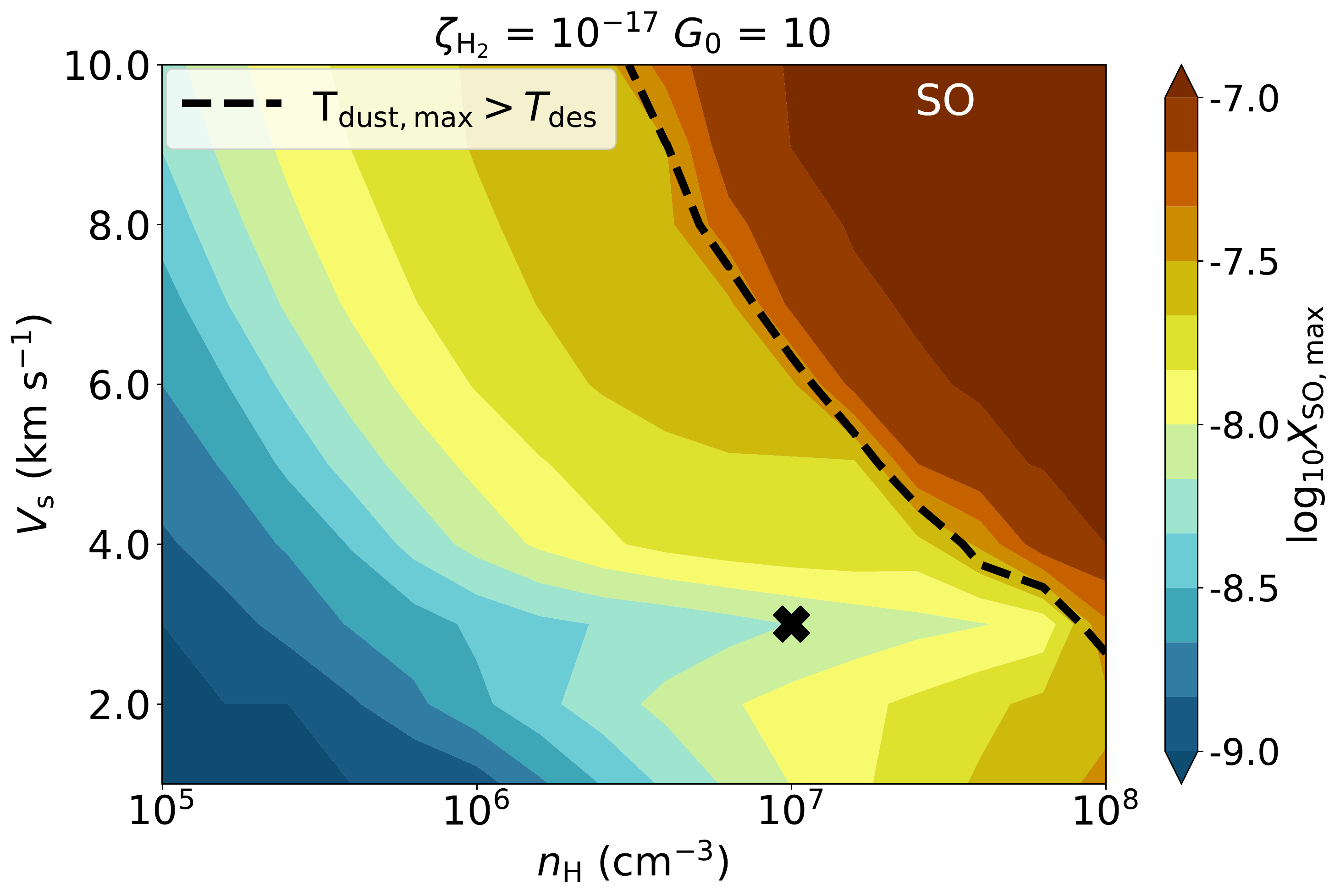}
\includegraphics[width=0.49\linewidth]{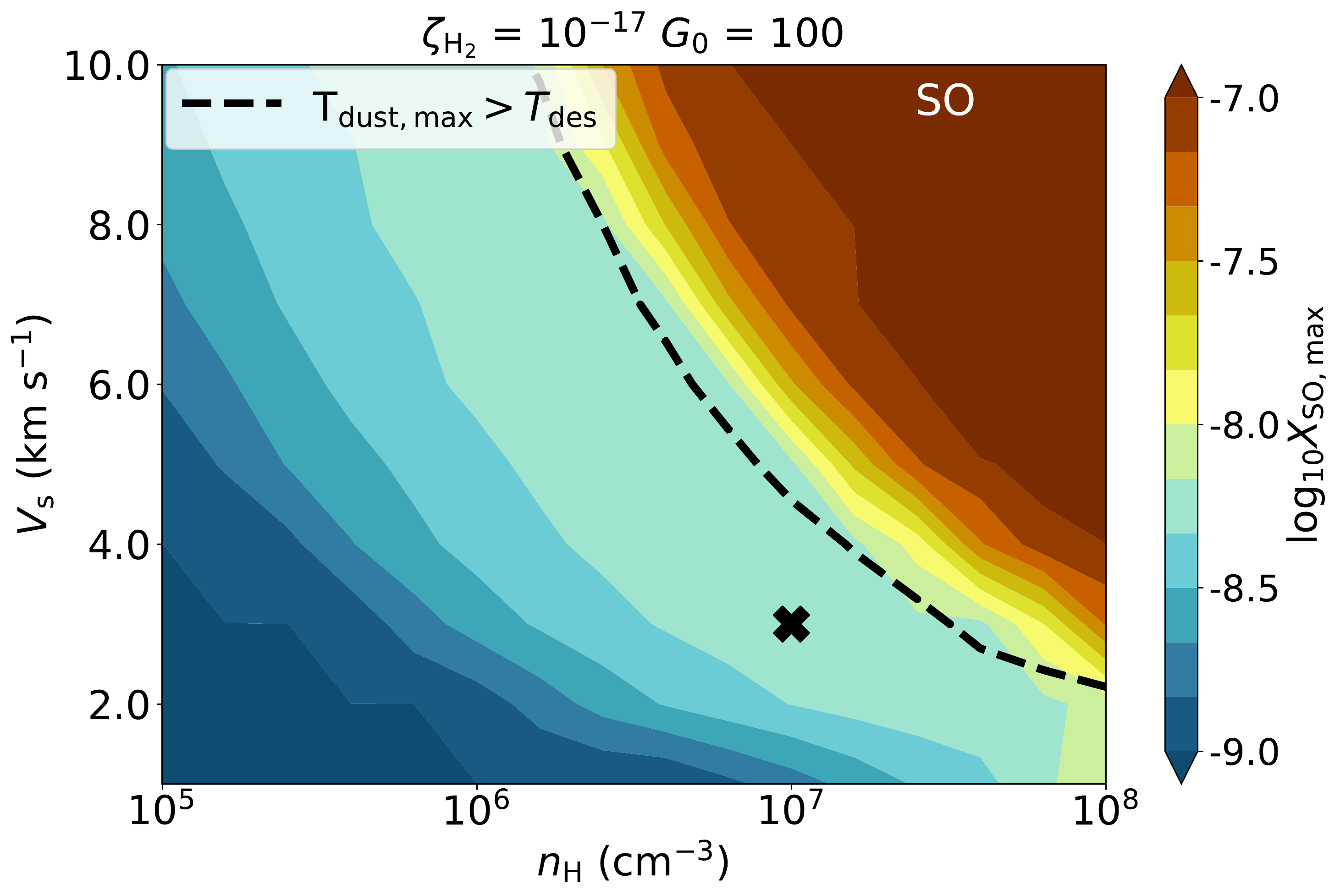}
\caption{Maximum abundance reached (in color) of SO in shock models as function of initial \nH\ and \Vs\ for $G_0 = 10^{-3}$ (top left), $G_0 = 10^{-2}$ (top right), $G_0 = 0.1$ (middle left), $G_0 = 1$ (middle right), $G_0 = 10$ (bottom left), and $G_0 = 100$ (bottom right). The black cross indicates the position of the fiducial model. The dashed black line shows the ice line, i.e., where 50\% of the ice is thermally desorbed into the gas in the shock. 
}
\label{fig:Vs_nH_SO_diffG0}
\end{figure*}

\begin{figure*}
\includegraphics[width=0.49\linewidth]{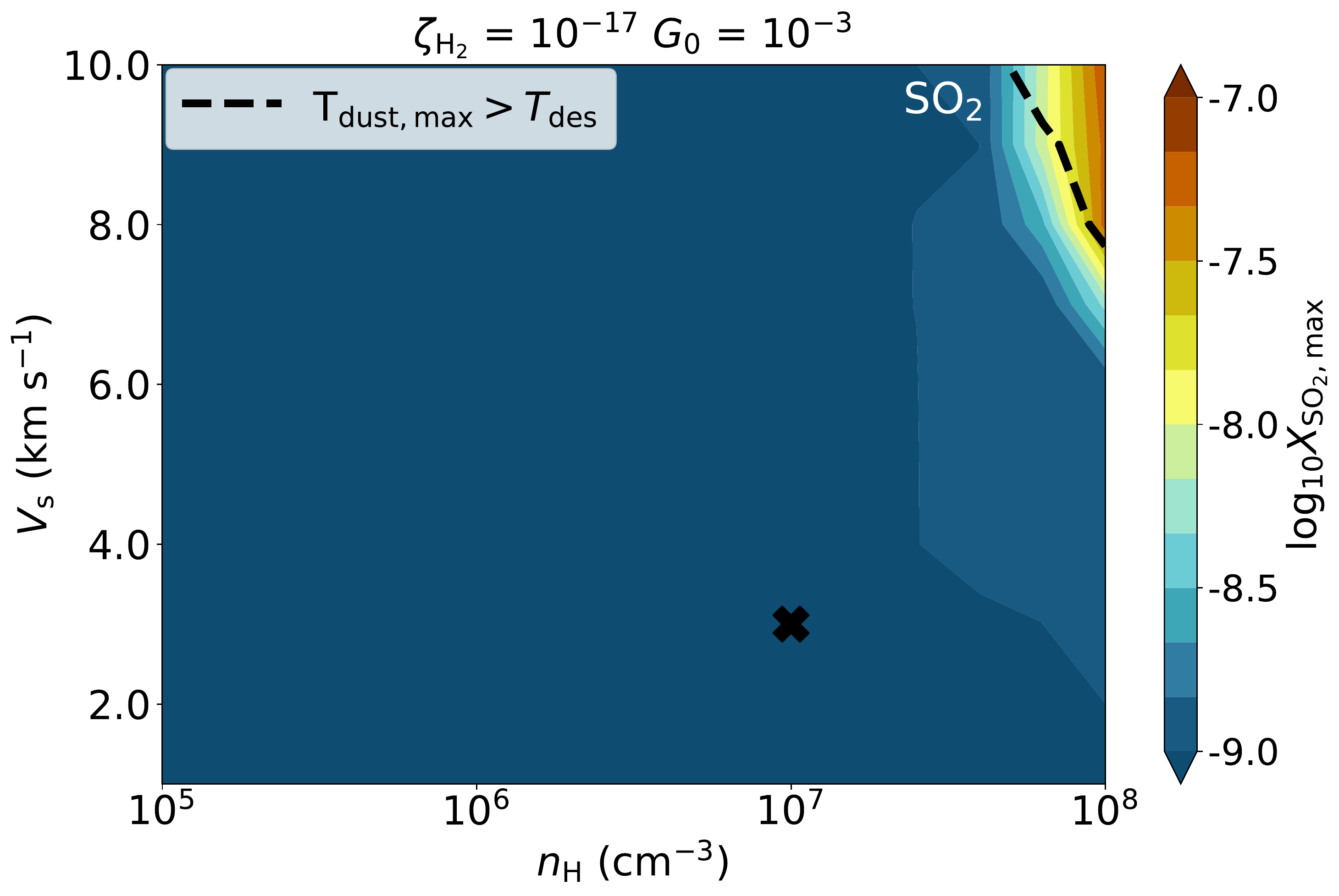}
\includegraphics[width=0.49\linewidth]{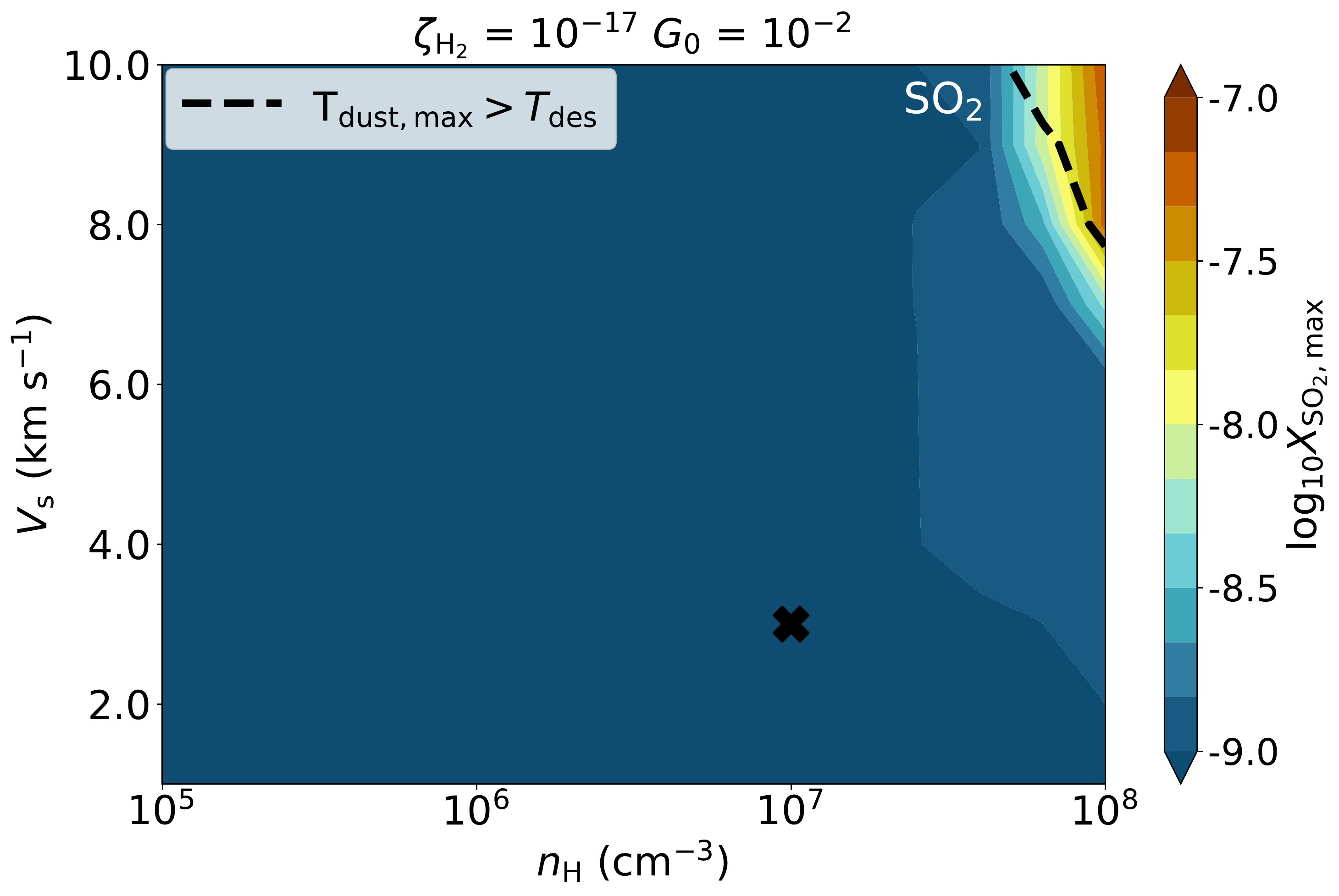}
\includegraphics[width=0.49\linewidth]{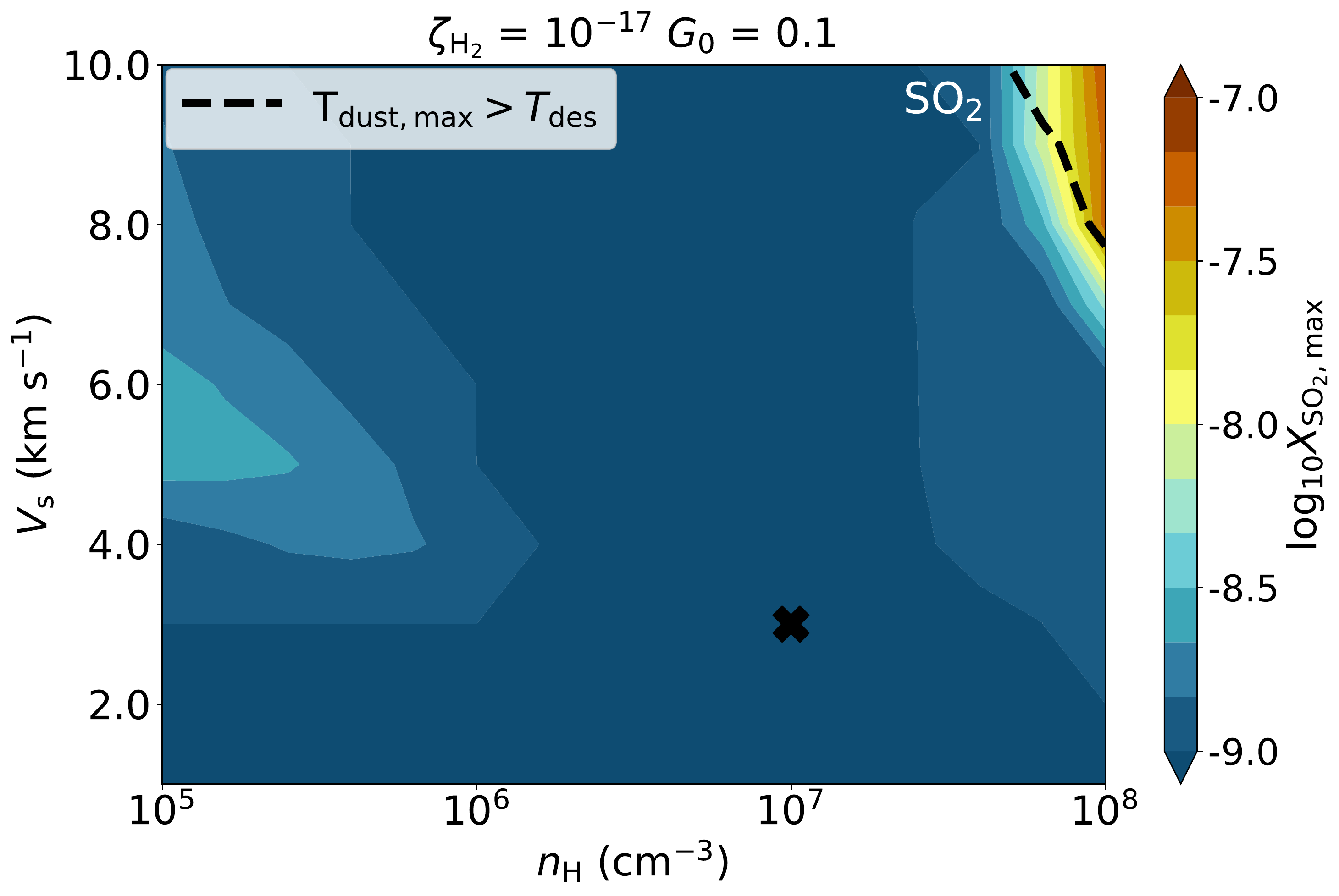}
\includegraphics[width=0.49\linewidth]{{SO2_abun_species_max_Vs_vs_nH_dz1e-17_G01e+00_bB0e+00}.pdf}
\includegraphics[width=0.49\linewidth]{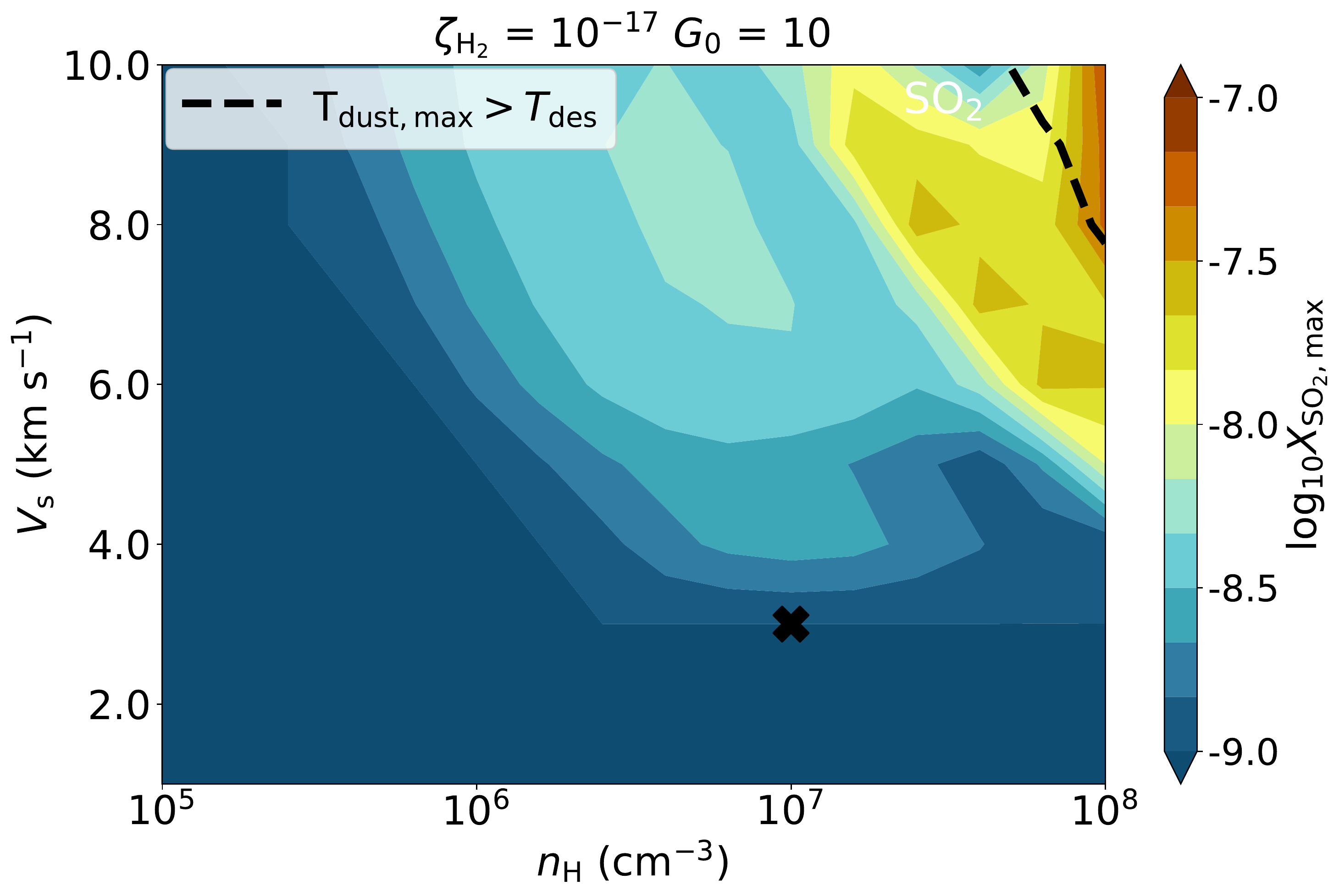}
\includegraphics[width=0.49\linewidth]{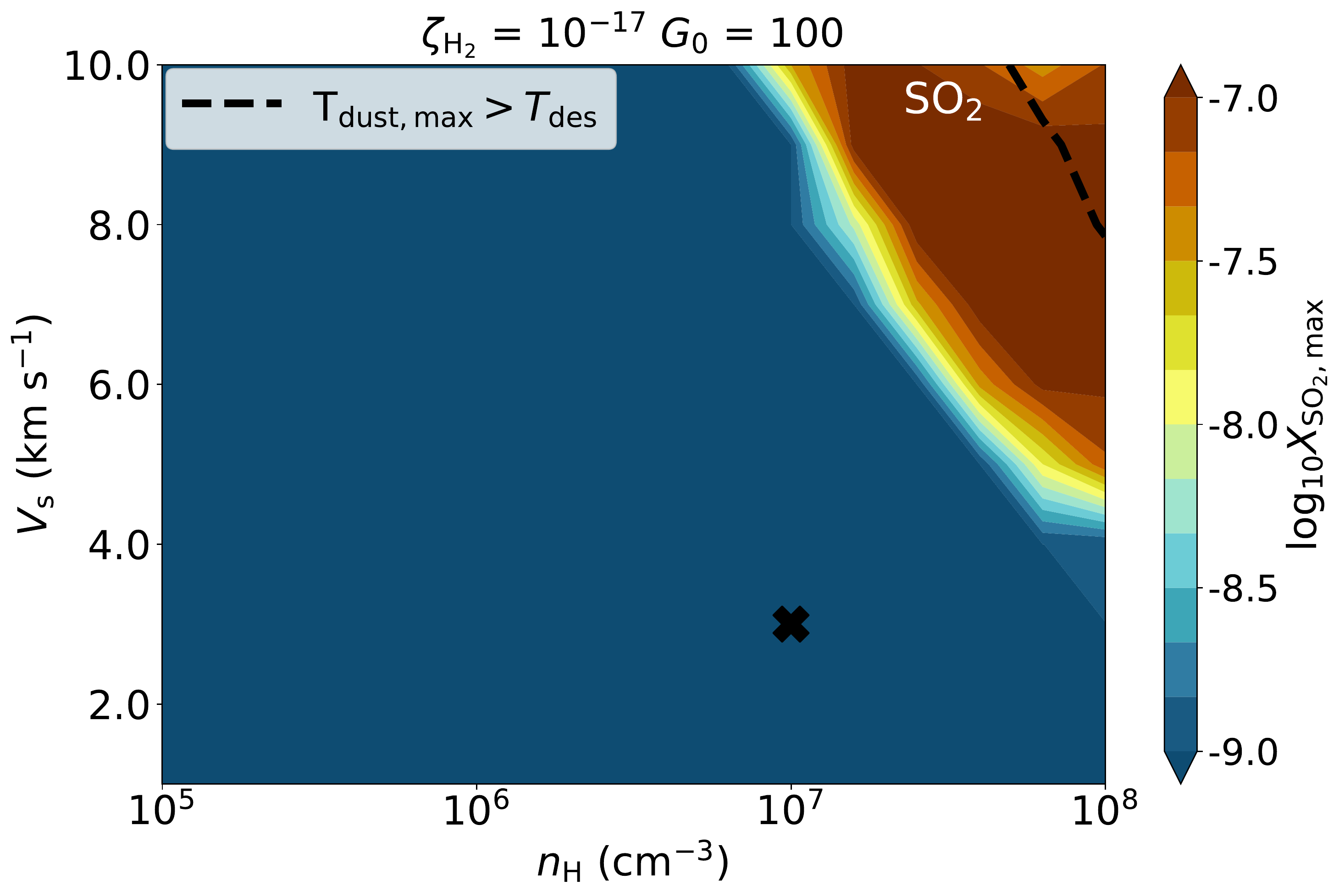}
\caption{Maximum abundance reached (in color) of SO$_2$ in shock models as function of initial \nH\ and \Vs\ for $G_0 = 10^{-3}$ (top left), $G_0 = 10^{-2}$ (top right), $G_0 = 0.1$ (middle left), $G_0 = 1$ (middle right), $G_0 = 10$ (bottom left), and $G_0 = 100$ (bottom right). The black cross indicates the position of the fiducial model. The dashed black line shows the ice line, i.e., where 50\% of the ice is thermally desorbed into the gas in the shock. 
}
\label{fig:Vs_nH_SO2_diffG0}
\end{figure*}

In Figs.~\ref{fig:Vs_nH_SO_diffG0} and \ref{fig:Vs_nH_SO2_diffG0}, the maximum abundance for respectively SO and SO$_2$ are presented as function of initial \nH\ and \Vs\ for various strengths of th UV radiation field (i.e., $G_0 = 10^{-3} - 10^{2}$).

%\subsection{Column densities of SO and SO$_2$}
%\begin{figure*}
%\includegraphics[width=0.49\linewidth]{{SO_Nx_Vs_vs_nH_dz1e-17_G01e+00_bB1e-04}.pdf}
%\includegraphics[width=0.49\linewidth]{{SO2_Nx_Vs_vs_nH_dz1e-17_G01e+00_bB1e-04}.pdf}
%\caption{
%The column density of SO (left) and SO$_2$ (right) in shock models as function of initial \nH\ and \Vs. All other physical parameters are kept constant to the fiducial values and listed on top of the figure. Only the region where $T_\mathrm{gas}>50$~K is included. The black cross indicates the position of the fiducial model. The dashed black line shows the ice line, i.e., where 50\% of the ice is thermally desorbed into the gas in the shock. 
%}
%\label{fig:Vs_nH_SO_SO2_column}
%\end{figure*} 

\section{Higher magnetized environments}
\label{app:mag_shock}
The focus of this paper is non-magnetized $J$-type shocks. However, in reality, magnetic fields are present in protostellar envelopes where observations hint at intermediate or strong magnetic field strengths \citep[i.e., $\sim0.1-1$~mG;][]{Hull2017}. The structure of shocks depends strongly on the strength of the magnetic field. In weakly magnetized regions, the emerging shock will remain a magnetized $J$-type shock with a similar structure to non-magnetized $J$-type shocks. However, in higher magnetized environments, the ions and electrons can decouple from the neutral species through the Lorentz force. While slowing down through the Lorentz force, the ions exert a drag force on the neutral fluid. Both fluids slow down and compress, resulting in an increase of the temperature and the emergence of a $C$-type (strong magnetic field strength) and $CJ$-type (intermediate magnetic field strength) shocks \citep{Draine1980,Flower2003,Godard2019}. Below, the results of $C$ and $CJ$-type accretion shocks are shortly discussed, where in both cases a pre-shock model was calculated over a timescale $t_\mathrm{pre-shock}=100$~years (see Appendix~\ref{app:pre-shock_mod} for details).

\subsection{$C$-type shocks}
In interstellar $C$-shocks, the mass of the ionized fluid is dominated by small dust grains (which couple well to ions) and ionized PAHs \citep[e.g.,][]{Godard2019}. However, since dust grows to larger grains (here assumed at 0.2~$\mu$m) in protostellar evelopes, it will likely be better coupled to the neutral fluid \citep{Guillet2007,Guillet2020}. Moreover, PAHs are lacking in the gas, likely frozen out onto dust grains \citep{Geers2009}. Hence, the mass of the ion fluid exerting a drag force on the neutrals is determined solely by small ions such as S$^+$, C$^+$, HCO$^+$, and N$_2$H$^+$. Having more small ions (e.g. due to a stronger UV radiation field) therefore strongly affects the structure of the $C$-type shock.

Since the decoupling between ions and neutrals is very slow, the length of $C$-type shocks is on envelope scales (i.e., $\gtrsim1000$~AU) for most of the parameter space considered (see Table~\ref{tab:phys_params}). Only for high velocity ($>8$~\kms) $C$-shocks at high densities ($>10^7$~cm$^{-3}$) the total length of the shock is on disk scales of $<100$~AU. For such shocks, the abundance of SO is slightly increased through gas-phase chemistry following the green route of Fig.~\ref{fig:S_chem_shock}, while the abundance of SO$_2$ is not increased in these $C$-type shocks. 

Contrary to $J$-type shocks, an increased PAHs abundance also affects the physical structure of a $C$-type shock as ionized PAHs carry most of the mass of the ion fluid if dust grains are assumed to be coupled to the neutrals \citep{Flower2003}. A higher abundance of PAHs results in a stronger decoupling between ions and neutrals and consequently a decrease in the shock length. Hence, in a larger part of the parameter space, $C$-type shocks on disk scales ($<100$~AU) are produced. However, given the lack of PAH emission from Class~0 and I sources, a high PAH abundance is not realistic.
%However, the effect of PAHs on $C$-type shocks in not discussed further in this work.

In reality, any small grains that are present in protostellar envelopes could be coupled to the ions \citep{Guillet2007,Godard2019}. 
%This increases the drag that the ions exert on the neutrals and therefore decreases the length of the shock. 
In turn, this also results in a larger fraction of the parameter space having a shock length of $<100$~AU. However, the full treatment of grain dynamics in $C$-type shocks would have to include a separate fluid for each grain size \citep{Ciolek2002,Guillet2007} with grain-grain interactions such as vaporization, shattering, and coagulation \citep{Guillet2009,Guillet2011}. This is beyond the scope of this work. 

\begin{figure*}
\includegraphics[width=0.49\linewidth]{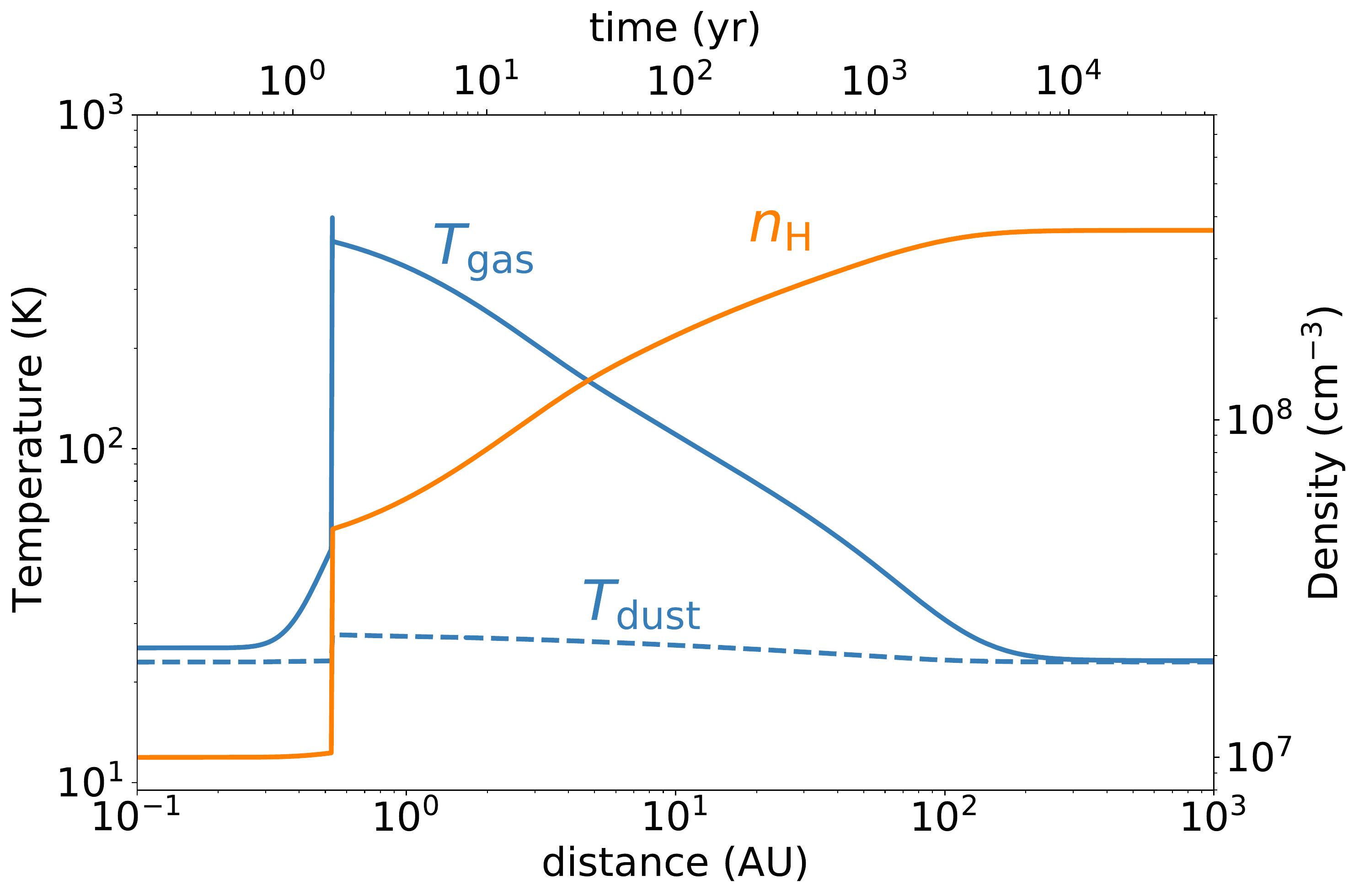}
\includegraphics[width=0.49\linewidth]{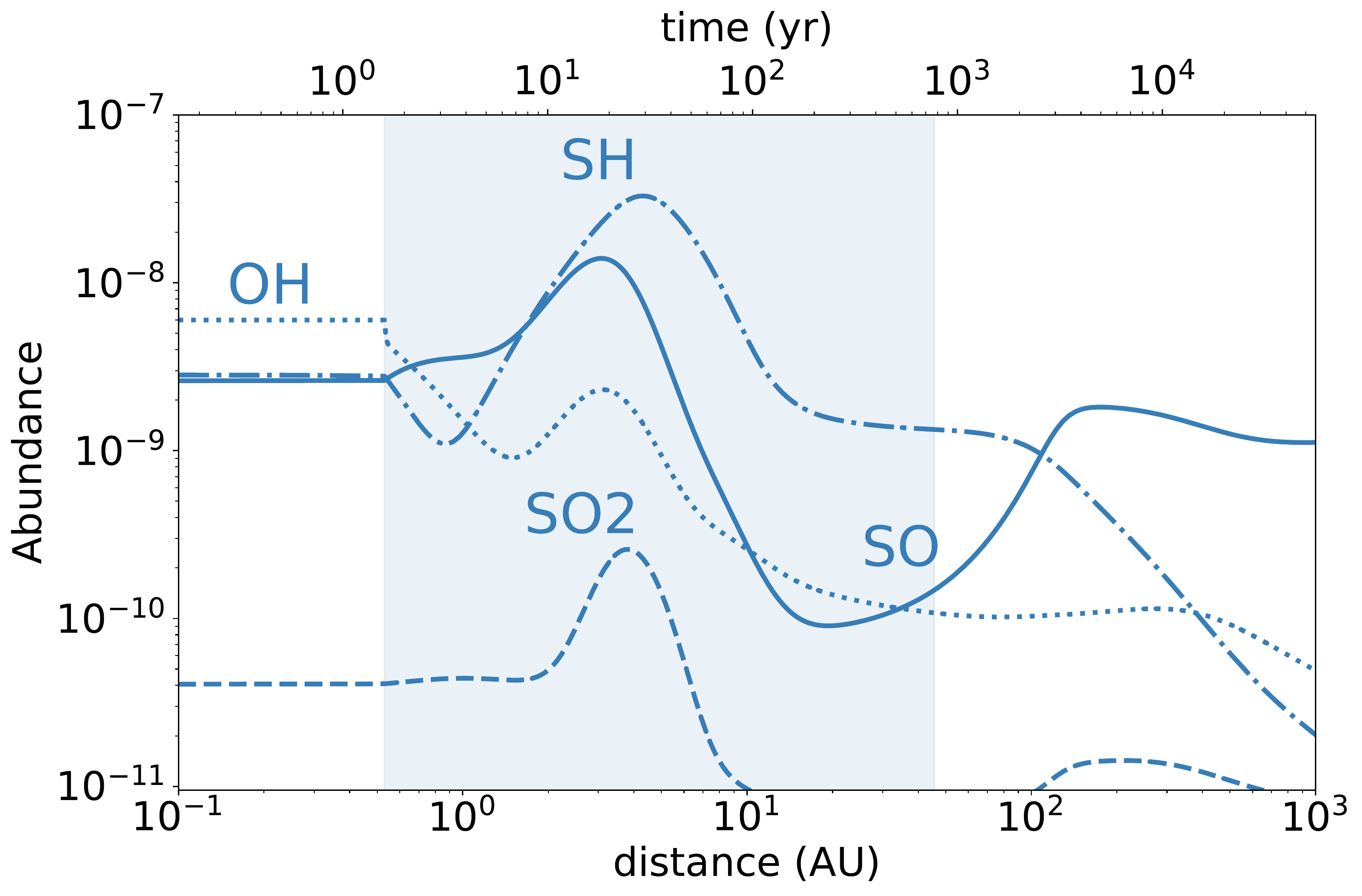}
\caption{
Same as Fig.~\ref{fig:fidu_shock_T_SO_SO2} but for a $CJ$-type shock with $B \sim 160$~$\mu$G. All other parameters are kept at their fiducial value (see Table~\ref{tab:phys_params}). 
%The shock starts at 0~AU and ends at about 50~AU. 
In the right panel, the region where $T_\mathrm{gas} > 50$~K is indicated with the shaded blue region.
}
\label{fig:CJ_shock_T_SO_SO2}
\end{figure*}

\subsection{$CJ$-type shocks}
With decreasing magnetic field strength, the decoupling between ions and neutrals becomes weaker. The motion of the neutral velocity is no longer completely dominated by the ion-neutral drag, but also by the viscous stresses and thermal pressure gradient \citep{Godard2019}. The resulting $CJ$-shock structure is presented in left part of Fig.~\ref{fig:CJ_shock_T_SO_SO2} for $B \sim 160$~$\mu$G. All other parameters are kept at their fiducial value (see Table~\ref{tab:phys_params}). Initially, the drag of the ions slows down the neutrals, increasing the gas temperature to $\sim50$~K. However, here the viscous stresses and thermal pressure take over, resulting in a jump in temperature and density as the neutral fluid jumps across the sonic point \citep{Godard2019}. Afterwards, the ions keep exerting a small drag force on the neutrals until the fluids recouple at about $\sim80$~AU. Since the heating induced by this drag is lower than the cooling of the gas, the shock gradually cools down. The abundances of SO and SO$_2$ do show an increase in the $CJ$-shock, see right panel of Fig.~\ref{fig:CJ_shock_T_SO_SO2}. Here, SO is predominantly formed through Reaction~\eqref{reac:SH_O} with SH coming from the green route of Fig.~\ref{fig:S_chem_shock}.

Given that the strength of the magnetic field on inner envelope scales might be high enough to create $CJ$-type shocks \citep{Hull2017}, these type of shocks are very relevant. 
However, in order to compute $CJ$-type shocks over the entire parameter space considered in this paper, the Paris-Durham shock code has to be tweaked to our specific parameter space. This is beyond the scope of this work.
%However, as the neutrals jump over the sonic point, the method used to compute the $CJ$-shock becomes numerical unstable \citep{Godard2019}, often resulting in the code not converging toward a solution. Therefore $CJ$-type shocks are not discussed further in detail. 
Nevertheless, as shown in Fig.~\ref{fig:CJ_shock_T_SO_SO2}, the chemical formation routes of Fig.~\ref{fig:S_chem_shock} are still valid for $CJ$-type shocks since similar physical conditions are attained.

\end{appendix}

\end{document}